\documentclass[10pt,a4paper]{article}
\usepackage[T1]{fontenc}
\usepackage[utf8]{inputenc}
\usepackage[ngerman,english]{babel}
\usepackage{cite}
\usepackage{amsmath}
\usepackage{amsfonts}
\usepackage{amssymb}
\usepackage{amsthm}
\usepackage{enumitem}
\usepackage{dcolumn}
\usepackage{bm}
\usepackage{bbm}
\usepackage{cancel}
\usepackage[pdftex]{graphicx}
\usepackage{tikz}
\usetikzlibrary{arrows}
\usetikzlibrary{patterns}
\usepackage{tabularx}
\usepackage{booktabs}
\usepackage{tensor}
\usepackage[small]{caption}
\usepackage{xcolor}
\usepackage{mathtools}
\usepackage{float}
\usepackage[pdfborder={0 0 0}]{hyperref}
\definecolor{darkblue}{rgb}{0,0,.5}
\hypersetup{colorlinks=true, breaklinks=true, linkcolor=darkblue, menucolor=darkblue, urlcolor=darkblue, linktocpage=true}
\usepackage{times}
\usepackage{geometry}
\makeatletter 
\newsavebox\myboxA 
\newsavebox\myboxB 
\newlength\mylenA 
 
\newcommand*\xoverline[2][0.75]{% 
    \sbox{\myboxA}{$\m@th#2$}%
    \setbox\myboxB\null% Phantom box 
    \ht\myboxB=\ht\myboxA% 
    \dp\myboxB=\dp\myboxA% 
    \wd\myboxB=#1\wd\myboxA% Scale phantom 
    \sbox\myboxB{$\m@th\overline{\copy\myboxB}$}%  Overlined phantom 
    \setlength\mylenA{\the\wd\myboxA}%   calc width diff 
    \addtolength\mylenA{-\the\wd\myboxB}% 
    \ifdim\wd\myboxB<\wd\myboxA% 
       \rlap{\hskip 0.5\mylenA\usebox\myboxB}{\usebox\myboxA}% 
    \else 
        \hskip -0.5\mylenA\rlap{\usebox\myboxA}{\hskip 0.5\mylenA\usebox\myboxB}% 
    \fi} 
\makeatother
\numberwithin{equation}{section}

\newcommand{\e}{\operatorname{e}}

\newcommand{\SO}[1]{\operatorname{SO}\left(#1\right)}

\newcommand{\Sph}[1]{\operatorname{S}^{#1}}
\newcommand{\of}[1]{\left(#1\right)}
\newcommand{\fof}[1]{\left[#1\right]}
\newcommand{\cof}[1]{\left\{#1\right\}}

\newcommand{\avof}[1]{\left\langle #1\right\rangle}

\newcommand{\eqb}{\begin{equation}}
\newcommand{\eqe}{\end{equation}}

\newcommand{\idd}[2]{\mathrm{d}^{#2}\,#1}

\newcommand{\DD}[1]{\mathcal{D}\left[#1\right]}

\newcommand{\partd}[2]{\frac{\partial #1}{\partial #2}}
\newcommand{\partdm}[3]{\frac{\partial^{#3} #1}{\partial #2^{#3}}}

\newcommand{\ford}{1^{\text{st}}\,\text{order}}
\newcommand{\sord}{2^{\text{nd}}\,\text{order}}

\newcommand{\nord}[1]{#1^{\text{th}}\,\text{order}}

\newcommand{\abs}[1]{\left|#1\right|}
\newcommand{\op}[1]{\operatorname{#1}}

\renewcommand*\[{\begin{equation}}
\renewcommand*\]{\end{equation}}
\renewcommand*\bar[1]{\xoverline{#1}}

\let\oldstackrel\stackrel
\renewcommand*\stackrel[2]{{\scriptstyle\oldstackrel{#1}{#2}}}

\definecolor{emphcol}{RGB}{0,0,0}
\let\oldemph\emph
\renewcommand*\emph[1]{\oldemph{\textcolor{emphcol}{#1}}}
\newlength{\hatchspread}
\newlength{\hatchthickness}
\newlength{\hatchshift}
\newcommand{\hatchcolor}{}
\tikzset{hatchspread/.code={\setlength{\hatchspread}{#1}},
         hatchthickness/.code={\setlength{\hatchthickness}{#1}},
         hatchshift/.code={\setlength{\hatchshift}{#1}},% must be >= 0
         hatchcolor/.code={\renewcommand{\hatchcolor}{#1}}}
\tikzset{hatchspread=3pt,
         hatchthickness=0.4pt,
         hatchshift=0pt,% must be >= 0
         hatchcolor=black}
\pgfdeclarepatternformonly[\hatchspread,\hatchthickness,\hatchshift,\hatchcolor]% variables
   {custom north west lines}% name
   {\pgfqpoint{\dimexpr-2\hatchthickness}{\dimexpr-2\hatchthickness}}% lower left corner
   {\pgfqpoint{\dimexpr\hatchspread+2\hatchthickness}{\dimexpr\hatchspread+2\hatchthickness}}% upper right corner
   {\pgfqpoint{\dimexpr\hatchspread}{\dimexpr\hatchspread}}% tile size
   {% shape description
    \pgfsetlinewidth{\hatchthickness}
    \pgfpathmoveto{\pgfqpoint{\dimexpr0pt}{\dimexpr\hatchspread+\hatchshift}}
    \pgfpathlineto{\pgfqpoint{\dimexpr\hatchspread+0.15pt}{\dimexpr\hatchshift-0.15pt}}
    \ifdim \hatchshift > 0pt
      \pgfpathmoveto{\pgfqpoint{\dimexpr-0.15pt}{\dimexpr\hatchshift+0.15pt}}
      \pgfpathlineto{\pgfqpoint{\dimexpr\hatchshift+0.15pt}{\dimexpr-0.15pt}}
    \fi
    \pgfsetstrokecolor{\hatchcolor}
    \pgfusepath{stroke}
   }

\pgfdeclarepatternformonly[\hatchspread,\hatchthickness,\hatchshift,\hatchcolor]% variables
   {custom north east lines}% name
   {\pgfqpoint{\dimexpr-2\hatchthickness}{\dimexpr-2\hatchthickness}}% lower left corner
   {\pgfqpoint{\dimexpr\hatchspread+2\hatchthickness}{\dimexpr\hatchspread+2\hatchthickness}}% upper right corner
   {\pgfqpoint{\dimexpr\hatchspread}{\dimexpr\hatchspread}}% tile size
   {% shape description
    \pgfsetlinewidth{\hatchthickness}
    \pgfpathmoveto{\pgfqpoint{0pt}{\dimexpr\hatchshift}}
    \pgfpathlineto{\pgfqpoint{\dimexpr\hatchspread+0.15pt}{\dimexpr\hatchspread+0.15pt+\hatchshift}}
    \ifdim \hatchshift > 0pt
      \pgfpathmoveto{\pgfqpoint{\dimexpr\hatchspread-\hatchshift-0.15pt}{\dimexpr-0.15pt}}
      \pgfpathlineto{\pgfqpoint{\dimexpr\hatchspread+0.15pt}{\dimexpr\hatchshift+0.15pt}}
    \fi
    \pgfsetstrokecolor{\hatchcolor}
    \pgfusepath{stroke}
   }

\pgfdeclarepatternformonly[\hatchspread,\hatchthickness,\hatchshift,\hatchcolor]% variables
   {custom vertical lines}% name
   {\pgfqpoint{\dimexpr-2\hatchthickness}{\dimexpr-2\hatchthickness}}% lower left corner
   {\pgfqpoint{\dimexpr\hatchspread+2\hatchthickness}{\dimexpr\hatchspread+2\hatchthickness}}% upper right corner
   {\pgfqpoint{\dimexpr\hatchspread}{\dimexpr\hatchspread}}% tile size
   {% shape description
    \pgfsetlinewidth{\hatchthickness}
    \pgfpathmoveto{\pgfqpoint{\dimexpr\hatchshift}{0pt}}
    \pgfpathlineto{\pgfqpoint{\dimexpr\hatchshift}{\dimexpr\hatchspread+0.15pt}}
    \pgfsetstrokecolor{\hatchcolor}
    \pgfusepath{stroke}
   }

\pgfdeclarepatternformonly[\hatchspread,\hatchthickness,\hatchshift,\hatchcolor]% variables
   {custom horizontal lines}% name
   {\pgfqpoint{\dimexpr-2\hatchthickness}{\dimexpr-2\hatchthickness}}% lower left corner
   {\pgfqpoint{\dimexpr\hatchspread+2\hatchthickness}{\dimexpr\hatchspread+2\hatchthickness}}% upper right corner
   {\pgfqpoint{\dimexpr\hatchspread}{\dimexpr\hatchspread}}% tile size
   {% shape description
    \pgfsetlinewidth{\hatchthickness}
    \pgfpathmoveto{\pgfqpoint{0pt}{\dimexpr\hatchshift}}
    \pgfpathlineto{\pgfqpoint{\dimexpr\hatchspread+0.15pt}{\dimexpr\hatchshift}}
    \pgfsetstrokecolor{\hatchcolor}
    \pgfusepath{stroke}
   }

\begin{document}\selectlanguage{english}
\newcounter{romanPagenumber}
\newcounter{arabicPagenumber}
\pagenumbering{Roman}
\title{Euclidean Dynamical Triangulation revisited: is the phase transition really 1st order?}

\author{Tobias Rindlisbacher \thanks{E-mail: rindlisbacher@itp.phys.ethz.ch}\\
Institute for Theoretical Physics,\\ ETH Z\"urich, Wolfgang-Pauli-Str. 27, Z\"urich, CH-8093 Switzerland\\
\and\\
Philippe de Forcrand \thanks{E-mail: forcrand@itp.phys.ethz.ch}\\
Institute for Theoretical Physics,\\ ETH Z\"urich, Wolfgang-Pauli-Str. 27, Z\"urich, CH-8093 Switzerland\\
and\\
CERN, Physics Department, TH Unit,\\ CH-1211 Gen\`eve 23, Switzerland}

\maketitle

\begin{abstract}
The transition between the two phases of 4D Euclidean Dynamical Triangulation \cite{Ambjorn} was long believed to be of second order until in 1996 first order behavior was found for
sufficiently large systems \cite{Bialas,deBakker}. However, one may wonder if this finding was affected by the numerical methods used: to control volume fluctuations, in both studies \cite{Bialas,deBakker} an artificial harmonic potential was added to the action and in \cite{deBakker} measurements were taken after a fixed number of \emph{accepted} instead of \emph{attempted} moves which introduces an additional error. Finally the simulations suffer from strong critical slowing down which may have been underestimated.\\
In the present work, we address the above weaknesses: we allow the volume to fluctuate freely within a fixed interval; we take measurements after a fixed number of attempted moves; and we overcome critical slowing down by using an optimized parallel tempering algorithm \cite{Bauer}. With these improved methods, on systems of size up to $N_{4}=64$k 4-simplices, we confirm that the phase transition is $\ford$.\\
In addition, we discuss a local criterion to decide whether parts of a triangulation are in the elongated or crumpled state and describe a new correspondence between EDT and the balls in boxes model. The latter gives rise to a modified partition function with an additional, third coupling.\\
Finally, we propose and motivate a class of modified path-integral measures that might remove the metastability of the Markov chain and turn the phase transition into $\sord$.\\
\end{abstract}
        
\newpage
\setcounter{romanPagenumber}{\value{page}} 
\pagenumbering{arabic}

\section{Introduction}\label{sec:intro}
Euclidean Dynamical Triangulation (EDT) in four dimensions, as introduced below in Sec. \ref{ssec:edtmodel}, was first studied by J. Ambjorn and J. Jurkiewicz back in 1992 \cite{Ambjorn}. They found that the model possesses two phases and the transition between them initially seemed to be of $\sord$, which is necessary for a continuum limit to be defined. In 1996 then, P. Bialas, Z. Burda, A. Krzywicki and B. Petersson reported for the first time the finding of some $\ford$ behavior in this phase transition for systems consisting of $N_{4}=32$k 4-simplices \cite{Bialas}. Shortly afterwards B. V. de Bakker verified this finding and extended the study to larger systems with $N_{4}=64$k. However, we were not completely convinced by the numerical methods used in the latter work. In particular, there were three things which disturbed us:
\vspace{-7pt}
\begin{enumerate}\itemsep -2pt
\item Measurements were taken after a fixed number of \emph{accepted} (instead of attempted) moves, which introduces a systematic error.
\item The use of an artificial harmonic potential to control volume fluctuations also introduces a systematic error.
\item Autocorrelation and thermalization times could easily have been underestimated.
\end{enumerate}
\vspace{-5pt}
Therefore we wanted to cross-check these old results with our own, hopefully correct code which satisfies detailed balance, uses a potential well instead of a harmonic potential to control volume fluctuations, and makes use of \emph{parallel tempering} to cope with critical slowing down.\\

The paper is organized as follows:\\
In the remainder of this section we give a brief overview of the EDT model and its phase diagram. In section \ref{sec:simulationmethods} we describe our simulation methods while in section \ref{sec:results} we present our results:
after having verified in part \ref{sec:orderofphasetransition} that the phase transition is $\ford$, we address in part \ref{ssec:microscopicprop} the question whether we can observe a coexistence of the two phases and present therefore a local criterion to determine whether a piece of triangulation is in a crumpled or elongated state and try to identify the nature of the metastability in the Markov chain that causes the $\ford$ transition. Finally in the appendix, we propose a modification of the path-integral measure, based on a counting of the number of \emph{possible moves} in each triangulation, which could weaken the $\ford$ nature of the phase transition.\\ 

\subsection{The EDT Model}\label{ssec:edtmodel}
In 4-dimensional Euclidean Dynamical Triangulation (EDT)\cite{Ambjorn} the formal path integral for \emph{Euclidean} (local $\SO{4}$ instead of $\SO{3,1}$ symmetry) \emph{gravity},
\[
Z\,=\,\int\DD{g\indices{_\mu_\nu}} \e^{-\,S_{EH}\fof{g\indices{_\mu_\nu}}},\label{eq:epf}
\]
with the \emph{Einstein-Hilbert action} $S_{EH}$:
\[
S_{EH}\,=\,-\frac{1}{16\,\pi\,G}\,\int\idd{x}{4}\,\sqrt{g}\,\of{R\,-\,2\,\Lambda},\label{eq:eha}
\]
is regularized by approximating the configuration space (space of all diffeomorphism inequivalent 4-metrics) with the space of simplicial piecewise linear (PL) manifolds consisting of equilateral 4-simplices with fixed edge length $a$ (such manifolds are also called \emph{abstract triangulations}). Under such a discretization, \eqref{eq:eha} turns into the \emph{Einstein-Regge action} $S_{ER}$ which for equilateral 4-simplices and a space-time of topology $\Sph{4}$ takes the simple form
\[
S_{ER}\,=\,-\,\kappa_{2}\,N_{2}\,+\,\kappa_{4}\,N_{4}.\label{eq:era}
\]
$N_{i}$ labels the number of $i$-simplices in the PL manifold and $\kappa_{2}$, $\kappa_{4}$ are related to the bare gravitational and cosmological couplings $G$, $\Lambda$ by
\[
\kappa_{2}\,=\,\frac{V_{2}}{4\,G}\quad,\quad\kappa_{4}\,=\,\frac{10\,\op{arccos}\of{1/4}\,V_{2}\,+\,\Lambda\,V_{4}}{8\,\pi\,G},
\]
with $V_{n}\,=\,a^{n}\,\frac{\sqrt{n+1}}{n!\,\sqrt{2^{n}}}$ being the volume of a $n$-simplex. 
The partition function \eqref{eq:epf} can now be written as
\[
Z\of{\kappa_{2},\kappa_{4}}\,=\,\sum\limits_{T}\,\frac{1}{C_{T}}\,\e^{\kappa_{2}\,N_{2}\of{T}\,-\,\kappa_{4}\,N_{4}\of{T}}\,=\,\sum\limits_{N_{4}}\,Z\of{\kappa_{2},N_{4}}\,\e^{-\kappa_{4}\,N_{4}},\label{eq:edtpf}
\]
where after the first equality sign the sum runs over all abstract triangulations $T$ of $\Sph{4}$ and $C_{T}$ is a symmetry or degeneracy factor (to avoid over-counting) which is assumed to be $\sim\,1$ for sufficiently large systems. After the second equality sign, the \emph{canonical partition function}
\[
Z\of{\kappa_{2},N_{4}}\,=\,\sum\limits_{\cof{T:N_{4}\of{T}=N_{4}}}\,\frac{1}{C_{T}}\,\e^{\kappa_{2}\,N_{2}\of{T}}\label{eq:edtcpf}
\]
was used.\\
The partition function \eqref{eq:edtpf} is suitable for use in a Markov chain Monte Carlo simulation with Metropolis updates consisting of the so-called \emph{Pachner moves} (see Sec. \ref{sec:pachnermoves}).  

\subsection{Phase Diagram}
The grand canonical partition function \eqref{eq:edtpf} is finite only if $\kappa_{4}\,>\,\kappa_{4}^{cr}\of{\kappa_{2}}$. We therefore have a critical line for convergence in the
$\of{\kappa_{2},\kappa_{4}}$-plane, given by $\kappa_{4}^{cr}\of{\kappa_{2}}$. To obtain the thermodynamic limit $\of{N_{4}\rightarrow\infty}$ we have to ensure that $\kappa_{4}\overset{\scriptscriptstyle N_{4}\rightarrow\infty}{\longrightarrow}\kappa_{4}^{cr}\of{\kappa_{2}}$.
For quasi-canonical simulations around some fixed volume $\bar{N}_{4}$\footnote{There is no set of ergodic moves known for fixed volumes, it is therefore necessary to let the volume fluctuate around
the desired value $\bar{N}_{4}$.}, we can use \eqref{eq:edtcpf} to define a pseudo-critical $\kappa_{4}^{pcr}$ by
\[
\kappa_{4}^{pcr}\of{\kappa_{2},\bar{N}_{4}}\,=\,\partd{\ln\of{Z\of{\kappa_{2},\,N_{4}}}}{N_{4}}\bigg|_{N_{4}=\bar{\scriptstyle N}_{4}},\label{eq:k4pcr}
\]
which corresponds to the value of $\kappa_{4}$ for which the $N_{4}$-distribution is flat around $\bar{N}_{4}$.\\
\begin{figure}[H]
\centering
\includegraphics[width=0.7\linewidth]{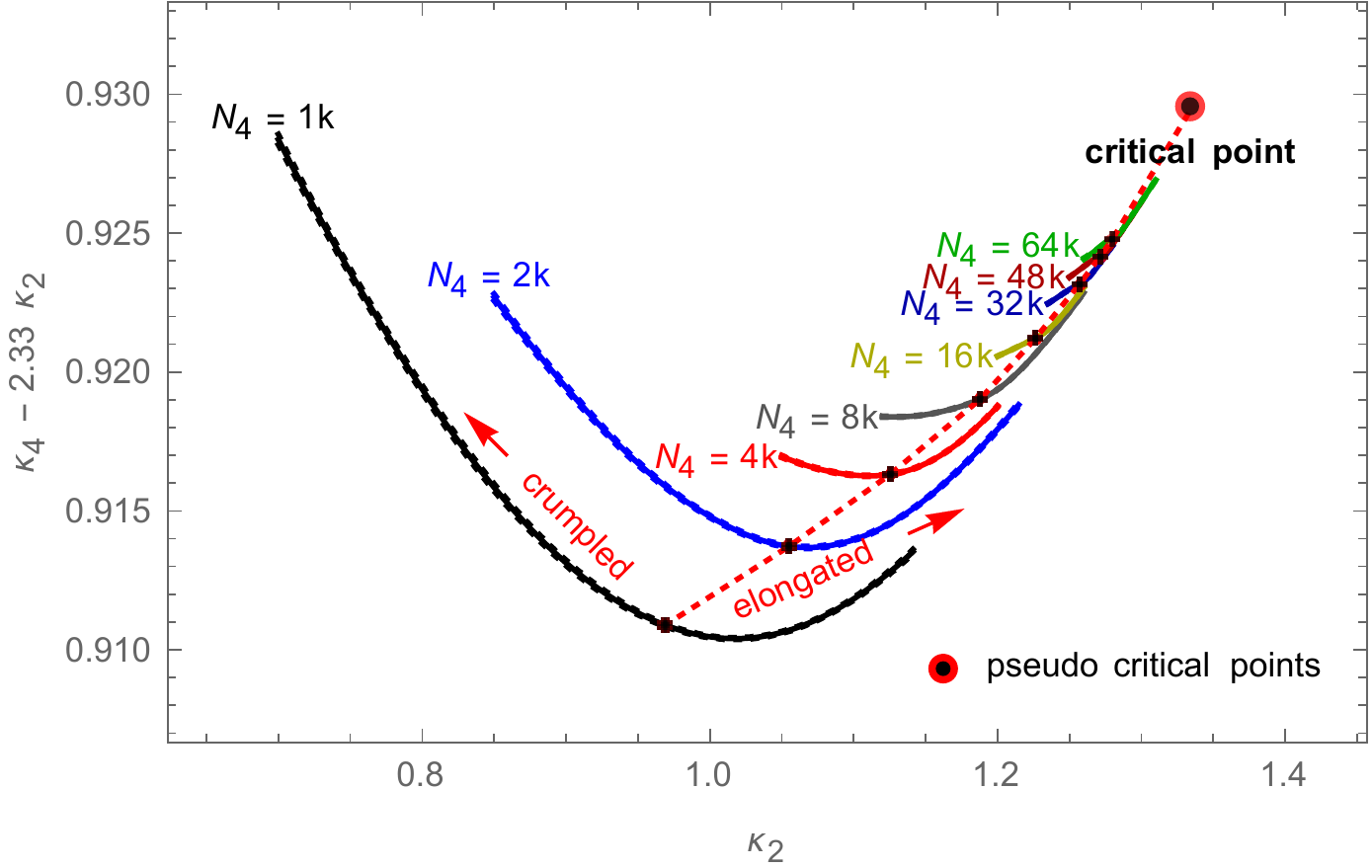}
\caption{Phase diagram for 4D EDT. The figure shows $\kappa_{4}^{pcr}\of{\kappa_{2},N_{4}}$ as a function of $\kappa_{2}$ for different $N_{4}$ together with the
corresponding pseudo-critical points $\of{\kappa_{2}^{pcr}\of{N_{4}},\kappa_{4}^{pcr}\of{\kappa_{2}^{pcr}\of{N_{4}},N_{4}}}$. The dotted red line separates the \emph{crumpled} from the
\emph{elongated} phase; in the limit $N_{4}\,\rightarrow\,\infty$ this line ends at the \emph{critical point}: $\of{\kappa_{2}^{cr},\kappa_{4}^{cr}}$. To improve readability, the
y-axis shows $\of{\kappa_{4}-2.33\,\kappa_{2}}$ instead of $\kappa_{4}$ itself.}
  \label{fig:edtpd}
\end{figure}
For constant $N_{4}$, we can define (see Fig. \ref{fig:edtpd}) a line $\kappa_{4}^{pcr}\of{\kappa_{2},N_{4}}$ as a function of $\kappa_{2}$, along which two phases are separated by a pseudo-critical point at $\kappa_{2}=\kappa_{2}^{pcr}\of{N_{4}}$. For $\kappa_{2}<\kappa_{2}^{pcr}\of{N_{4}}$ we are in the \emph{crumpled phase} where a typical configuration is highly collapsed in the sense that the distance between any two 4-simplices is very short, leading to a large (infinite) Hausdorff dimension. For $\kappa_{2}>\kappa_{2}^{pcr}\of{N_{4}}$ we are in the \emph{elongated phase} with Hausdorff dimension $\sim\,2$, where a typical configuration consists of a so-called \emph{baby-universe tree}: the total volume is subdivided into smaller parts, the \emph{baby-universes}\footnote{We consider a \emph{baby-universe} as a collection of 4-simplices which could all be pairwise connected by a path on the dual lattice that does not pass through any 3-simplex that belongs to a \emph{minimal neck}\label{fn:babyuniverse}.}, which are pairwise connected by only a small \emph{minimal neck}\footnote{In four dimensions, a \emph{minimal neck} consists of five 3-simplices forming a closed hyper-surface that looks like the boundary of a 4-simplex, but without a corresponding 4-simplex being present in the triangulation. Intuitively, a minimal neck is something like the bottleneck of a sand glass (but an extremely narrow bottle neck).}. This structure is hierarchical in a treelike manner: consider the largest baby-universe as ``mother'' with outgrowing smaller ``babies'' which in turn give birth to their own ``babies'', and so on (see right-hand side of Fig. \ref{fig:butrees}).\\
The true \emph{critical point} in the thermodynamic limit is obtained as
\[
\of{\kappa_{2}^{cr},\kappa_{4}^{cr}}\,=\,\lim\limits_{N_{4}\rightarrow\infty}\,\of{\kappa_{2}^{pcr}\of{N_{4}},\kappa_{4}^{pcr}\of{\kappa_{2}^{pcr}\of{N_{4}},N_{4}}}.
\]
\begin{figure}[H]
\centering
\begin{minipage}[t]{0.49\linewidth}
\centering
\includegraphics[angle=90,origin=c,width=0.85\linewidth]{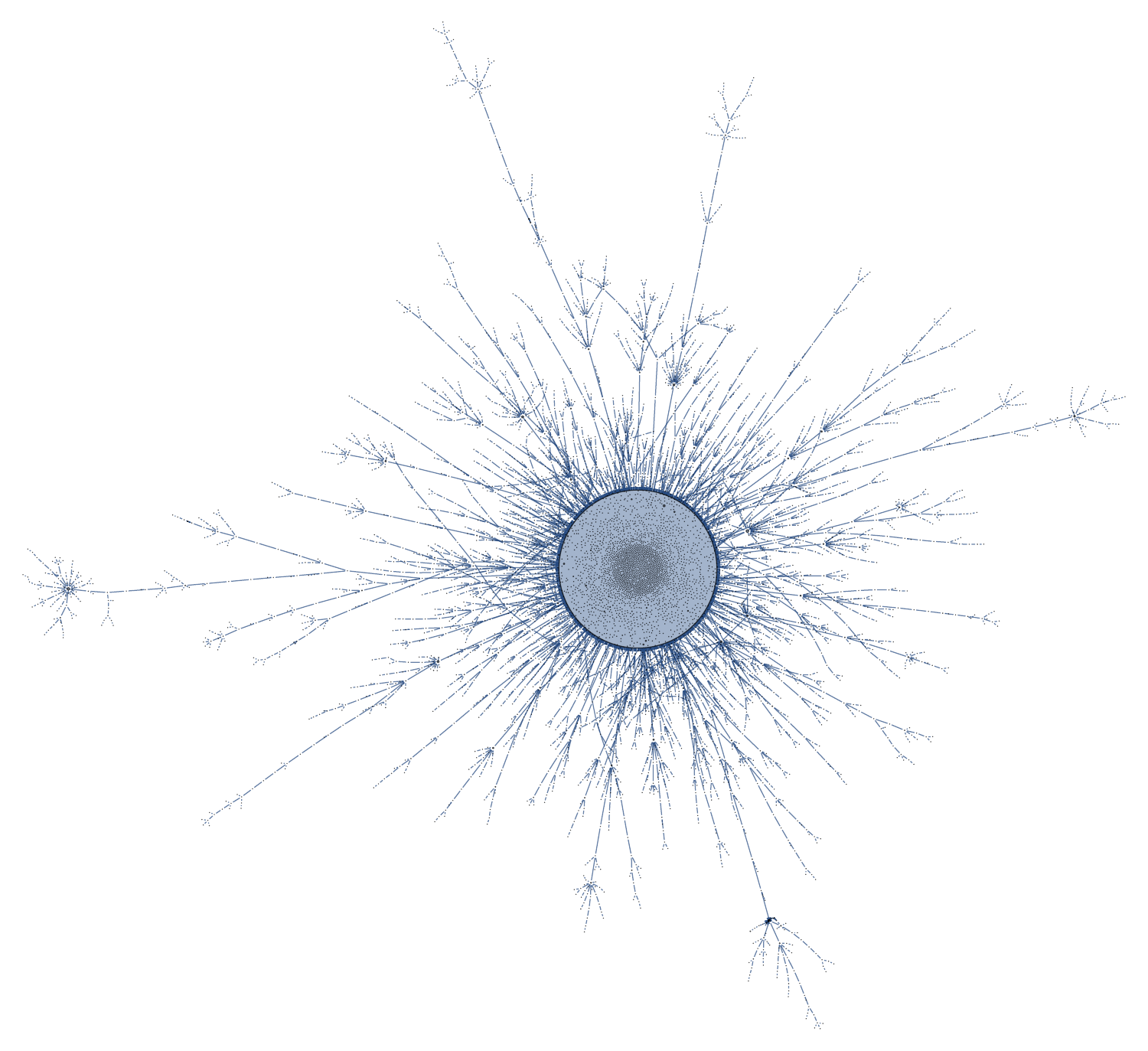}
\end{minipage}\hfill
\begin{minipage}[t]{0.49\linewidth}
\centering
\includegraphics[angle=90,origin=c,width=0.85\linewidth]{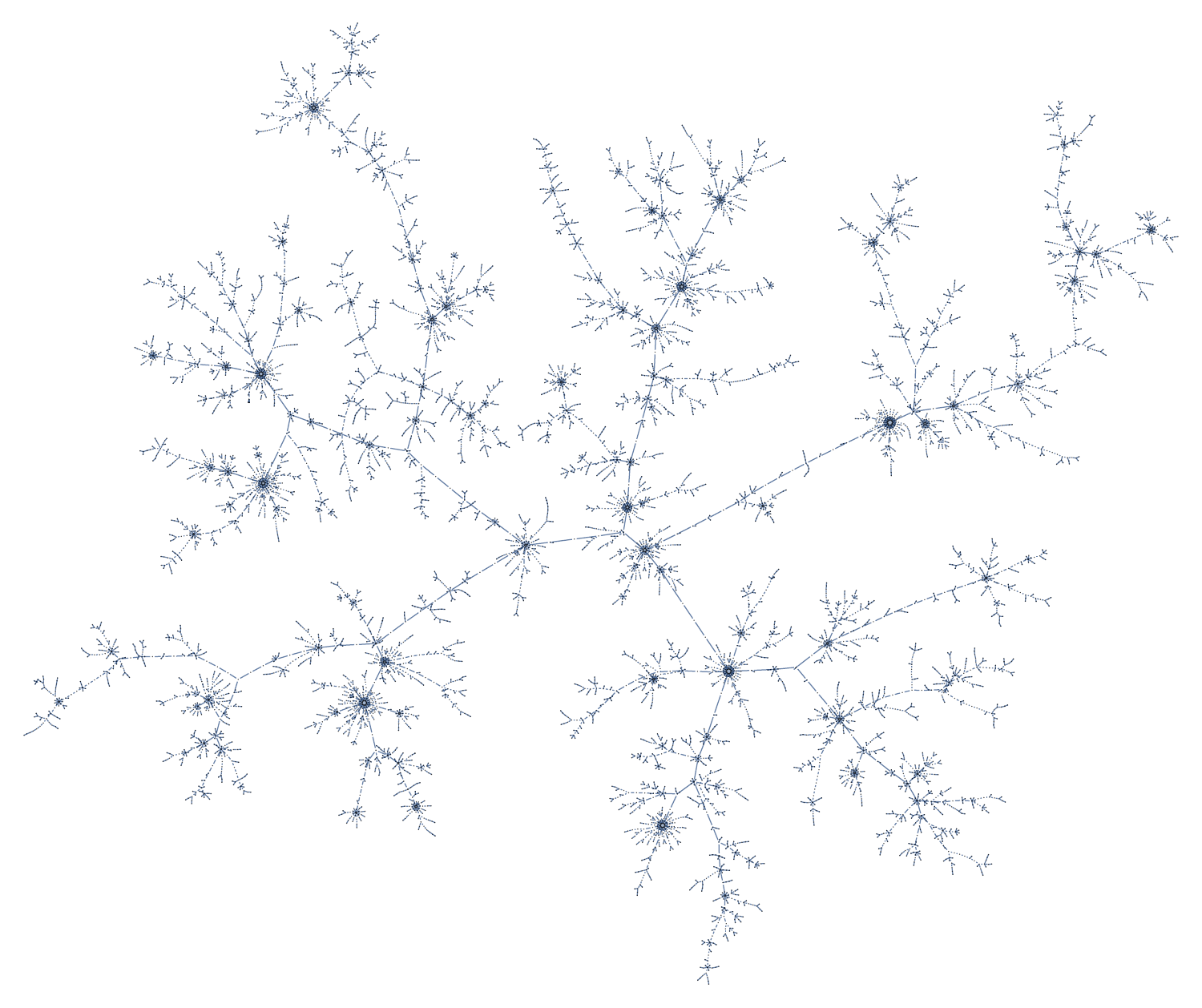}
\end{minipage}
\caption{Representative configurations in the crumpled (left, $\kappa_{2}=1.26$) and elongated (right, $\kappa_{2}=1.30$) phase at system size $N_{4}\approx 64$k: in the
crumpled phase, the triangulation consists of one large, highly connected bunch with outgrowths which are at least an order of magnitude smaller. In the elongated phase on the other hand, although a
largest component still exists and may be called ``mother universe'', it is much smaller than in the crumpled phase and some of its outgrowths (the ``babies'') are of comparable size.}
\label{fig:butrees}
\end{figure}

\section{Simulation Methods}\label{sec:simulationmethods}

\subsection{Pachner Moves}\label{sec:pachnermoves}
\begin{figure}[H]
\centering
\begin{minipage}[t]{0.25\linewidth}
\centering
	\begin{tikzpicture}[scale=0.2,baseline=-35]
    \node (a) at (0,0) {};
    \draw (a) node[below] {\tiny 100000};
    \node (b) at (-2.8,7) {};
    \draw (b) node[below left] {\tiny 000100};
    \node (c) at (5,15) {};
    \draw (c) node[above] {\tiny 010000};
    \node (d) at (7.7,4.3) {};
    \draw (d) node[below right] {\tiny 001000};
    \node (e) at (-2,14) {};
    \draw (e) node[above left] {\tiny 000010};
    \node (f) at (10,11) {};
    \draw (f) node[above right] {\tiny 000001};
    \draw (a.center) -- (b.center);
    \draw (a.center) -- (d.center);
    \draw (a.center) -- (e.center);
    \draw (a.center) -- (f.center);
    \draw (c.center) -- (e.center);
    \draw (c.center) -- (f.center);
    \draw[dashed] (c.center) -- (b.center);
    \draw[dashed] (c.center) -- (d.center);
 	  \draw[dashed,thick,red] (b.center) -- (d.center);
 	  \draw[thick,red] (d.center) -- (f.center) -- (e.center) -- (b.center);
   	\draw[dashed,thick,red] (b.center) -- (f.center);
 	  \draw[dashed,thick,red] (d.center) -- (e.center);
  \end{tikzpicture}
\end{minipage}\hspace{35pt}
\begin{minipage}[t]{0.25\linewidth}
\centering
	\begin{tikzpicture}[scale=0.2,baseline=-35]
    \node (a) at (0,0) {};
    \draw (a) node[below] {\tiny 100000};
    \node (b) at (-2.8,7) {};
    \draw (b) node[below left] {\tiny 000100};
    \node (c) at (5,15) {};
    \draw (c) node[above] {\tiny 010000};
    \node (d) at (7.7,4.3) {};
    \draw (d) node[below right] {\tiny 001000};
    \node (e) at (-2,14) {};
    \draw (e) node[above left] {\tiny 000010};
    \node (f) at (10,11) {};
    \draw (f) node[above right] {\tiny 000001};
    \draw (a.center) -- (b.center);
    \draw (a.center) -- (d.center);
    \draw (a.center) -- (e.center);
    \draw (a.center) -- (f.center);
    \draw (c.center) -- (e.center);
    \draw (c.center) -- (f.center);
    \draw[dashed] (c.center) -- (b.center);
    \draw[dashed] (c.center) -- (d.center);
 	  \draw[dashed] (b.center) -- (d.center);
 	  \draw (d.center) -- (f.center) -- (e.center) -- (b.center);
   	\draw[dashed] (b.center) -- (f.center);
 	  \draw[dashed] (d.center) -- (e.center);
 	  \draw[dashed,thick,blue] (a.center) -- (c.center);
  \end{tikzpicture}
\end{minipage}
\vspace{-1.5pt}\caption{Illustration of a Pachner 3-move in four dimensions: on the left-hand side we have two 4-simplices which share a common 3-simplex (red, projected onto two dimensions). The 3-move replaces this complex with the one shown on the right-hand side of the figure where four 4-simplices share a common 1-simplex (blue). The 1-move is just the inverse of the 3-move.}
\label{fig:pachnermoves}
\end{figure}
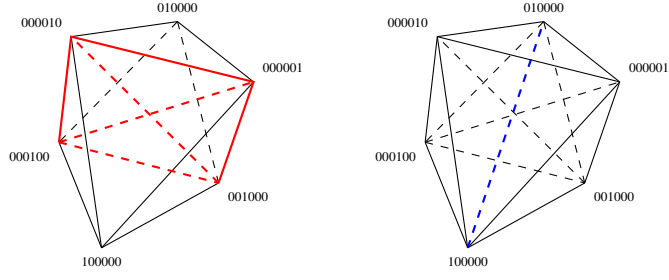

In $d$ dimensions, there exist $\of{d+1}$ Pachner moves. They form an ergodic set of local updates \cite{Pachner} in the space of abstract triangulations of fixed topology without boundary. A \emph{$n$-move} ($n\in\cof{0,\ldots,d}$) consists
of the following procedure: pick a $n$-simplex which is contained in $\of{d+1-n}$ $d$-simplices. The complex consisting of these $\of{d+1-n}$ $d$-simplices has the same boundary as a corresponding \emph{dual} complex spanned by $\of{n+1}$ $d$-simplices that share a common $\of{d-n}$-simplex. We can therefore just replace the complex around the selected $n$-simplex with its dual (see Fig. \ref{fig:pachnermoves} for an illustration in four dimensions). The only additional constraint is the so called \emph{manifold constraint}, that is: the $\of{d-n}$-simplex shared
by the $\of{n+1}$ newly created $d$-simplices of the dual complex must not already exist in the triangulation as this could result in topology changes. From now on we will consider only the
4-dimensional case.

\subsection{Detailed Balance}\label{ssec:detailedbalance}
Calling $T_{k}$ the current triangulation in our Markov chain, we obtain $T_{k+1}$ as follows: 
\vspace{-5pt}
\begin{enumerate}\itemsep -2pt
 \item randomly choose a move type $n\in\cof{0,\ldots,4}$,
 \item randomly choose one of the $N_{4}$ $4$-simplices of $T_{k}$ and call it $D$,
 \item randomly choose one of the $\binom{5}{n+1}$ $n$-simplices contained in $D$ and call it $S$,
 \item\label{it:acctest} perform a Metropolis test with acceptance probability $p_{n}\of{T_{k},S}$:
\vspace{-5pt}
 \begin{itemize}\itemsep -2pt
  \item accept: $T_{k+1}$ is obtained from $T_{k}$ by applying the $n$-move at $S$,
  \item reject: $T_{k+1} = T_{k}$.
 \end{itemize}
\vspace{-5pt}
\end{enumerate}
\vspace{-4pt}
The acceptance probability at step \ref{it:acctest} is given by \cite{deBakker2}
\[
p_{n}\of{T,S}=
\begin{cases}
p_{n}\of{N_{4}\of{T}}\quad &\text{\small
if}\quad\text{\small $n$-move possible at $S\,\in\,T$}\\
0\quad &\text{\small else}
\end{cases},\label{eq:transprobab}
\]
where $p_{n}\of{N_{4}}=\textstyle\min\cof{1,\frac{\scriptstyle N_{4}}{\scriptstyle N_{4}+\Delta N_{4}\of{n}}\,\e^{\kappa_{2}\Delta N_{2}\of{n}-\kappa_{4}\Delta N_{4}\of{n}}}$ is the so-called \emph{reduced transition probability}, $\Delta N_{i}\of{n}$ labels the change of $N_{i}$ under a $n$-move, and a $n$-move is considered as possible at $S$ if $S$ is contained in $\of{5-n}$ $4$-simplices and the application of the move does not violate the manifold constraint mentioned above in Sec. \ref{sec:pachnermoves}.\\
Equation \eqref{eq:transprobab} can be derived from the detailed balance condition: assume first we have two valid 4-dimensional triangulations $T,\,T'$, where $T'$ can be obtained from $T$ by applying a $n$-move at a specific $n$-simplex $S$ of $T$. The detailed balance equation then reads
\[
\textstyle
\rho\of{T}\,P\of{T\,\stackrel{n}{\rightarrow}\,T'}\,=\,\rho\of{T'}\,P\of{T'\,\stackrel{4-n}{\rightarrow}\,T},\label{eq:detailedbalance}
\]
where $\rho\of{T}\,=\,\e^{\kappa_{2}\,N_{2}\of{T}\,-\,\kappa_{4}\,N_{4}\of{T}}$ and $P\of{T\,\stackrel{n}{\rightarrow}\,T'}$ is the transition probability. The latter can be written as
\[
P\of{T\,\stackrel{n}{\rightarrow}\,T'}\,=\,\frac{5-n}{5\,N_{4}\of{T}\,\binom{5}{n+1}} p_{n}\of{N_{4}\of{T}},
\]
where $p_{n}\of{N_{4}}$ is again the reduced transition probability and the factor in front of it is the probability for selecting (with the update scheme mentioned above) the $n$-simplex $S$ through which $T$ and $T'$ can be related by applying a $n$-move: $1/5$ is the probability for choosing the correct move type $n$, $\frac{5-n}{N_{4}}$ the one for selecting a 4-simplex $D$ which contains $S$ and $\frac{1}{\binom{5}{n+1}}$ is the probability for selecting $S$ out of the $\binom{5}{n+1}$ $n$-simplices of $D$. Note that $\frac{5-n}{\binom{5}{n+1}}$ is the local 4-volume of a $n$-simplex\footnote{The 4-volume containing all points which are closer to the $n$-simplex under consideration than to any other $n$-simplex.} that allows for a $n$-move (in units of $V_{4}$) and $p_{n}\of{N_{4}}$ can therefore be interpreted as the transition probability for a $n$-move \emph{per unit volume}, as the term \emph{reduced} suggests.\\
For the inverse transition probability on the right hand side of \eqref{eq:detailedbalance} we have analogously:
\[
P\of{T'\,\stackrel{4-n}{\rightarrow}\,T}\,=\,\frac{n+1}{5\,N_{4}\of{T'}\,\binom{5}{5-n}} p_{4-n}\of{N_{4}\of{T'}}\,=\,\frac{5-n}{5\,N_{4}\of{T'}\,\binom{5}{n+1}} p_{4-n}\of{N_{4}\of{T'}}.
\]
Equation \eqref{eq:detailedbalance} reduces therefore to
\[
\textstyle
\rho\of{T}\,\frac{1}{N_{4}\of{T}}\,p_{n}\of{N_{4}\of{T}}\,=\,\rho\of{T'}\,\frac{1}{N_{4}\of{T'}}\,p_{4-n}\of{N_{4}\of{T'}},\label{eq:detailedbalancered}
\]
which, by noting that $N_{4}\of{T'}\,=\,N_{4}\of{T}\,+\,\Delta N_{4}\of{n}$, is satisfied by setting
\[
p_{n}\of{N_{4}}\,=\,\min\cof{1,\frac{N_{4}}{N_{4}\,+\,\Delta N_{4}\of{n}}\,\e^{\kappa_{2}\,\Delta N_{2}\of{n}\,-\,\kappa_{4}\,\Delta N_{4}\of{n}}}.\label{eq:redtransprobab} 
\]
This gives the upper part of \eqref{eq:transprobab}. For the lower part, when $T'$ does not exist, we have $\rho\of{T'}\,=\,0$ and in order to satisfy \eqref{eq:detailedbalance}, we have to set
\[
P\of{T\,\stackrel{n}{\rightarrow}\,T'}\,=\,\frac{5-n}{5\,N_{4}\of{T}\,\binom{5}{n+1}} p_{n}\of{T,\,S}\,=\,0.
\]

We now have a prescription for how to produce a Markov chain containing the configurations required to evaluate \eqref{eq:edtpf}. As neighboring elements in our Markov chain are highly correlated, it
is appropriate to take measurements only on every $k^{\text{th}}$ element in the chain, where $k$ must be a constant in order to preserve the probability distribution. As already mentioned at the
beginning, this was not always respected in previous work, as e.g. in \cite{deBakker} the measurements were separated by a fixed number of \emph{accepted} moves, which turns the true separation $k$ between measurements into a random variable whose value depends on what kind of configurations is currently sampled by the Markov chain.

\subsection{Controlling the Volume}
As $\kappa_{4}^{pcr}\of{\kappa_{2},\bar{N}_{4}}$ is monotonically growing with $N_{4}$, it is practically impossible to run fully grand canonical MC simulations for the EDT model. But canonical simulations are anyway better suited to investigate finite size scaling. Unfortunately, as already mentioned before, there is no set of ergodic moves known for the space of triangulations of fixed $N_{4}$ and it is therefore not possible to run canonical simulations either. The best we can do is to run \emph{quasi canonical} simulations based on \eqref{eq:edtpf} but with $N_{4}$ constrained to fluctuate around some desired $\bar{N}_{4}$. In previous work \cite{Ambjorn,Bialas,deBakker}, this was often achieved by adding a harmonic potential,
\[
U\of{N_{4},\bar{N}_{4},\delta}\,=\,\frac{\delta}{2}\,\of{N_{4}\,-\,\bar{N}_{4}}^{2},\label{eq:harmonicpot}
\]
to the action \eqref{eq:era}. This of course introduces a systematic error for all moves which change $N_{4}$. We therefore decided to rather use a infinite potential well of some reasonable width $w\,\approx\,2\,\sigma\of{N_{2}}\,/\,2.5$, where $2.5\,=\,\max\limits_{n}\,\cof{\frac{\Delta N_{2}\of{n}}{\Delta N_{4}\of{n}}}$ and $\sigma\of{N_{2}}$ is the square root of the $N_{2}$-variance.\\

\begin{figure}[htbp]
\centering
\begin{minipage}[t]{0.49\linewidth}
\centering
\includegraphics[width=\linewidth]{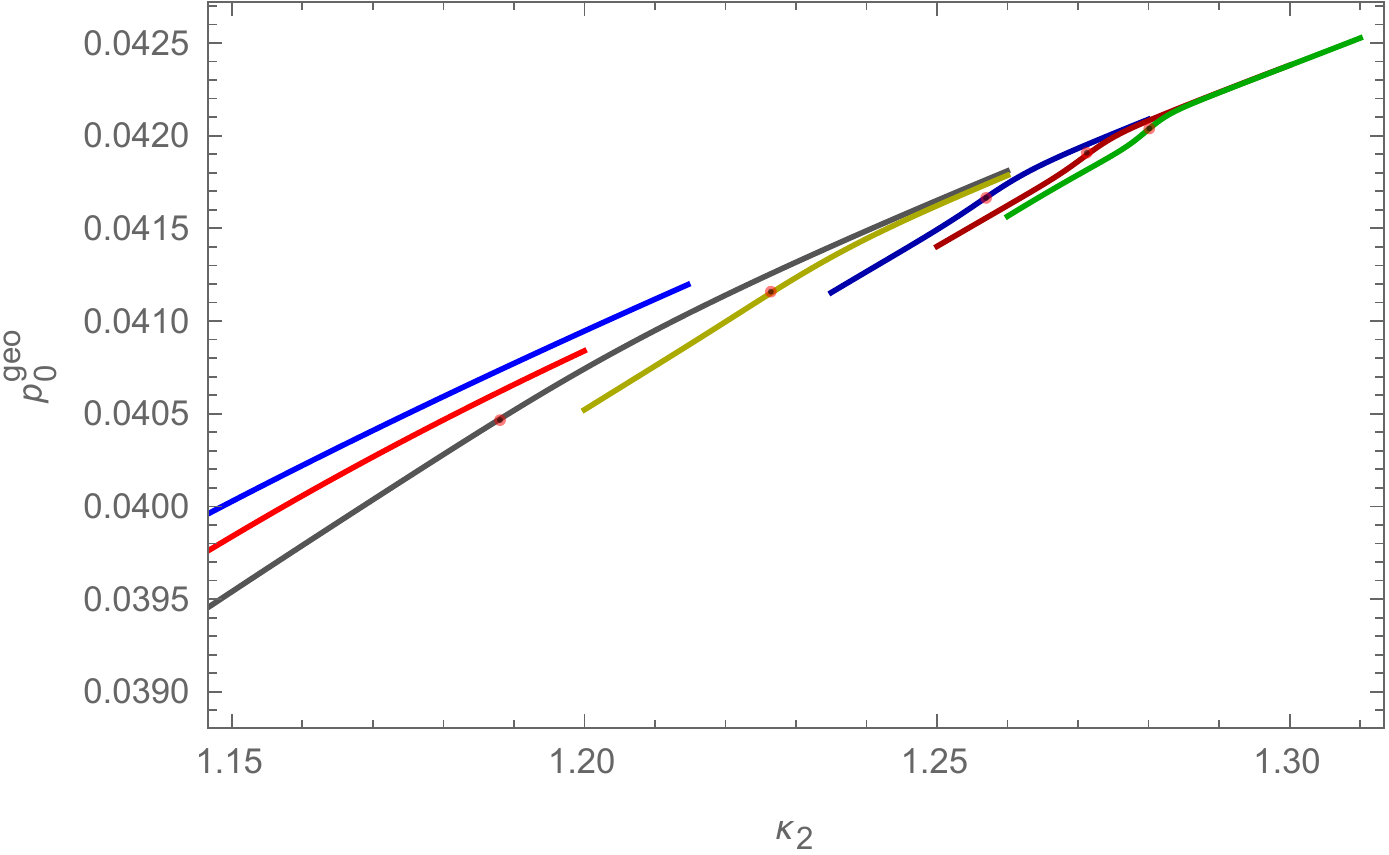}
\end{minipage}\hfill
\begin{minipage}[t]{0.49\linewidth}
\centering
\includegraphics[width=\linewidth]{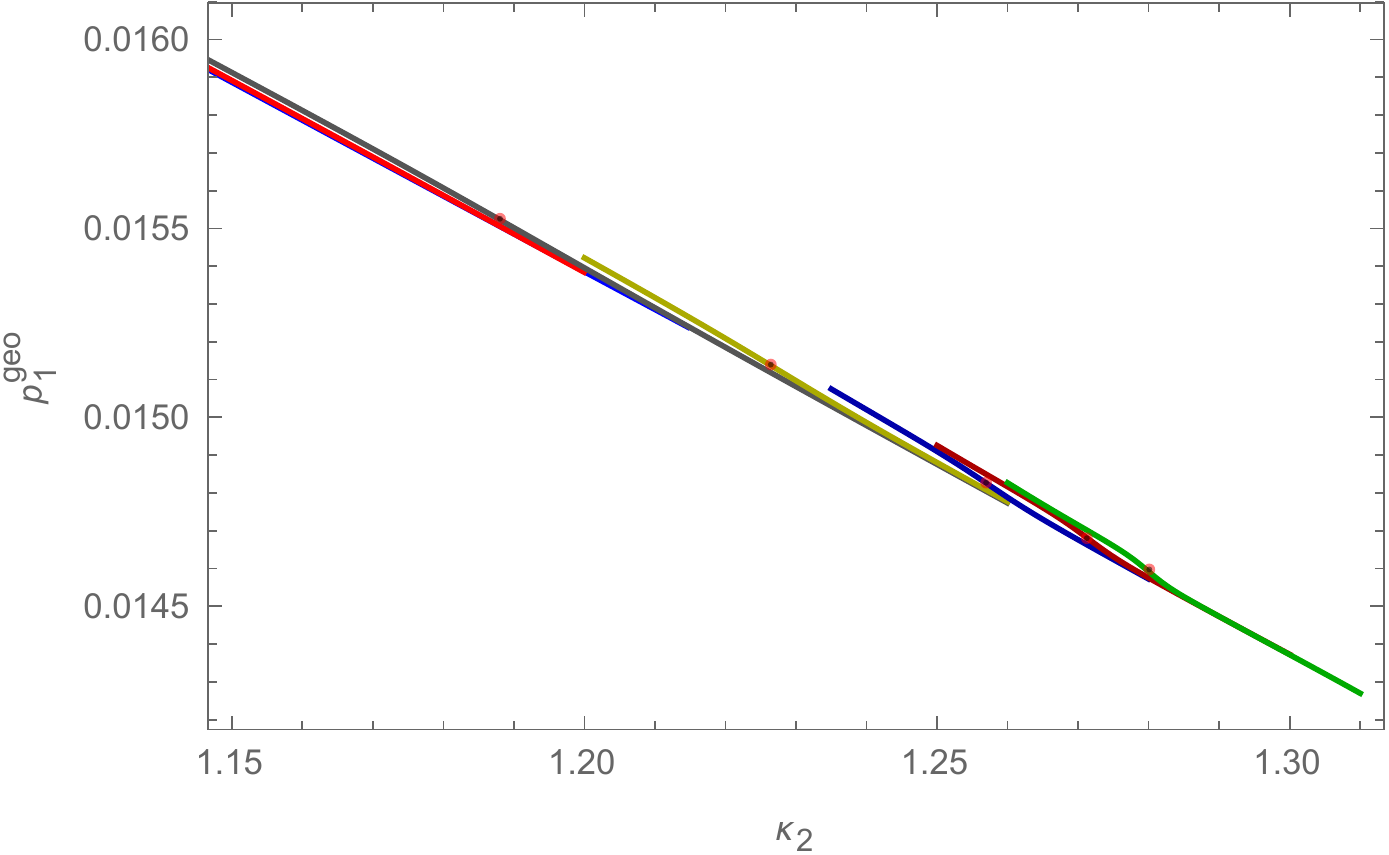}
\end{minipage}
\begin{minipage}[t]{0.49\linewidth}
\centering
\includegraphics[width=\linewidth]{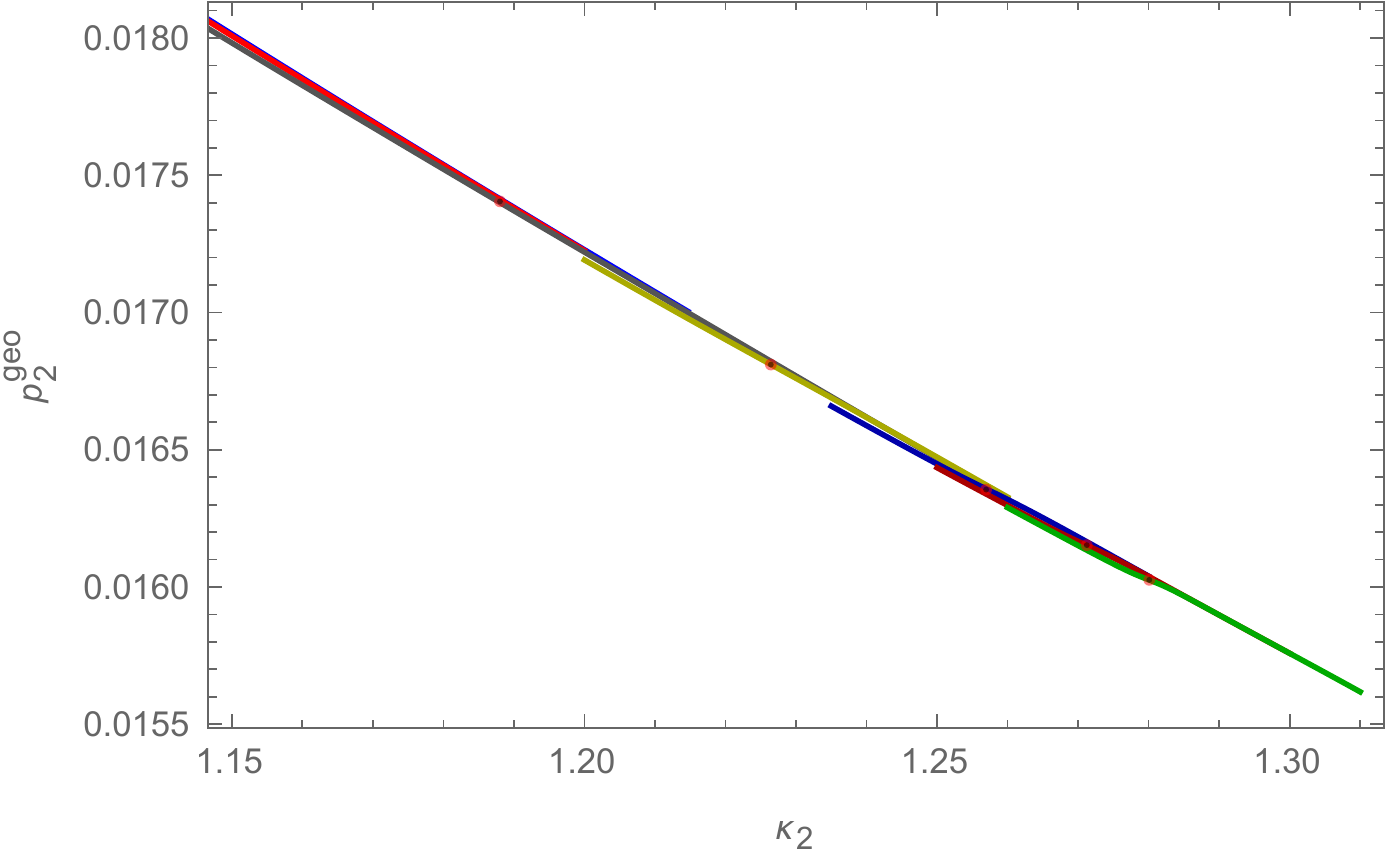}
\end{minipage}\hfill
\begin{minipage}[t]{0.49\linewidth}
\centering
\includegraphics[width=\linewidth]{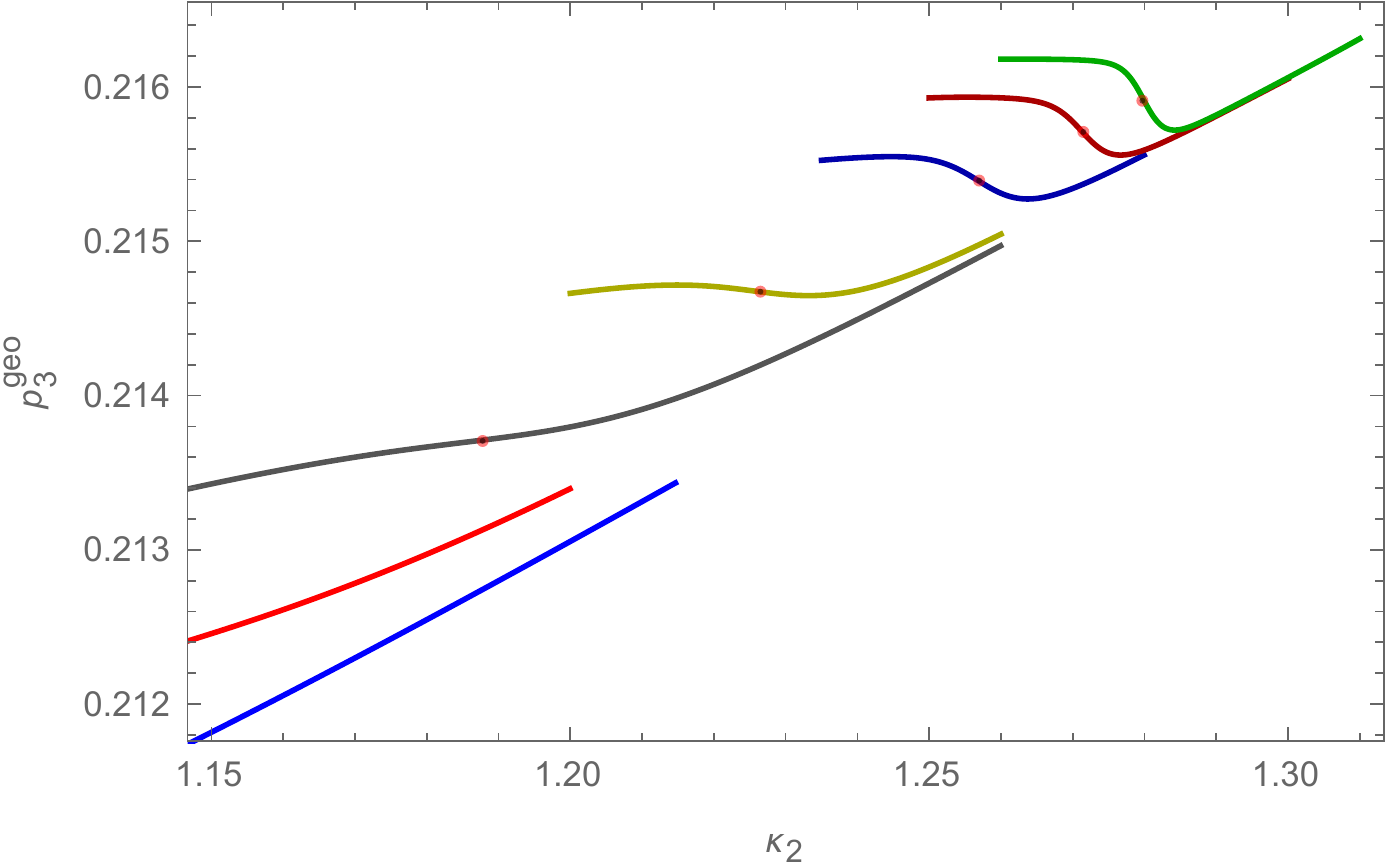}
\end{minipage}
\begin{minipage}[t]{0.49\linewidth}
\centering
\includegraphics[width=\linewidth]{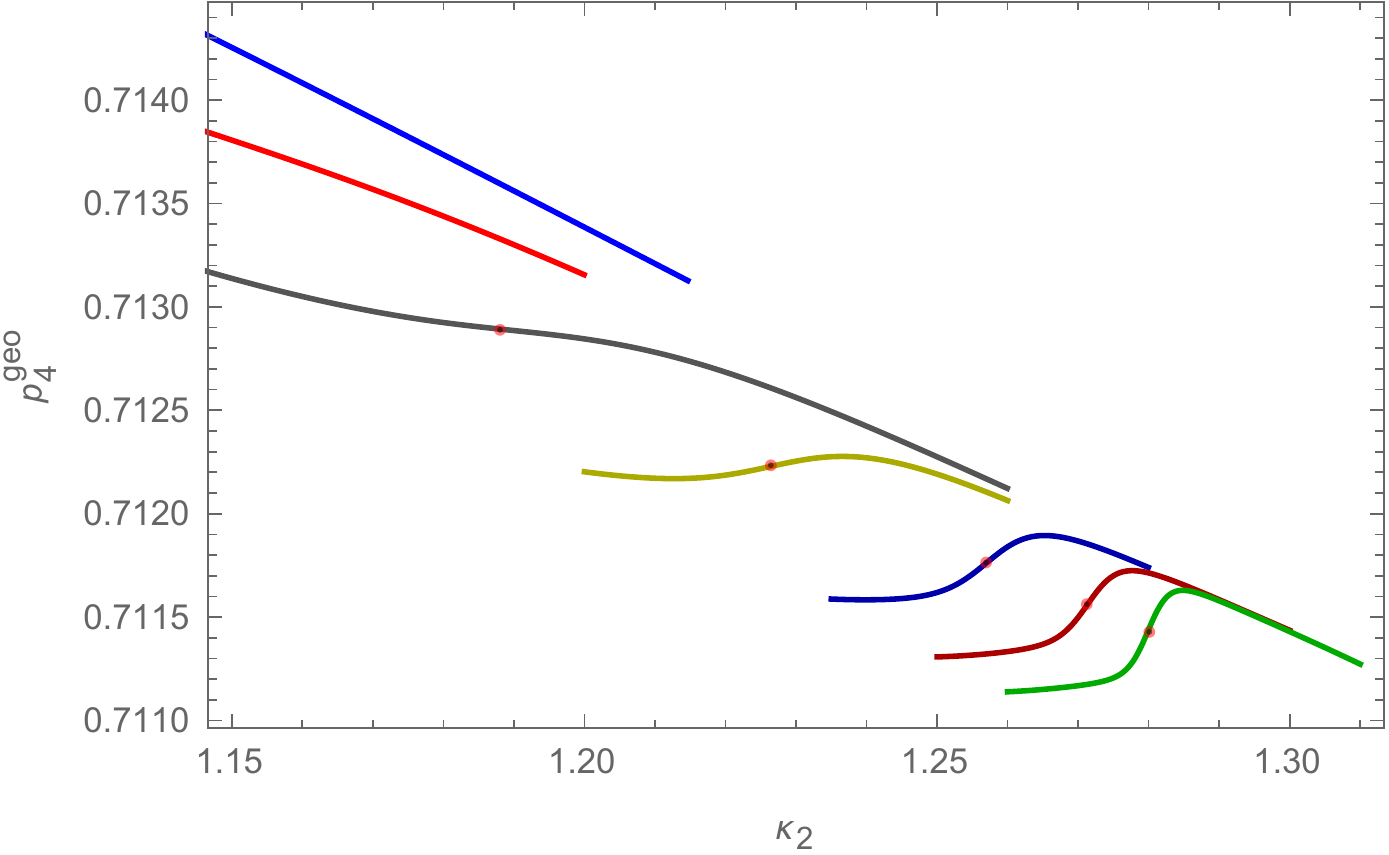}
\end{minipage}\hfill
\begin{minipage}[t]{0.49\linewidth}
\centering
\includegraphics[width=\linewidth]{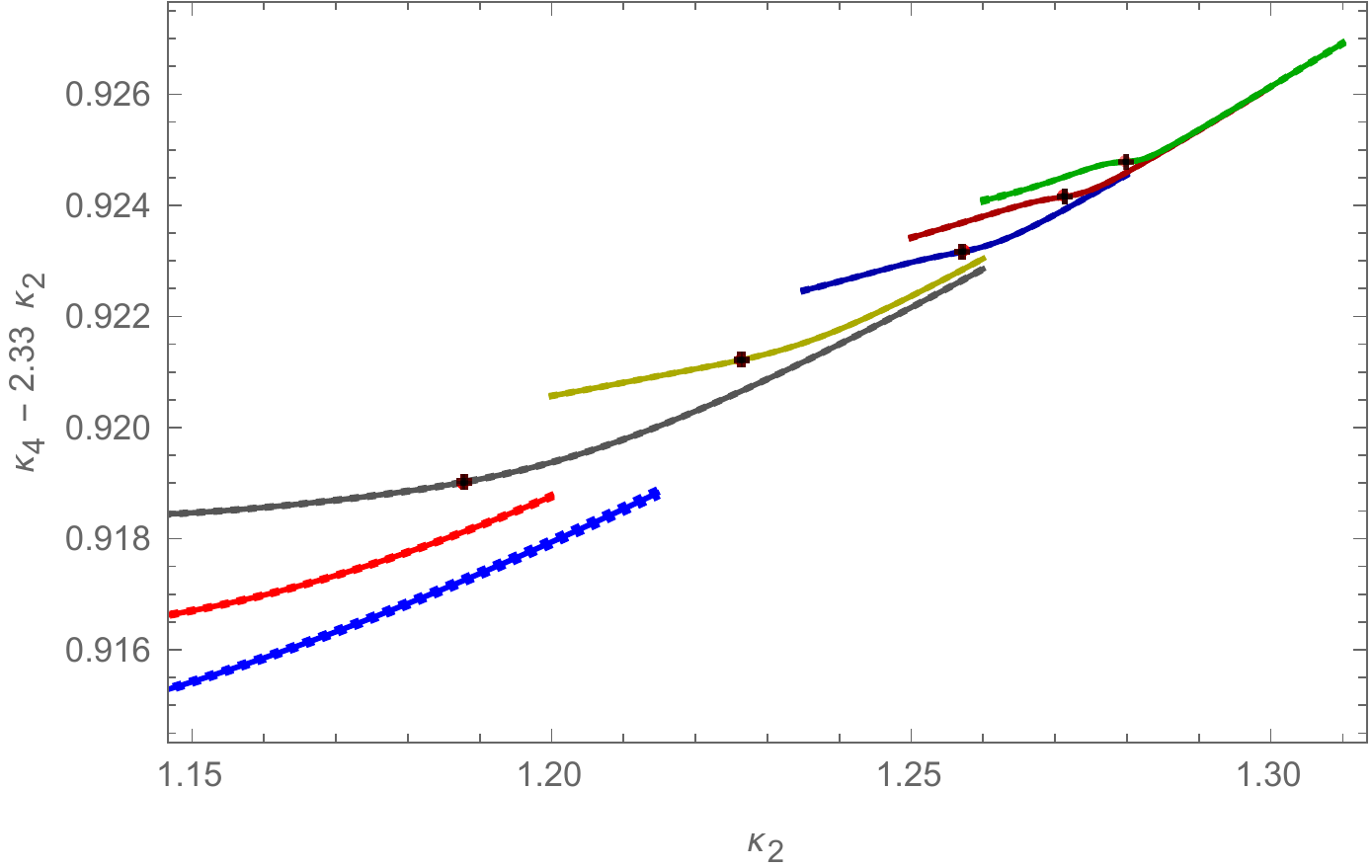}
\end{minipage}
\caption{The average \emph{geometric probabilities} $\bar{p}_{n}^{geo}\of{\bar{N}_{4}}$, $n\in\cof{0,\ldots,4}$, are plotted as functions of $\kappa_{2}$ for different system sizes $N_{4} = 2$k (blue), $4$k (red), $8$k (dark gray), $16$k (dark yellow), $32$k (dark blue), $48$k (dark red), $64$k (dark green). Note that, as the $4$-move is always possible, the change in $\bar{p}_{4}^{geo}\of{\bar{N}_{4}}$ is just due to the normalization chosen in \eqref{eq:geoprobab}.\\
The last plot at bottom right shows the pseudo-critical $\kappa_{4}^{pcr}\of{\kappa_{2},\bar{N}_{4}}$ corresponding to $\bar{p}_{0}^{geo}\of{\bar{N}_{4}}$ and $\bar{p}_{4}^{geo}\of{\bar{N}_{4}}$ as given by \eqref{eq:geok4pcr}. It is a close up version of Fig. \ref{fig:edtpd}, showing in more detail the behavior of $\kappa_{4}^{pcr}\of{\kappa_{2},\bar{N}_{4}}$ close to the pseudo critical point (to improve readability, the y-axis shows again $\of{\kappa_{4}-2.33 \kappa_{2}}$ instead of $\kappa_{4}$ itself). The small red dots indicate the pseudo-critical points.}
\label{fig:geoprobabs}
\end{figure}

As with such a potential well we cannot use the saddle point expansion method from \cite{deBakker2} to tune $\kappa_{4}$ to its pseudo-critical value $\kappa_{4}^{pcr}\of{\kappa_{2},\bar{N}_{4}}$, we instead made use of a method mentioned in \cite{Bruegmann}: as the $N_{4}$-histogram has to be flat around $\bar{N}_{4}$ if $\kappa_{4}\,=\,\kappa_{4}^{pcr}\of{\kappa_{2},\bar{N}_{4}}$, we have that
\[
\bar{p}_{4}^{geo}\of{\bar{N}_{4}}\,p_{4}^{pcr}\of{\bar{N}_{4}}\,=\,\bar{p}_{0}^{geo}\of{\bar{N}_{4}+\Delta N_{4}\of{4}}\,p_{0}^{pcr}\of{\bar{N}_{4}+\Delta N_{4}\of{4}},\label{eq:totmoveequil}
\]
where $p_{n}^{pcr}\of{N_{4}}$ is the reduced transition probability \eqref{eq:redtransprobab} with $\kappa_{4}\,=\,\kappa_{4}^{pcr}\of{\kappa_{2},\bar{N}_{4}}$ and $\bar{p}_{n}^{geo}\of{N_{4}}$ is the average \emph{geometric probability} for a $n$-move, i.e. the fraction of the total volume that allows for a change (through one of the Pachner moves), to which a $n$-move could be applied\footnote{Alternatively, one could define $p_{n}^{geo}\of{T}$ as the fraction of the \emph{overall total volume} of a triangulation $T$, to which a $n$-move could be applied, which would lead to a different normalization: $p_{n}^{geo}\of{T}\,=\,\frac{1}{5 N_{4}}\frac{5-n}{\binom{5}{n+1}}\,N_{n}\of{T}\,f_{n}^{legal}\of{T}$, such that $p_{n}^{geo}\of{T}$ would coincide with the probability to select a good location for a $n$-move within the update scheme described above in Sec. \ref{ssec:detailedbalance}.} (see Fig. \ref{fig:geoprobabs}):
\[
p_{n}^{geo}\of{T}\,=\,\frac{\frac{5-n}{\binom{5}{n+1}}\,N_{n}\of{T}\,f_{n}^{legal}\of{T}}{\sum\limits_{m=0}^{4}\,\frac{5-m}{\binom{5}{m+1}}\,N_{m}\of{T}\,f_{m}^{legal}\of{T}},\label{eq:geoprobab}
\]
where $f_{n}^{legal}\of{T}$ is the fraction of $n$-simplices in the triangulation $T$ where a $n$-move is possible. One can then solve for
$\kappa_{4}^{pcr}\of{\kappa_{2},\bar{N}_{4}}$ which leads to
\begin{multline}
\kappa_{4}^{pcr}\of{\kappa_{2},\bar{N}_{4}}\,=\\
\frac{1}{\Delta N_{4}\of{4}}\fof{\ln\of{\frac{\bar{p}_{4}^{geo}\of{\bar{N}_{4}}}{\bar{p}_{0}^{geo}\of{\bar{N}_{4}+\Delta
N_{4}\of{4}}}}\,-\,\ln\of{1+\frac{\Delta
N_{4}\of{4}}{\bar{N}_{4}}}}\,+\,\frac{\Delta N_{2}\of{4}}{\Delta N_{4}\of{4}}\,\kappa_{2}.\label{eq:geok4pcr}
\end{multline}
Note that the 4-move is always possible and one therefore effectively only has to measure the average fraction of vertices which allow for a 0-move in order to determine $\kappa_{4}^{pcr}\of{\kappa_{2},\bar{N}_{4}}$. Furthermore, as the average geometric probabilities vary slowly with $N_{4}$, we can, as long as the width of the potential well for $N_{4}$ is much smaller than $\bar{N}_{4}$, just use $\frac{\avof{p_{4}^{geo}}}{\avof{p_{0}^{geo}}}$ instead of $\frac{\bar{p}_{4}^{geo}\of{\bar{\scriptstyle
N}_{4}}}{\bar{p}_{0}^{geo}\of{\bar{\scriptstyle N}_{4}+\Delta N_{4}\of{4}}}$ to set $\kappa_{4}$ to its pseudo-critical value.

\subsection{Autocorrelation Time}
The autocorrelation time $\tau$ is a measure for the typical distance between two uncorrelated elements in a Markov chain. It can be thought of as the time it takes for a change to propagate through the typical volume of the system over which the degrees of freedom are correlated. One therefore writes $\tau\,\propto\,\xi^{z}$, where $\xi$ is the correlation length and the \emph{dynamical critical exponent} $z$ is expected to be $z\,\approx\,2$ for a local update scheme; i.e. updated information propagates in a diffusion-like manner.\\

For a $2^{\text{nd}}$ order transition where the correlation length diverges when approaching the critical point, $\xi$ is truncated by the linear system size $L$ and we have $\tau\,\propto\,L^{z}\,=\,V^{z/d_{H}}$ with $d_{H}$ being the Hausdorff dimension of the system. In this case the autocorrelation time obviously diverges as a power of the system size.\\

For a $1^{\text{st}}$ order transition $\xi$ remains finite at the transition point for all system sizes. Nevertheless the autocorrelation time can diverge even more dramatically as transitions between the two phases that coexist at the transition point become \emph{exponentially} suppressed with increasing system size. The autocorrelation function should then consist of two parts: a relatively steep first one corresponding to the decay of autocorrelations within a single phase, as well as a second, much less steep part which reflects the fact that the system remains in one and the same phase for a rather long time. Unfortunately it is almost impossible to verify this as it would need ridiculously long simulations to obtain the required accuracy on the auto-correlation function.\\

In both cases, for $1^{\text{st}}$ and $\sord$ phase transitions, \emph{parallel tempering} can be used to reduce autocorrelations \cite{Troyer,Bauer}. The idea is to run $K$ simulations for different values of the couplings in parallel where the couplings are chosen such that they lie on a line in coupling space which connects a region with slow relaxation with another where relaxation is fast. One now periodically attempts to swap configurations between neighboring sets (called \emph{replicas}) with an acceptance probability given by
\begin{multline}
p^{swap}\fof{\cof{\of{\scriptstyle\kappa_{i},N_{i}},\of{\scriptstyle\kappa'_{i},N'_{i}}}\rightarrow
\cof{\of{\scriptstyle\kappa_{i},N'_{i}},\of{\scriptstyle\kappa'_{i},N_{i}}}}\\
=\,\min\cof{1,\e^{S\of{\kappa_{i},N_{i}}-S\of{\kappa_{i},N'_{i}}+S\of{\kappa'_{i},N'_{i}}-S\of{\kappa'_{i},N_{i}}}},
\end{multline}
where $\of{\kappa_{i},N_{i}},\of{\kappa'_{i},N'_{i}}$ are the sets of couplings and configuration variables for the two neighboring replicas and $S\of{\kappa_{i},N_{i}}$ is the action of a configuration with variables $N_{i}$ at couplings $\kappa_{i}$. The advantage of this procedure is twofold: first, we no longer have just one Markov chain per simulation point in coupling space that has to build up the whole corresponding statistical ensemble but now all $K$ chains alternately contribute to all of the statistical ensembles at different couplings. Second, if every Markov chain frequently reaches regions in coupling space where relaxation is fast before passing again through the critical region, then the configurations which the chain contributes to the statistical ensembles in the critical region, are on average much less correlated than corresponding configurations of a Markov chain that remains all the time in the critical region. For more details see \cite{Troyer,Bauer} where it is also explained how 
this procedure can be optimized. Especially in \cite{Bauer} the application of parallel tempering to $\ford$ transitions is discussed.\\

\begin{figure}[H]
\begin{minipage}[t]{0.49\linewidth}
\centering
\includegraphics[width=\linewidth]{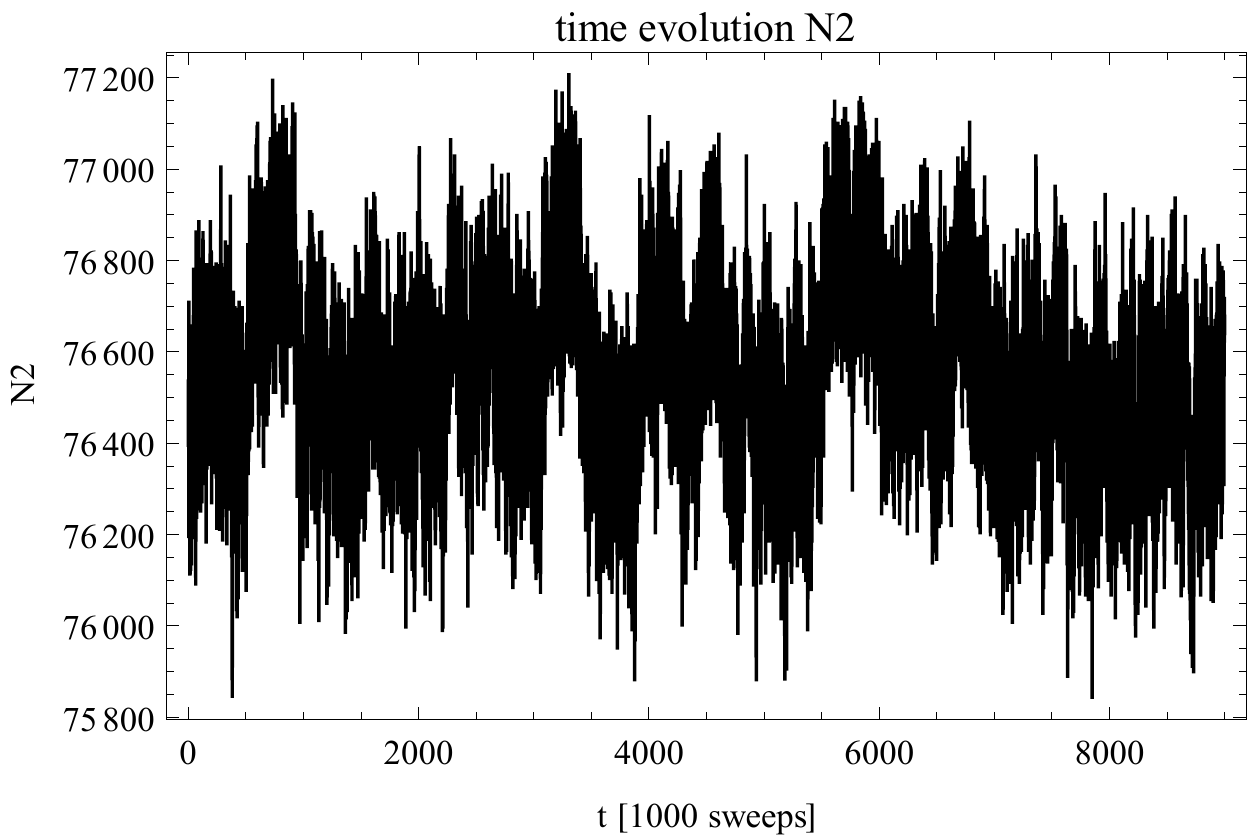}
\end{minipage}\hfill
\begin{minipage}[t]{0.49\linewidth}
\centering
\includegraphics[width=\linewidth]{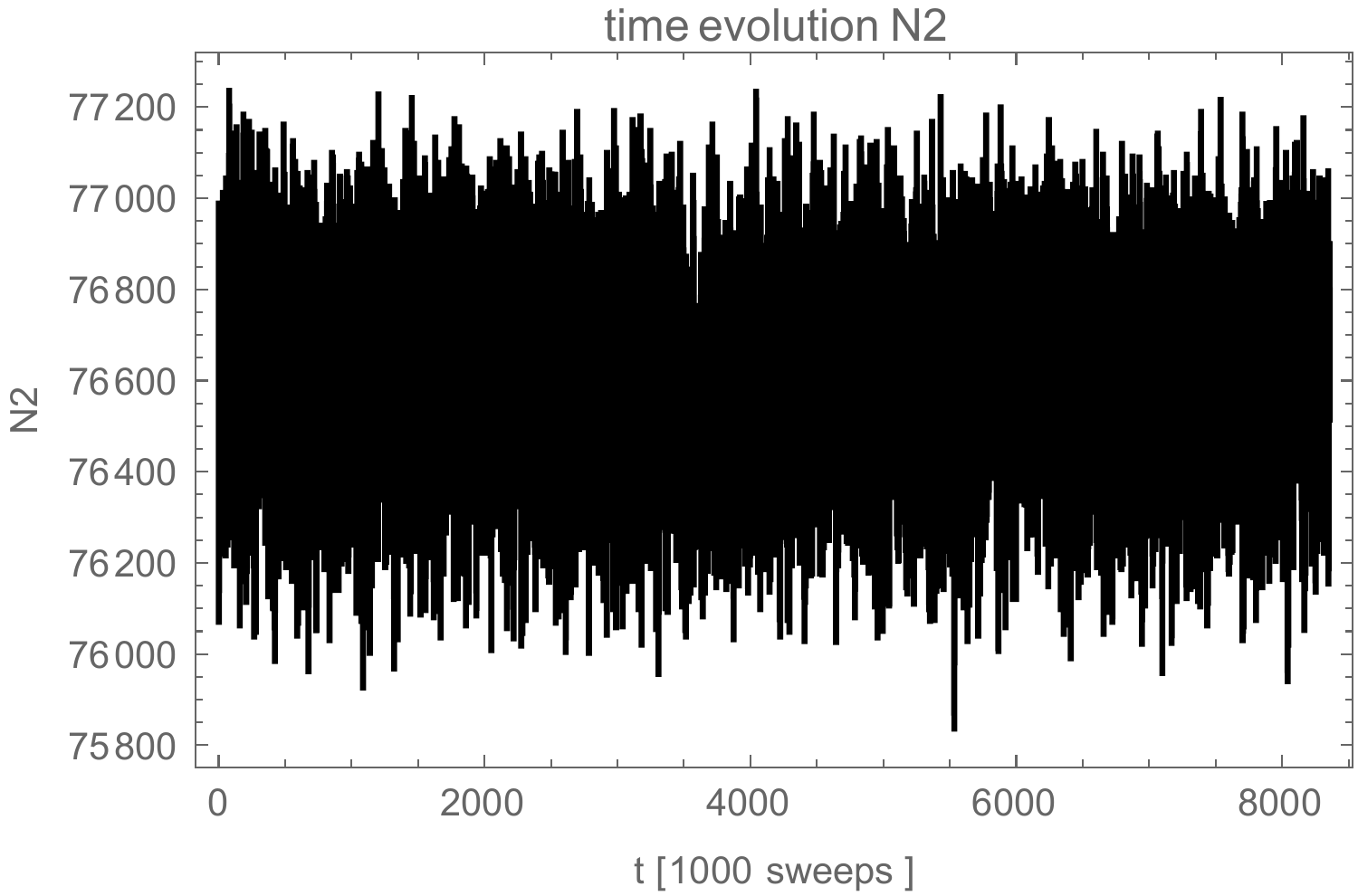}
\end{minipage}
\caption{The figures show Monte Carlo time histories for the observable $N_{2}$ (number of triangles) for systems of size $N_{4}=32k$ with $\kappa_{2}$ set to the pseudo-critical value $\kappa_{2}^{pcr}\of{32\text{k}}\approx 1.258$. The left-hand figure corresponds to the result of an ordinary simulation whereas the right-hand figure was obtained using parallel tempering with 24 replicas.}
\label{fig:n2timehist}
\end{figure}

In our implementation we chose, for a fixed average volume $\bar{N}_{4}$, 24 or 48 equally spaced (w.r.t. the $\kappa_{2}$ direction) couplings along the pseudo-critical line $\kappa_{4}^{pcr}\of{\kappa_{2},\bar{N}_{4}}$, such that they join a region in the crumpled phase where relaxation is fast, with one in the elongated phase where relaxation is also fast (compared to the critical region), and thereby pass through the critical region around $\kappa_{2}^{pcr}\of{\bar{N}_{4}}$. After some runtime, the optimization procedure of \cite{Troyer,Bauer} is applied which gives us a new set of couplings for which the configuration exchange between replicas is more frequent.\\
In Fig. \ref{fig:n2timehist}, two Monte Carlo time histories for the observable $N_{2}$ (number of triangles) are shown for comparison, both for a system of size $N_{4}=32$k at the pseudo-critical point $\kappa_{2}=\kappa_{2}^{pcr}\of{32\text{k}}\approx 1.258$. The left one stems from an ordinary simulation while the right one was obtained using parallel tempering.

\subsection{Data Analysis}
\begin{figure}[H]
\centering
\includegraphics[width=0.7\linewidth]{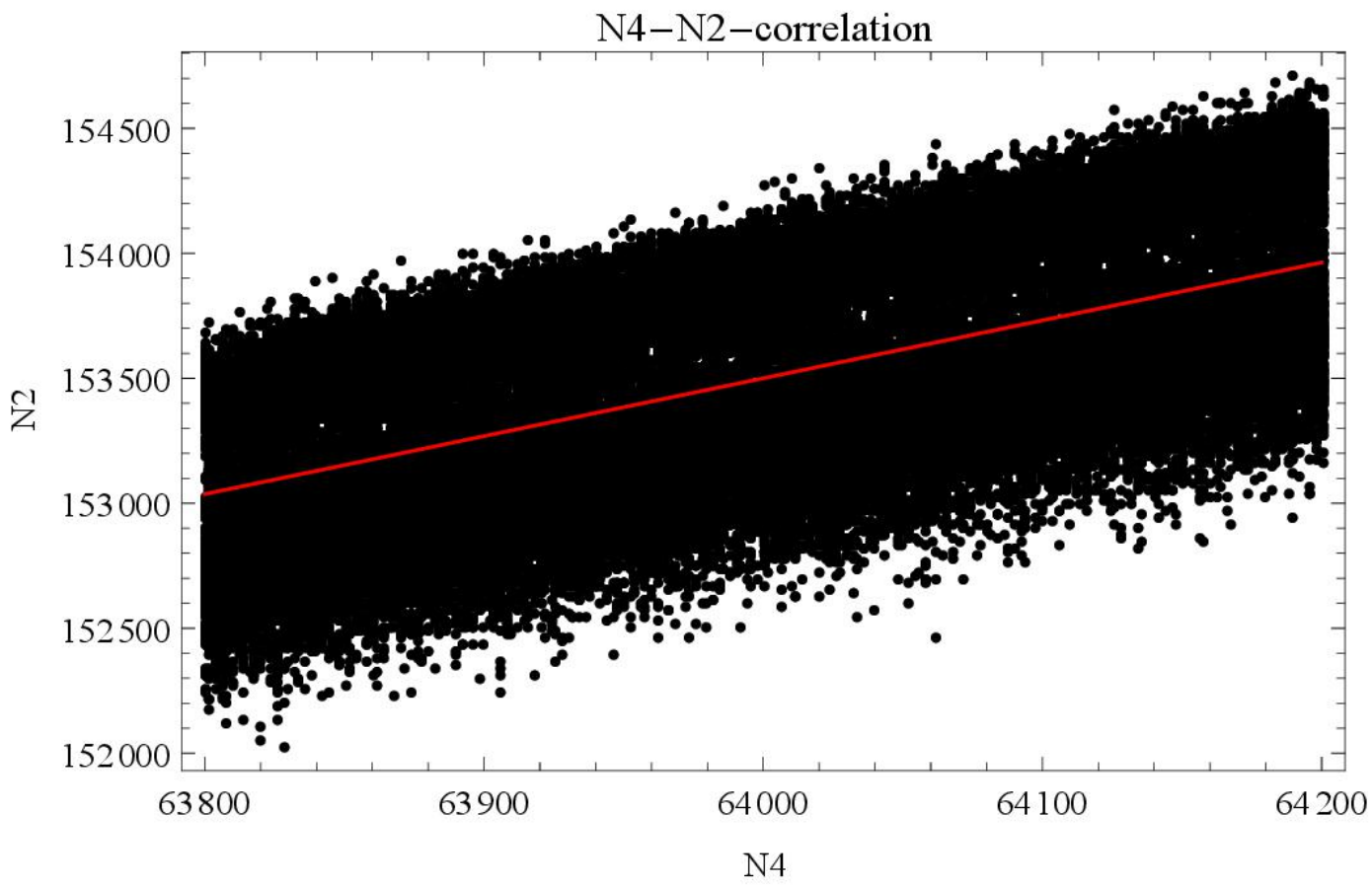}
\caption{The figure shows the $N_{2}$-$N_{4}$ distribution at $\kappa_{2}\,=\,2.8014$ for a system of average size $\bar{N}_{4}\,=\,64$k. As long as the fluctuations in $N_{4}$ are forced to be much smaller than the average system size $\bar{N}_{4}$ itself, we can project the data along the $N_{2}$-$N_{4}$-correlation direction (indicated by a red line) and evaluate observables as if we had a true canonical simulation at system size $\bar{N}_{4}$ and fluctuating triangle number $\bar{N}_{2}$ as given by \eqref{eq:n2eff} and \eqref{eq:n2eff2}.}
  \label{fig:n2proj}
\end{figure}

Due to the use of a potential well instead of a harmonic potential to control the system volume and due to the tuning of $\kappa_{4}$ to its pseudo-critical value, we have significant volume fluctuations in the data which also affect for example the $N_{2}$ distribution. To take this into account, we project the data in the $\of{N_{2},N_{4}}$-plane along the "correlation direction" before evaluating any observables (see Fig. \ref{fig:n2proj}), i.e. instead of $N_{2}$ we use
\[
\bar{N}_{2}\,=\,N_{2}\,-\,f\of{N_{4}}\label{eq:n2eff}
\]
to evaluate observables depending on $N_{2}$, where
\[
f\of{N_{4}}\,=\,\frac{\avof{\of{N_{2}-\avof{N_{2}}}\of{N_{4}-\avof{N_{4}}}}}{\avof{N_{4}-\avof{N_{4}}}^{2}}\,\of{N_{4}\,-\,\avof{N_{4}}}.\label{eq:n2eff2}
\]
We checked that this leads to the same results as when evaluating the observables only on data subsets corresponding to single, fixed $N_{4}$ values.\\

After that, we use \emph{multi-histogram reweighting} \cite{Ferrenberg} with respect to $\kappa_{2}$. The parallel tempering optimization procedure mentioned above also leads to a good distribution of simulation points for the reweighting. The errors are determined by the Jack-Knife method with 20 sets. In multi-histogram reweighting, these sets consist of the simultaneous data of all the simulations at different $\kappa_{2}$ values (i.e. to form the Jack-Knife sets we consider as a measurement all the measurements at different $\kappa_{2}$ values which correspond to the same Monte Carlo time), therefore cross correlations should automatically be taken into account. 

\section{Results}\label{sec:results}
\subsection{Order of Phase Transition}\label{sec:orderofphasetransition}
\subsubsection{\texorpdfstring{$N_{2}$}{TEXT} Distribution}
\begin{figure}[H]
\centering
\begin{minipage}[t]{0.49\linewidth}
\centering
\includegraphics[width=\linewidth]{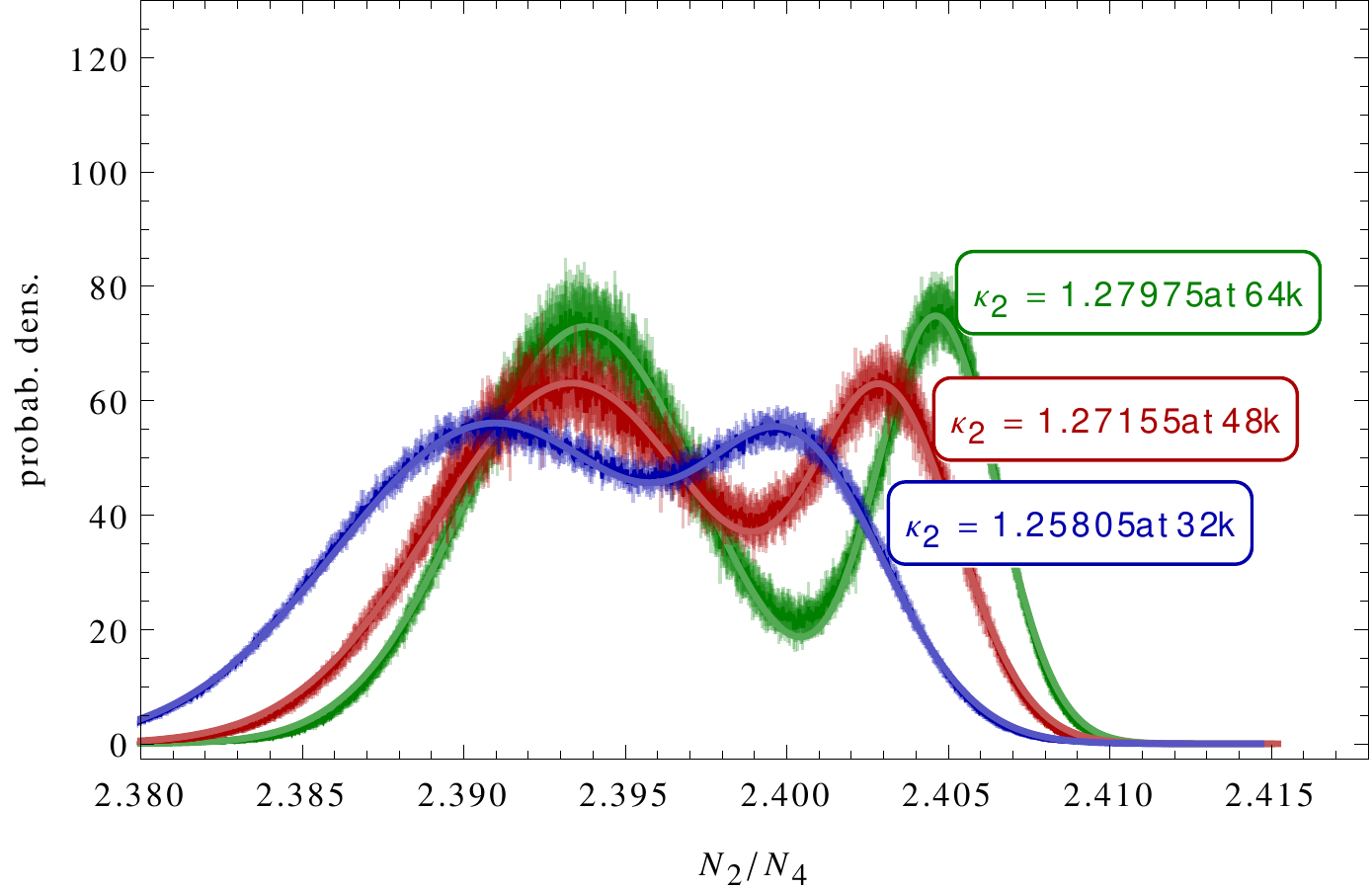}
\end{minipage}\hfill
\begin{minipage}[t]{0.49\linewidth}
\centering
\includegraphics[width=\linewidth]{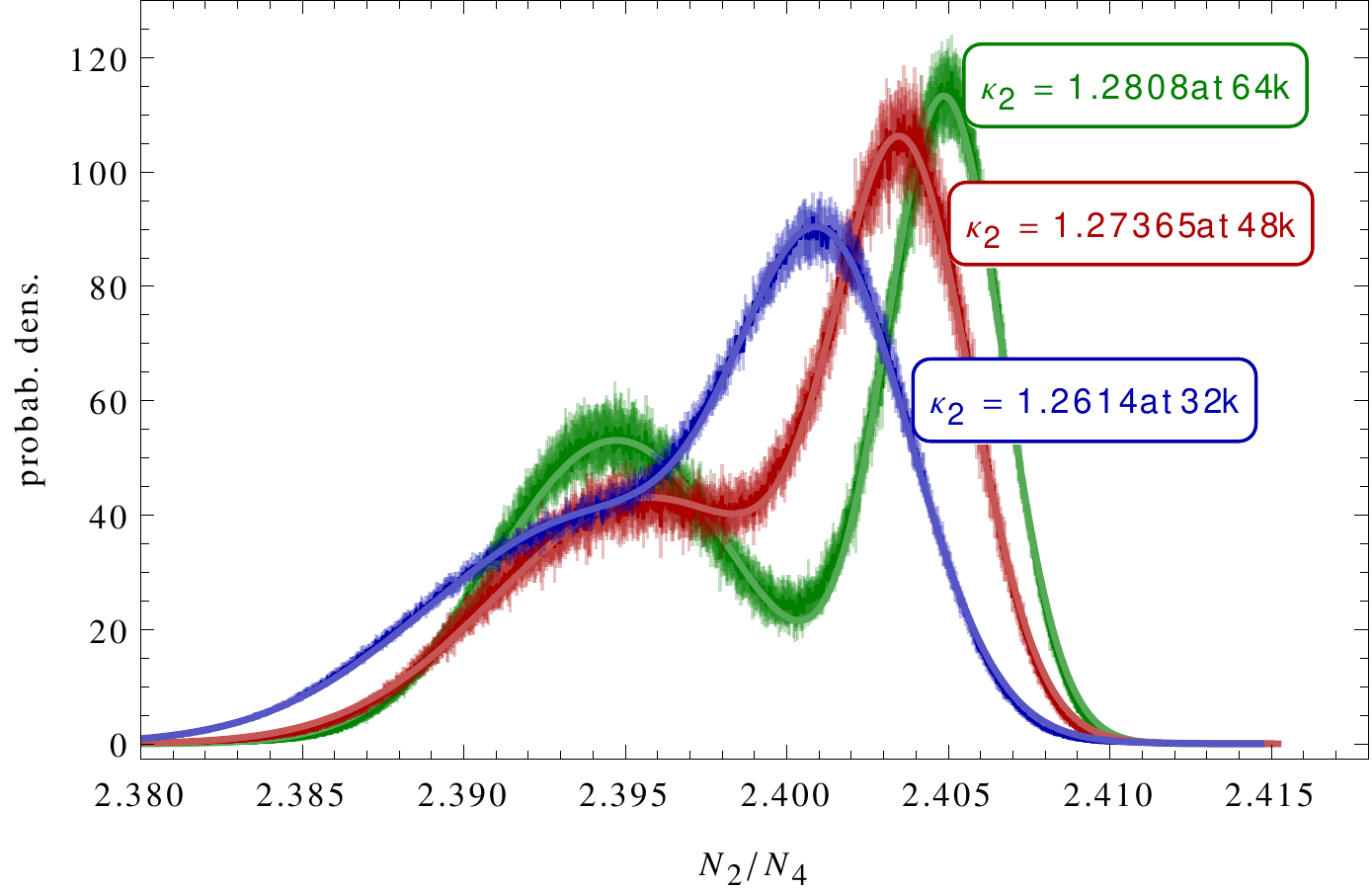}
\end{minipage}
\caption{Normalized $N_{2}$-distribution for systems of size $N_{4}\,=\,$ 32k (blue), 48k (red) and 64k (green): the solid lines are double-Gaussian fits to the data. To the left, the values of $\kappa_2$ were chosen such that the two Gaussian parts of the distribution function have the same height, whereas on the right-hand side the $\kappa_2$ values are such that the two Gaussians have the same area, i.e. the two states are equally probable. It can be seen that the double peak structure becomes more pronounced with increasing system size and there is no sign that the peaks will merge again in the thermodynamic limit. This is characteristic of a $\ford$ transition.}
\label{fig:N2dist}
\end{figure}

As stated in \cite{deBakker}, the pseudo-critical $N_{2}$-distribution starts to be visibly double-Gaussian for systems consisting of more than about $32000$ 4-simplices. For systems containing 32k, 48k and 64k 4-simplices, these distributions are shown in Figure \ref{fig:N2dist}, where $\kappa_{2}$ was tuned to produce peaks of equal height (left) or equal area (right). Either way it can be seen that the double peak structure becomes more pronounced with increasing system size and that there is so far no sign that the peaks will merge again in the thermodynamic limit. This behavior is characteristic of a $\ford$ transition. 

\begin{figure}[H]
\centering
\begin{minipage}[t]{0.49\linewidth}
\centering
\includegraphics[angle=90,origin=c,width=0.85\linewidth]{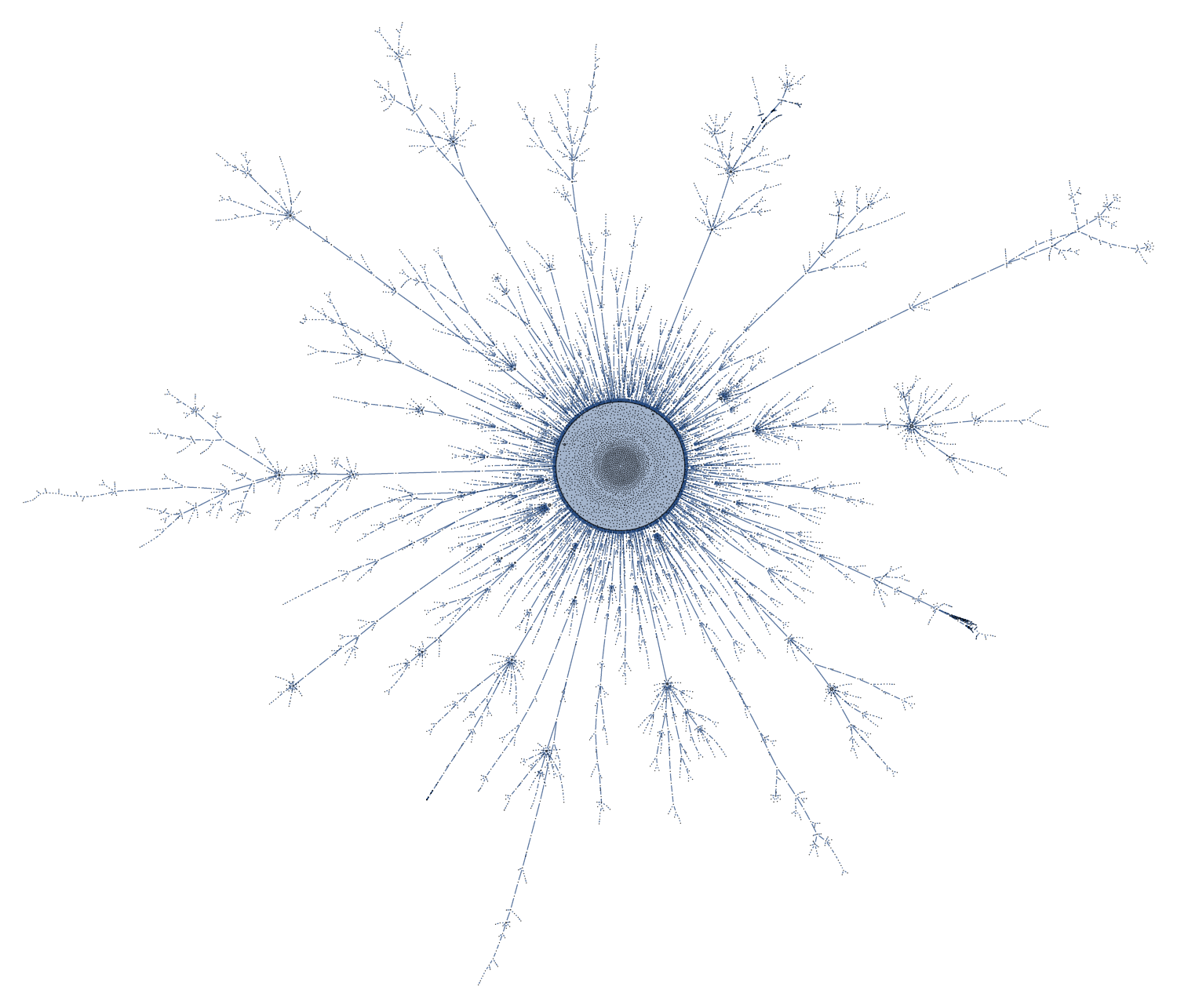}
\end{minipage}\hfill
\begin{minipage}[t]{0.49\linewidth}
\centering
\includegraphics[angle=90,origin=c,width=0.85\linewidth]{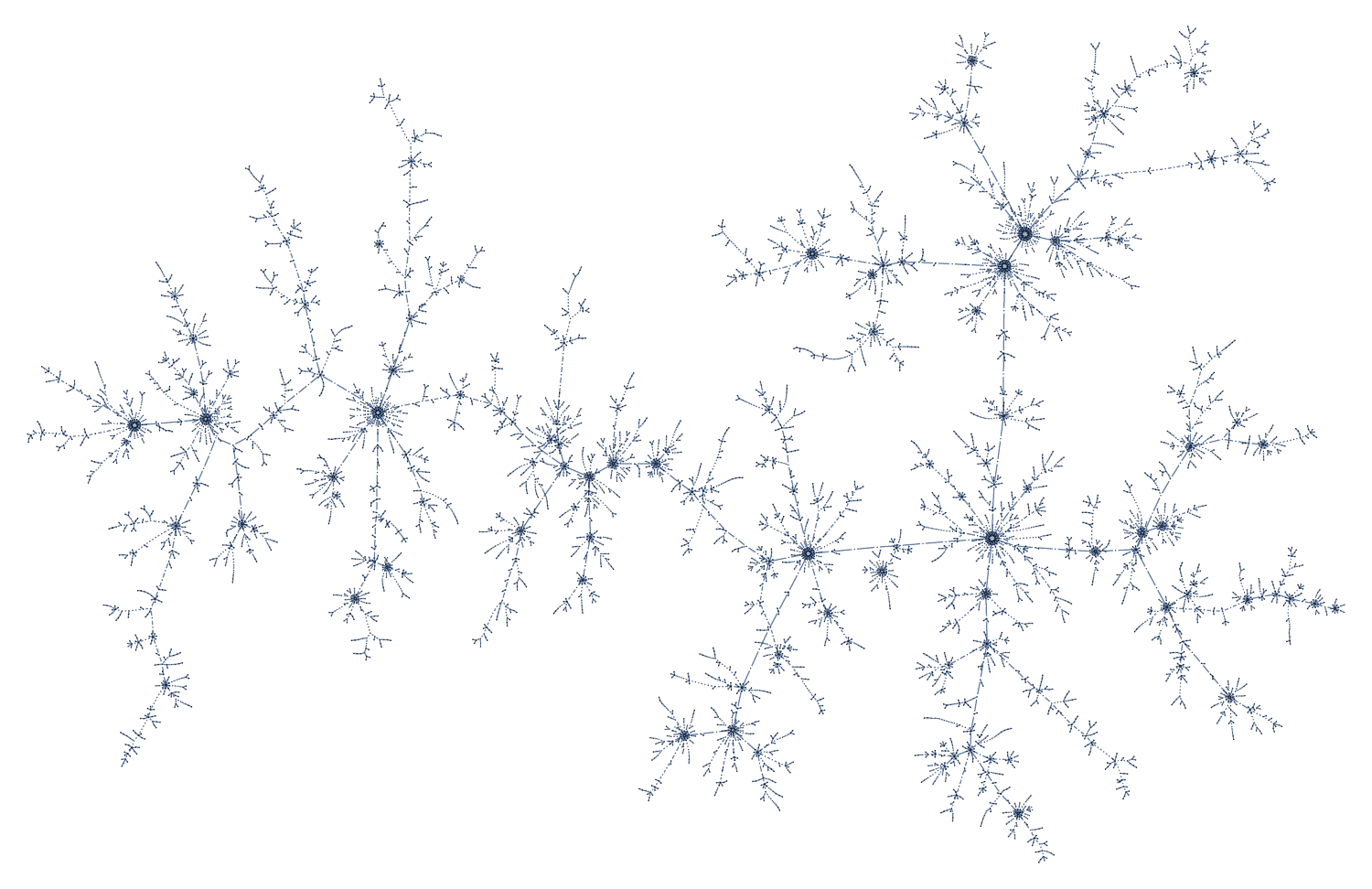}
\end{minipage}
\caption{Example of a crumpled (left) and an elongated (right) configuration with $N_{4}\approx 64$k, both recorded at $\kappa_{2}=1.28$, which is approximately the pseudo-critical point.}
\label{fig:butreesfixedkappa2}
\end{figure}

The reason for the double peak structure in the $N_{2}$-distribution is, that at a $\ford$ transition point, the two phases can coexist (see Fig. \ref{fig:butreesfixedkappa2}) while $N_{2}$ takes on different average values in each of these phases.   

\subsubsection{Scaling of \texorpdfstring{$B_{4}$}{TEXT}}
\begin{figure}[H]
\begin{minipage}[t]{0.49\linewidth}
\centering
\includegraphics[width=0.965\linewidth]{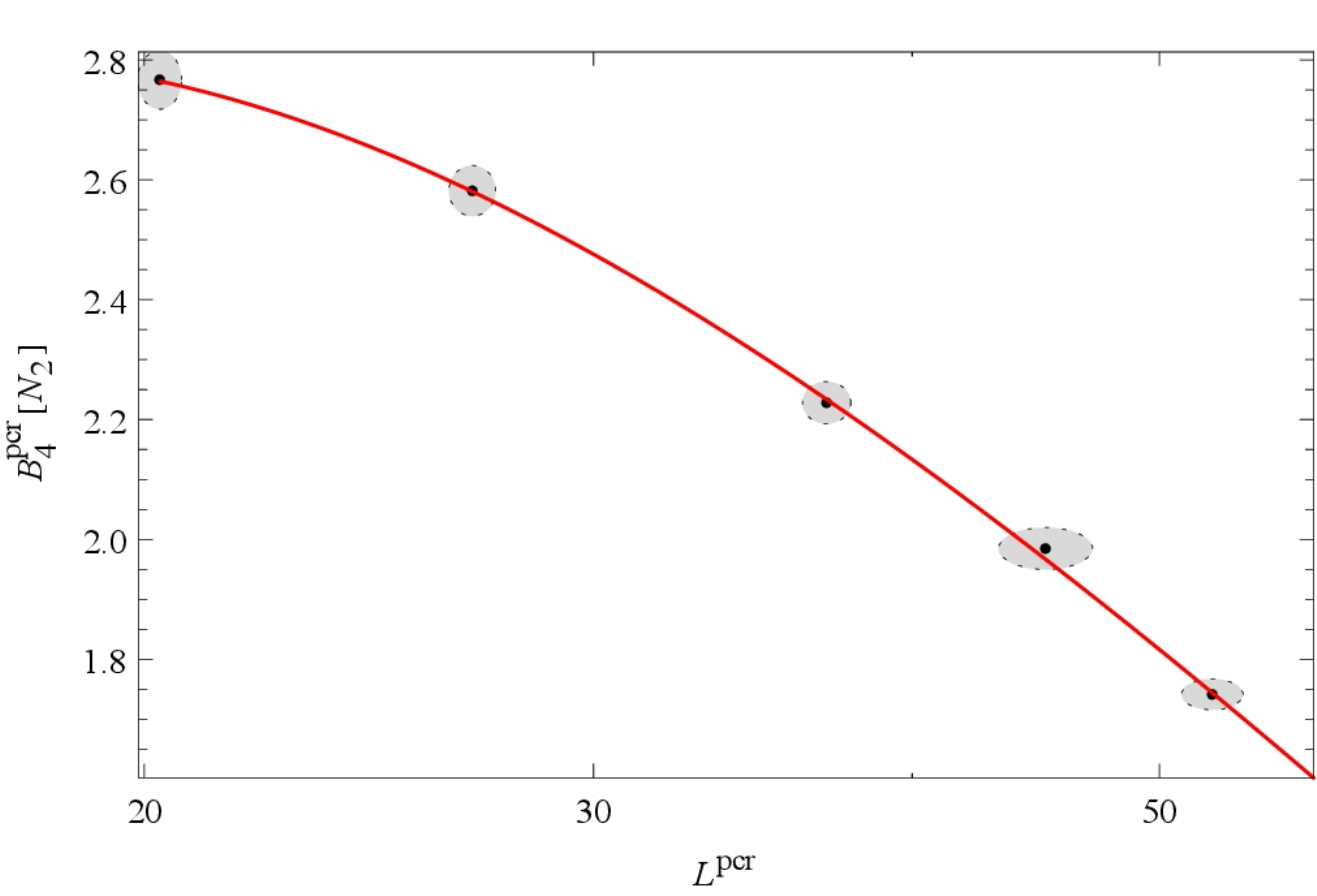}
\caption{Binder cumulant $B_{4}^{pcr}\fof{N_{2}}$ as a function of average linear system size $L^{pcr}\of{N_{4}}$ \eqref{eq:avlpcr}, assuming a $\sord$ transition, together with a fit of the form
\eqref{eq:b4fss} where $N_{4}^{\omega}=\of{L^{pcr}}^{1/\nu}$. We also included higher order corrections. It can be seen that the fit seems to work fine, but the obtained values
$B_{4}^{cr}\fof{N_{2}}\,=\,-4.2\,\pm\,5.0$ and $\nu\,=\,2.01\,\pm\,0.74$ do not make much sense.}
  \label{fig:B4scale2}
\end{minipage}\hfill
\begin{minipage}[t]{0.49\linewidth}
\centering
\includegraphics[width=\linewidth]{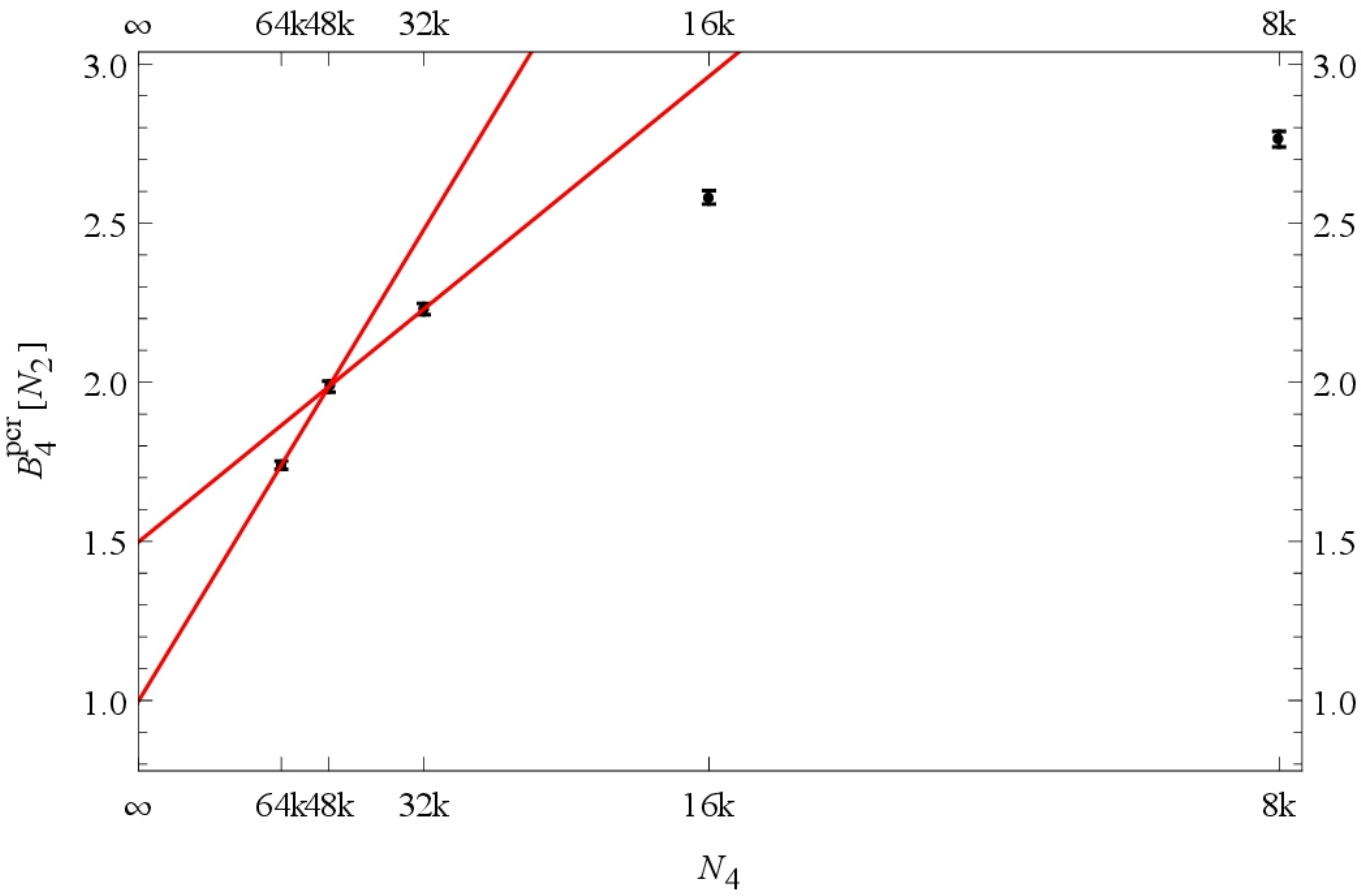}
\caption{Binder cumulant $B_{4}^{pcr}\fof{N_{2}}$ as a function of $1/N_{4}$. The red lines correspond to fits of the form \eqref{eq:b4fss} with $\omega=1$ ($\ford$ transition) to the
data of the largest and second largest pair of systems. The value $B_{4}^{cr}\fof{N_{2}}= 1.00\pm 0.08$ obtained from the largest pair is compatible with the expected value of 1 for a $\ford$ transition. The fit for the second largest pair yields a
too large value: $B_{4}^{cr}\fof{N_{2}} 1.50\pm 0.07$. This large change indicates that we are still far away from the thermodynamic limit.}
  \label{fig:B4scale1}
\end{minipage}
\end{figure}

A more quantitative method to determine the order of a phase transition is to study finite-size scaling of the $\nord{4}$ Binder cumulant (Kurtosis) of the $N_{2}$ distribution,
$B_{4}\fof{N_{2}}=\frac{\avof{\of{N_{2}-\avof{N_{2}}}^{4}}}{\avof{\of{N_{2}-\avof{N_{2}}}^{2}}^{2}}$.
According to \cite{Binder}, for large enough $N_{4}$ this quantity should scale like
\[
B_{4}^{pcr}\fof{N_{2}}\of{N_{4}}\,\approx\,B_{4}^{cr}\fof{N_{2}}\,+\,c_{1}\,N_{4}^{-\omega},\label{eq:b4fss}
\]
where $B_{4}^{cr}\fof{N_{2}}$ is the critical, infinite volume value of the Binder cumulant. For a $\sord$ transition one should get $1<B_{4}^{cr}\fof{N_{2}}<3$ and $\omega\,=\,1/d_{H}\nu$, where
$\nu$ is the critical exponent of the correlation length $\xi_{N_{2}}\approx \abs{\kappa_{2}^{cr}-\kappa_{2}}^{-\nu}$ and $d_{H}$ the \emph{Hausdorff dimension}, whereas for a $\ford$ transition we should obtain $B_{4}^{cr}\fof{N_{2}}=1$ and
$\omega=1$.\\
We tried to fit our data for $B_{4}^{pcr}\fof{N_{2}}\of{N_{4}}$ assuming $1^{\text{st}}$ and $\sord$ scaling ansaetze (see figures \ref{fig:B4scale1} and \ref{fig:B4scale2}). For the $\sord$ ansatz, we looked at $B_{4}^{pcr}$ as a function of the average linear system size $L^{pcr}$ instead of volume $N_{4}$, as a diverging correlation length will be truncated to this $L^{pcr}$, which is defined as
\[
L^{pcr}\of{N_{4}}\,=\,\sum\limits_{r}\,r\,n\of{r;N_{4},\kappa_{2}^{pcr}\of{N_{4}}},\label{eq:avlpcr}
\]
with $n\of{r;N_{4},\kappa_{2}}$ being the average volume profile of a triangulation of size $N_{4}$ at a given value of $\kappa_{2}$, i.e.:
\[
n\of{r;N_{4},\kappa_{2}}\,=\,\avof{\frac{1}{N_{4}^{2}}\sum\limits_{s_{1}}\,\sum\limits_{s_{2}}\,\delta\of{d_{\Delta}\of{s_{1},s_{2}}-r}}_{\kappa_{2}},
\]
where $s_{1}$, $s_{2}$ run over all $N_{4}$ 4-simplices and $d_{\Delta}\of{s_{1},s_{2}}$ is the geodesic distance between the simplices $s_{1}$ and $s_{2}$ in the triangulation $\Delta$ and $\avof{\ldots}_{\kappa_{2}}$ refers to the average over triangulations at $\kappa_{2}$. 
As is typical for a weak $\ford$ transition, the $\sord$ fit seems to work fine, but the obtained values $B_{4}^{cr}\fof{N_{2}}\,=\,-4.2\,\pm\,5.0$ and $\nu\,=\,1/d_{H}\omega\,=\,2.01\,\pm\,0.74$ do not make much sense.
Instead fixing $\omega=1$ for the $\ford$ ansatz, and using only the data points from the largest two simulated systems ($N_{4}=64$k and $48$k), we obtain $B_{4}^{cr}\fof{N_{2}}= 1.00\pm 0.08$ which is the expected value for a $\ford$ transition.
A fit to the data of the next smaller pair of systems, i.e. those consisting of 48k and 32k 4-simplices, leads to $B_{4}^{cr}\fof{N_{2}}=1.50\pm 0.07$. This large change in the value of the extrapolated $B_{4}^{cr}\fof{N_{2}}$ indicates that we are still far away from the thermodynamic limit.\\
\newpage
\subsection{Coexistence of Phases}\label{ssec:microscopicprop}
\begin{figure}[h]
\begin{minipage}[t]{0.485\linewidth}
\centering
\begin{tikzpicture}
\node[inner sep=0pt,above right] {\includegraphics[width=\linewidth]{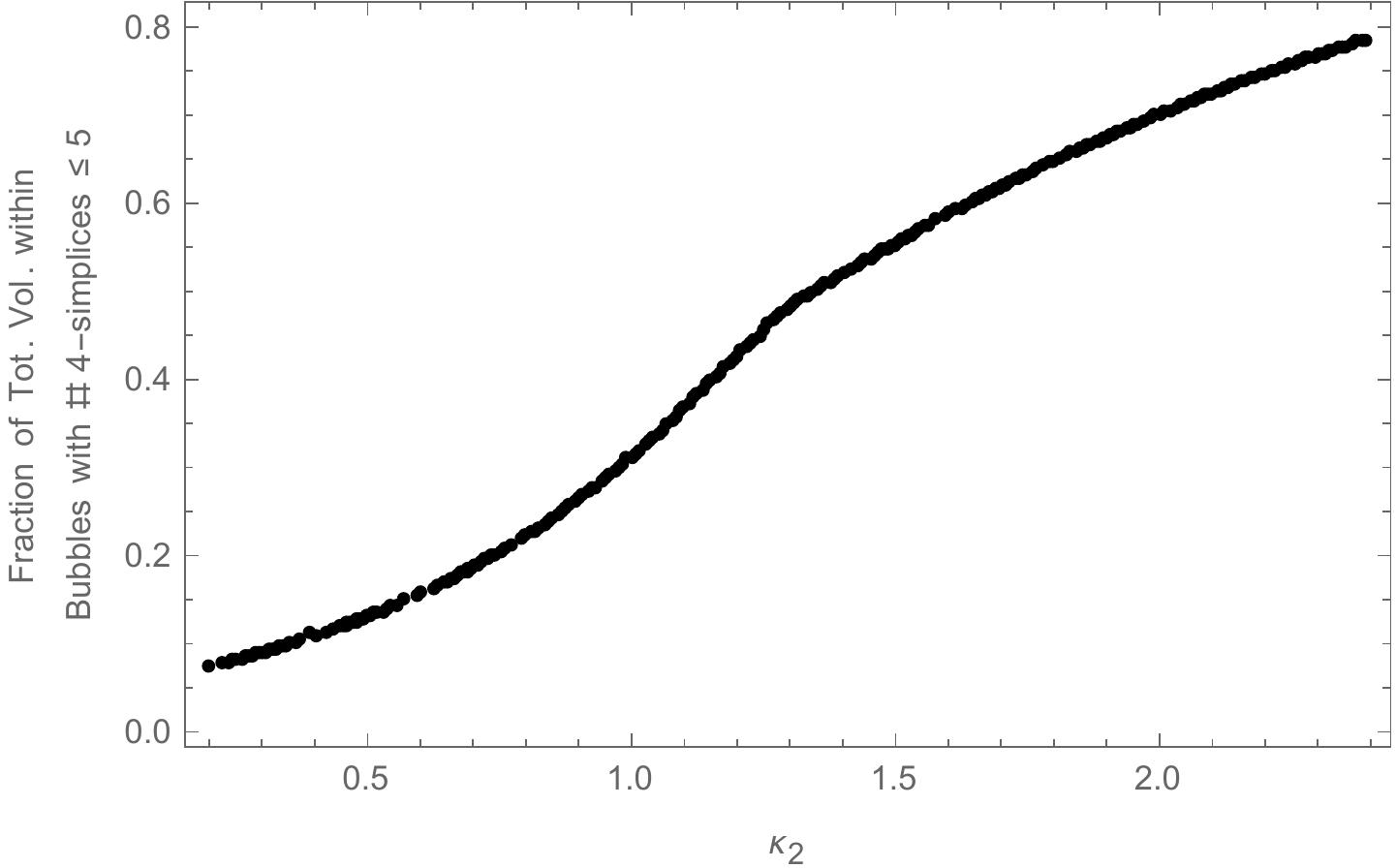}};
\draw[-,line width=0.1pt,red] (0.9,2.8125) -- (3.94,2.8125) -- (3.94,0.6) node[below] {\tiny $\kappa_{2}^{ds}$};
\end{tikzpicture}
\end{minipage}\hfill
\begin{minipage}[t]{0.495\linewidth}
\centering
\includegraphics[width=\linewidth]{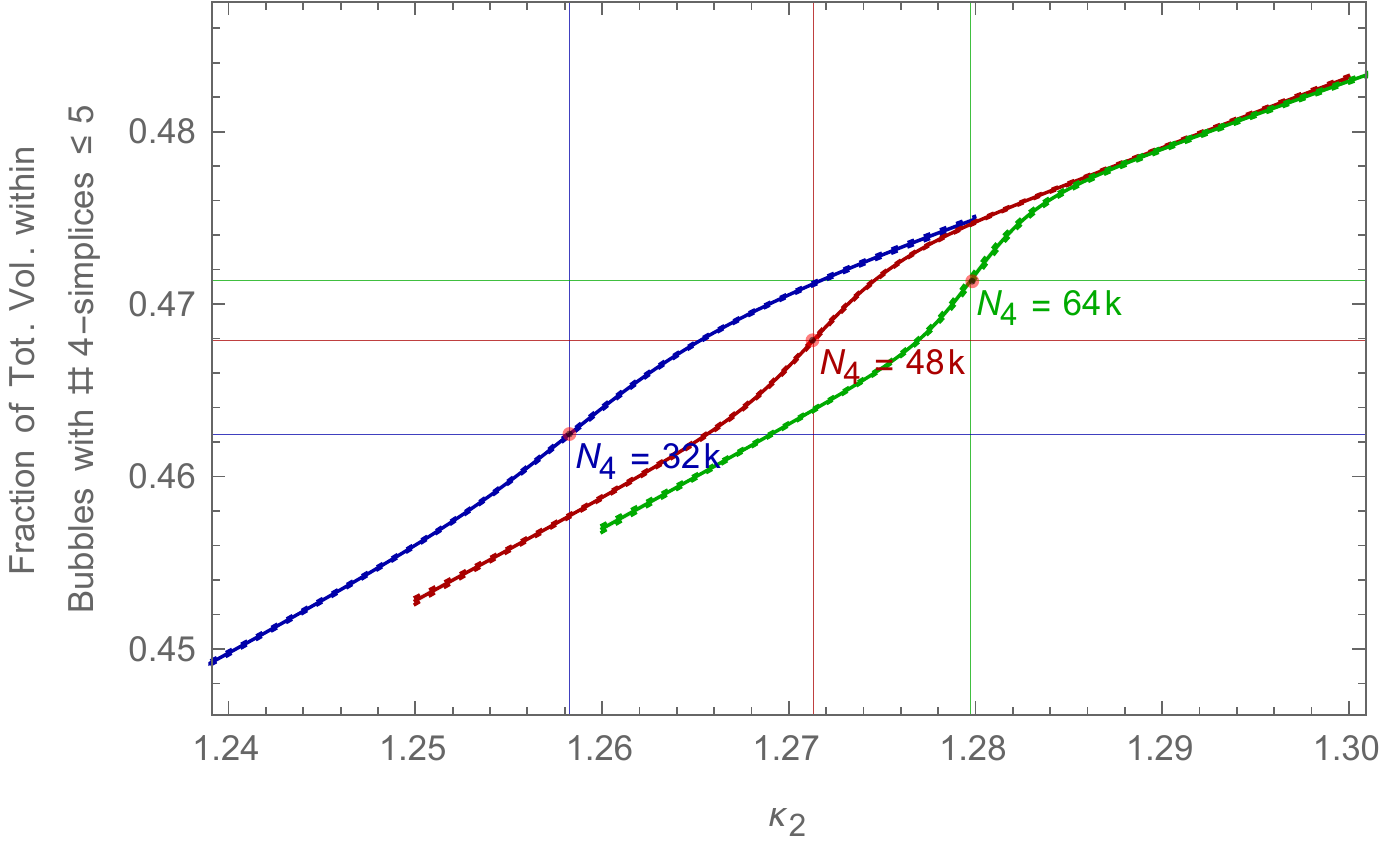}
\end{minipage}
\caption{The left-hand figure shows the ensemble averaged fractional volume contained in bubbles consisting of five or less $4$-simplices (i.e. bubbles of size 6) as a function of $\kappa_{2}$ at system size $N_{4}=32$k. With increasing $\kappa_{2}$, it can be seen that at some point $\kappa_{2}=\kappa_{2}^{ds}\of{N_{4}}$, the curve goes above $0.5$ and therefore the two classes of bubbles, those consisting of more than five and those consisting of five or less $4$-simplices, change their roles as dominant and non-dominant parts of the system. The right-hand figure shows the same quantity but for three different system sizes, $N_{4}=32$k (dark blue), $48$k (dark red), $64$k (dark green), in a neighbourhood of the corresponding pseudo-critical points: $\kappa_{2}^{pcr}\of{\cof{32\text{k},48\text{k},64\text{k}}}=\cof{1.258,1.271,1.280}$. Due to the finite volumes, $\kappa_{2}^{pcr}\of{N_{4}}$ is smaller than $\kappa_{2}^{ds}\of{N_{4}}$ which seems to be volume independent and is very close to the expected infinite volume critical value of the coupling $\kappa_{2}^{cr}\approx 1.33$.}
  \label{fig:bubblesizeevol}
\end{figure}

\begin{figure}[H]
\centering
\includegraphics[width=0.6\linewidth]{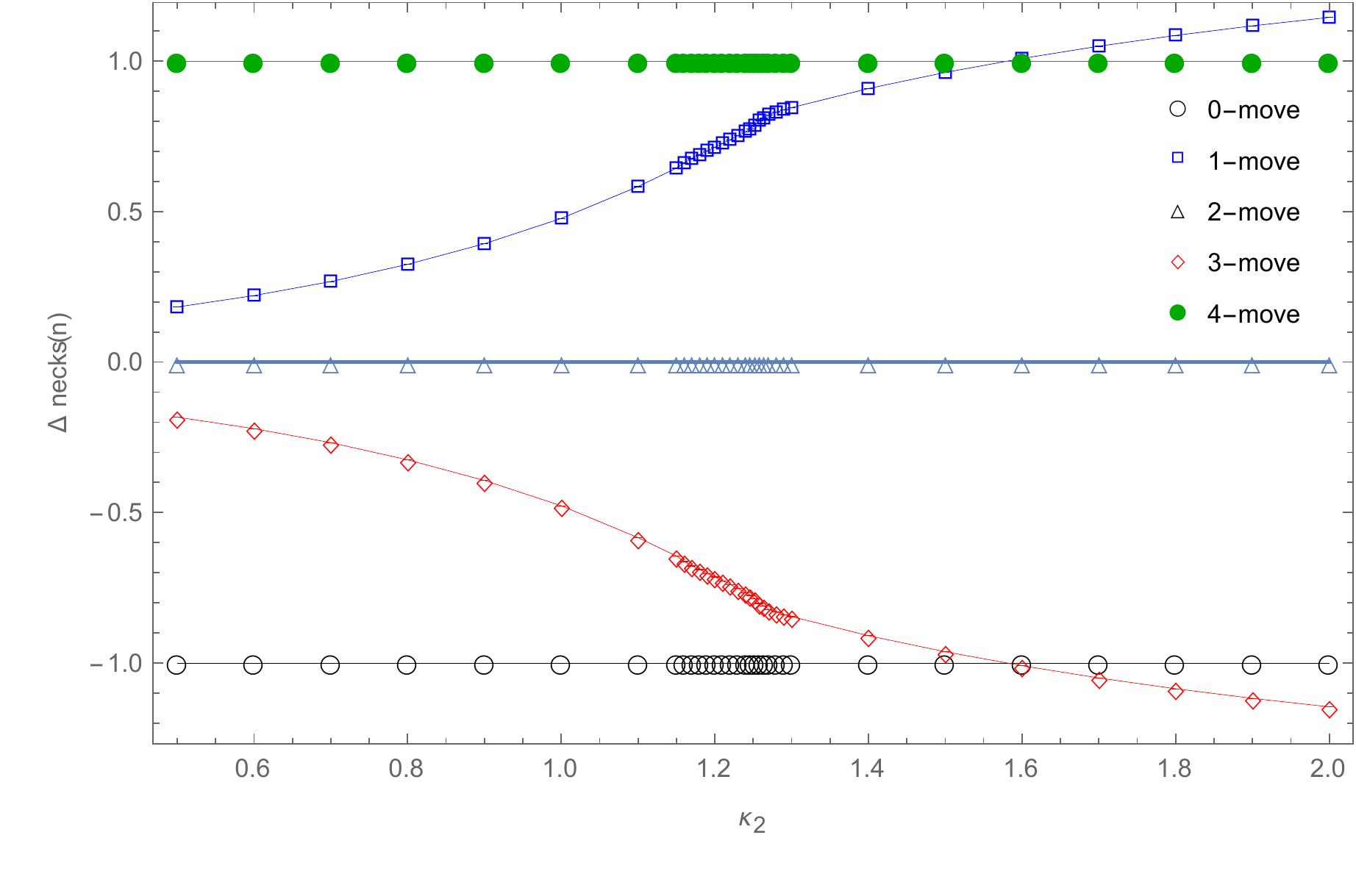}
\caption{The figure shows for a system of size $N_{4}\,=\,32$k, as a function of $\kappa_{2}$, the average change of the number of necks caused by the next $n$-move. The 0- and 4-moves always remove or add a "volume 5" bubble and a corresponding neck, while the 2-move never changes the number of necks. For the 1- and 3-moves $\Delta necks\of{n}$ changes as function of $\kappa_{2}$ as the fraction of 3-simplices, which allow for a 3-move and are also part of a minimal neck changes. For $\kappa_{2}>1.6$, it seems that the average 3-simplex which allows for a move, is rather part of more than one neck than of no neck, which is why $\Delta necks\of{3}$ drops below -1. The behaviour of $\Delta necks\of{1}$ follows from the fact that the 1-moves is the inverse of the 3-move. Thus for $\kappa_{2}>1.6$, the triangle that will be created by applying a 1-move to one of the 1-simplices that allow for such a move, will be rather part of more than one neck than of no neck. At the pseudo-critical point $\kappa_{2}^{pcr}=1.258$, we have that $\Delta necks\of{\cof{1,3}}\approx \cof{0.8,-0.8}$.}
  \label{fig:deltanecks}
\end{figure}

As the phase transition is $\ford$ we expect for finite systems a coexistence of the elongated and crumpled phases in some neighborhood of the pseudo-critical point. Instead of speaking of crumpled and elongated configurations as in Fig. \ref{fig:butreesfixedkappa2}, we should rather speak of configurations in which the crumpled or the elongated part dominates. For example, the left-hand graph in Fig. \ref{fig:butreesfixedkappa2} is dominated by the large bubble\footnote{\emph{Bubble} is just another word for \emph{baby-universe} as defined above in footnote \ref{fn:babyuniverse}.} in the middle that is in the crumpled state, but attached to that large bubble are baby-universe trees which are in the elongated state. Similarly, the right-hand graph in Fig. \ref{fig:butreesfixedkappa2}, whose strong branching indicates that the configuration is mainly in the elongated state, also contains some larger bubbles that presumably are in the crumpled state.\\
Although criteria like "strong branching" and "large bubbles" seem to work well to decide if a piece of triangulation corresponds to the elongated or crumpled phase, there is some ambiguity in what "strong branching" should mean or what bubble sizes should be considered as small. As a first attempt, we could define the elongated phase as consisting of bubbles of size 6 and everything else as belonging to the crumpled phase. By the "size" of a bubble, we mean from now on the number of the bubble's 4-simplices plus the number of its minimal necks\footnote{A bubble of size 6 (which is the smallest possible size) can consist of five 4-simplices plus a minimal neck or four 4-simplices plus two minimal necks, and so on.} (i.e. the volume the bubble would have after replacing all minimal necks by ordinary 4-simplces). In contrast: the "volume" of a bubble still refers to just the number of 4-simplices of that bubble. Figure \ref{fig:bubblesizeevol} shows, as a function of $\kappa_{2}$, the average fractional volume of a triangulation contained in size 6 bubbles. The complementary fractional volume would therefore be the one contained in bubbles of size larger than 6. It can be seen that at small values of $\kappa_{2}$, deep in the crumpled phase, the volume is dominated by contributions from large bubbles (in fact one very large bubble containing almost all the $4$-simplices) whereas at large values of $\kappa_{2}$, deep in the elongated phase, the dominant contribution to the total volume comes from the size 6 bubbles. The swap in the dominance occurs not exactly at the pseudo-critical point but at a somewhat larger value of $\kappa_{2}\,=\,\kappa_{2}^{ds}$ which, however, seems to coincide with the expected infinite volume critical value of the coupling $\kappa_{2}^{cr}\approx 1.33$.\\
As the curves in Fig. \ref{fig:bubblesizeevol} show no volume dependency in the elongated phase and may get close to unity only in the limit $\kappa_{2}\rightarrow\infty$, it should be clear that the characterisation of this phase as consisting of just size 6 bubbles is not adequate. The reason is, that if a system would consist of size 6 bubbles only, Pachner 3-moves could only be applied to 3-simplices which are part of minimal necks, such that these moves would necessarily destroy those necks and thereby produce bubbles of size larger than 6. As can be seen in Fig. \ref{fig:deltanecks}, which shows as a function of $\kappa_{2}$ the quantity $\Delta necks\of{n}$, i.e. the average change of the number of necks under a $n$-move\footnote{The quantity $\Delta necks\of{n}$ is computed by counting for each location where a $n$-move could be applied, the number of necks that would be created or destroyed by such a move, and taking the average.}, indeed, more and more 3-moves change the number of necks as $\kappa_{2}$ increases. But nevertheless, the size 6 bubbles seem to be the dominant building blocks of the elongated phase.

\begin{figure}[htbp]
\centering
\begin{minipage}[t]{0.45\linewidth}
\centering
\includegraphics[width=\linewidth]{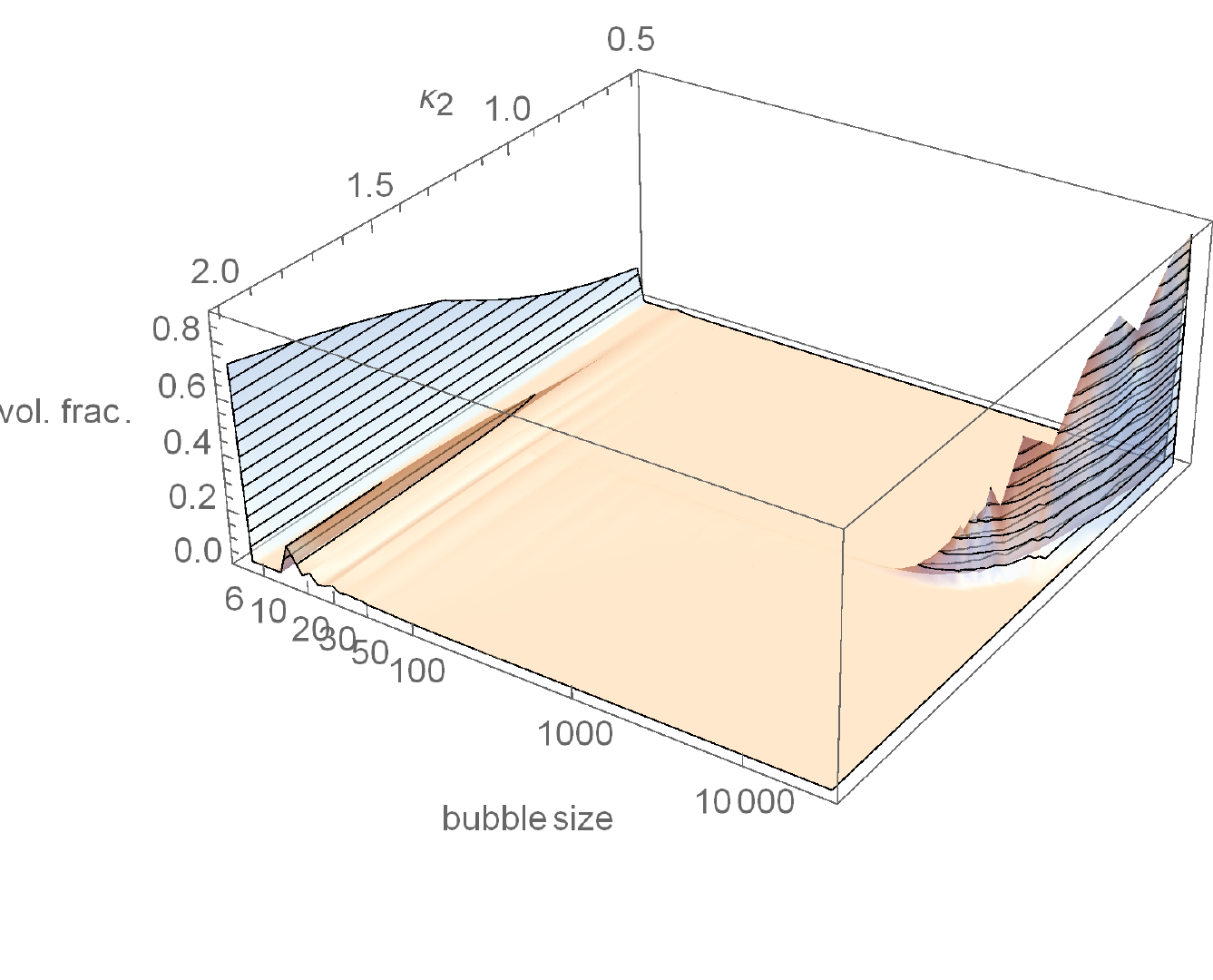}
\end{minipage}\hfill
\begin{minipage}[t]{0.54\linewidth}
\centering
\includegraphics[width=\linewidth]{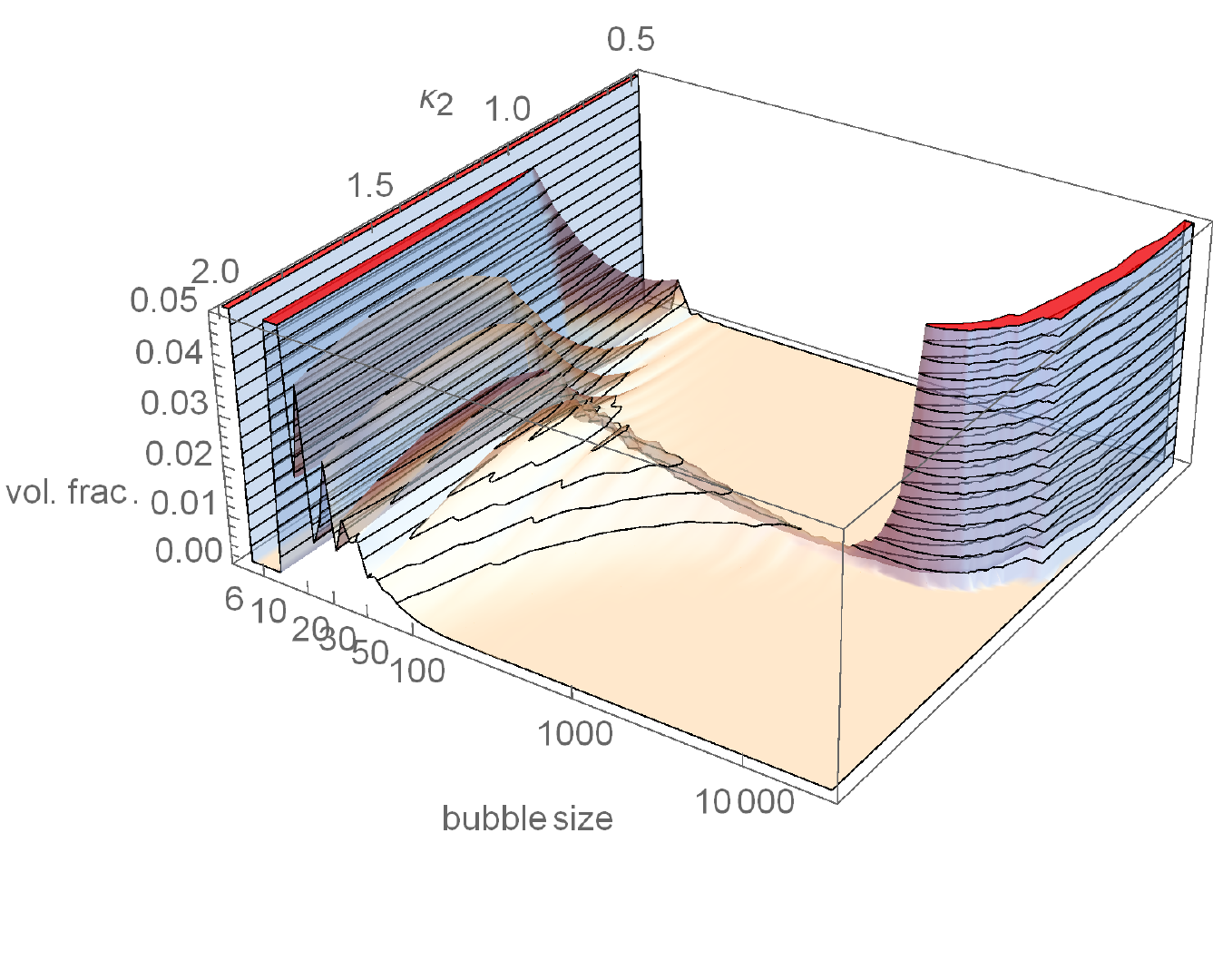}
\end{minipage}
\caption{The figures both show (on different scales) as a function of $\kappa_{2}$ how the total volume of a $N_{4}\,=\,32$k system is distributed, on average, over bubbles of different sizes, where the size is given by the number of 4-simplices plus the number of minimal necks: for $\kappa_{2}<<\kappa_{2}^{pcr}\of{N_{4}}$, almost the whole volume is concentrated in just one large bubble, but with increasing $\kappa_{2}$ the volume distributes over more and more (and larger) small bubbles until the (distinguishably) largest bubble disappears for $\kappa_{2}\approx\kappa_{2}^{pcr}\of{N_{4}}$. At this point, the largest bubble contains only about $20\%$ of the total volume.}
  \label{fig:busizehistvsk2}
\end{figure}

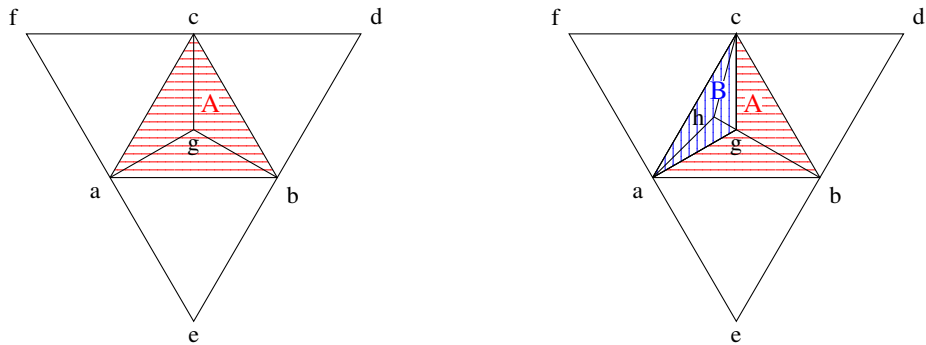
\begin{figure}[htbp]
\centering
\begin{minipage}[t]{0.49\linewidth}
\centering
  \begin{tikzpicture}[scale=1.1,baseline=-35]
    \node (a) at (-1,0) {};
    \node (b) at (1,0) {};
    \node (c) at (0,1.73205) {};
    \node (f) at (-2,1.73205) {};
    \node (d) at (2,1.73205) {};
    \node (e) at (0,-1.73205) {};
    \node (g) at (0,0.57735) {};
    %\filldraw[pattern=horizontal lines,pattern color=black,fill opacity=0.5] (a.center) -- (b.center) -- (c.center)  -- (a.center);
    \filldraw[pattern=custom horizontal lines,hatchspread=3pt,hatchthickness=0.25pt,hatchshift=1pt,hatchcolor=red] (a.center) -- (b.center) -- (c.center)  -- (a.center);
    \node[rectangle,red,inner sep=-0.5pt,fill=white,fill opacity=1] (al) at (0.2,0.9) {A};
    \draw (a.center) -- (f.center) -- (c.center);
    \draw (c.center) -- (d.center) -- (b.center);
    \draw (b.center) -- (e.center) -- (a.center);
    \draw (g.center) -- (a.center);
    \draw (g.center) -- (b.center);
    \draw (g.center) -- (c.center);
    \draw (a) node[below left] {\small a};
    \draw (b) node[below right] {\small b};
    \draw (c) node[above] {\small c};
    \draw (f) node[above left] {\small f};
    \draw (d) node[above right] {\small d};
    \draw (e) node[below] {\small e};
    \draw (g) node[below] {\small g};
  \end{tikzpicture}
\end{minipage}\hfill
\begin{minipage}[t]{0.49\linewidth}
\centering
  \begin{tikzpicture}[scale=1.1,baseline=-35]
    \node (a) at (-1,0) {};
    \node (b) at (1,0) {};
    \node (c) at (0,1.73205) {};
    \node (f) at (-2,1.73205) {};
    \node (d) at (2,1.73205) {};
    \node (e) at (0,-1.73205) {};
    \node (g) at (0,0.57735) {};
    \node (h) at (-0.266,0.730925) {};
    \filldraw[pattern=custom horizontal lines,hatchspread=3pt,hatchthickness=0.25pt,hatchshift=1pt,hatchcolor=red] (a.center) -- (b.center) -- (c.center)  -- (a.center);
    \filldraw[fill=white,fill opacity=1] (a.center) -- (g.center) -- (c.center)  -- (a.center);
    \filldraw[pattern=custom vertical lines,hatchspread=3pt,hatchthickness=0.25pt,hatchshift=1pt,hatchcolor=blue] (a.center) -- (g.center) -- (c.center)  -- (a.center);
    \draw (a.center) -- (f.center) -- (c.center);
    \draw (c.center) -- (d.center) -- (b.center);
    \draw (b.center) -- (e.center) -- (a.center);
    \draw (g.center) -- (a.center);
    \draw (g.center) -- (b.center);
    \draw (g.center) -- (c.center);
    \draw (h.center) -- (a.center);
    \draw (h.center) -- (g.center);
    \draw (h.center) -- (c.center);
    \draw (a) node[below left] {\small a};
    \draw (b) node[below right] {\small b};
    \draw (c) node[above] {\small c};
    \draw (f) node[above left] {\small f};
   	\draw (d) node[above right] {\small d};
    \draw (e) node[below] {\small e};
    \draw (g) node[below] {\small g};
    \draw (h) node[left] {\small h};
    \node[rectangle,red,inner sep=-0.5pt,fill=white] at (0.2,0.9) {A};
    \node[rectangle,blue,inner sep=0pt,fill=white] at (-0.21,1.05) {B};
  \end{tikzpicture}
\end{minipage}
\captionsetup{singlelinecheck=off}
\caption[Illustration in two dimensions]{Illustration in two dimensions: in the left Figure, by applying a 2-move to the general triangle (abc) (which is not part of a "volume 3"-baby-universe), we have replaced the triangle by a baby-universe "A" (horizontally shaded) consisting of the three triangles (abg), (bcg) and (cag). We could continue by applying 1-moves to all the 1-simplices of the neck (abc) which would replace the link (ac) by (gf), (cb) by (gd) and (ba) by (ge). In the right hand figure, we have instead produced another "volume 3"-baby-universe "B" (vertically shaded) with neck (agc) inside the already existing baby-universe "A". In this case, we can only apply a 1-move either to (ac) and (ag) or to (ac) and (gc), as applying a 1-move to both (ag) and (gc) would result in a double link (hb).\\
It is also clear that the insertion of the vertex "h" into one of the triangles of "A" makes it impossible to remove the vertex "g" by a 0-move. If "h" had instead been inserted into e.g. the triangle (acf), not just "h", but also "g" could still be removed.\\
Similarly in the four dimensional case, a piece of triangulation produced by applying a 4-move inside a "volume 5"-baby-universe "A" to produce a new "volume 5"-baby "B" of "A", has the following effects:
\begin{itemize}\itemsep-5pt
\item[-] the number of possible 0-moves is not increased, as the 4-move that created "B" has also destroyed an already existing location where a 0-move could have been applied: the vertex that was in the center of "A". This does not happen, if a 4-move is applied to a 4-simplex that is not part of a "volume-5"-bubble,\\
\item[-] all the 3-simplices that are only part of the neck between "A" and "B" (but not of the neck between "A" and the rest of the triangulation), would allow for a 3-move, but by performing one of these 3-moves, the remaining ones become impossible. If "B" were not the baby of a "size 5,4 or 3"-bubble, 3-moves could in general be applied to all of its neck's 3-simplices,\\
\item[-] and finally, in four dimensions, there is also an effect on the 2-move: the application of a 3-move to one of the 3-simplices that are part of only the neck between  "A" and "B", removes another location where a 0-move had been possible (as it destroys "B"), but at the same time, generates three new locations where a 2-move could be applied.
\end{itemize}
}
\label{fig:doublenecks}
\end{figure}

\begin{figure}[htbp]
\centering
\begin{minipage}[t]{0.33\linewidth}
\centering
\includegraphics[width=\linewidth]{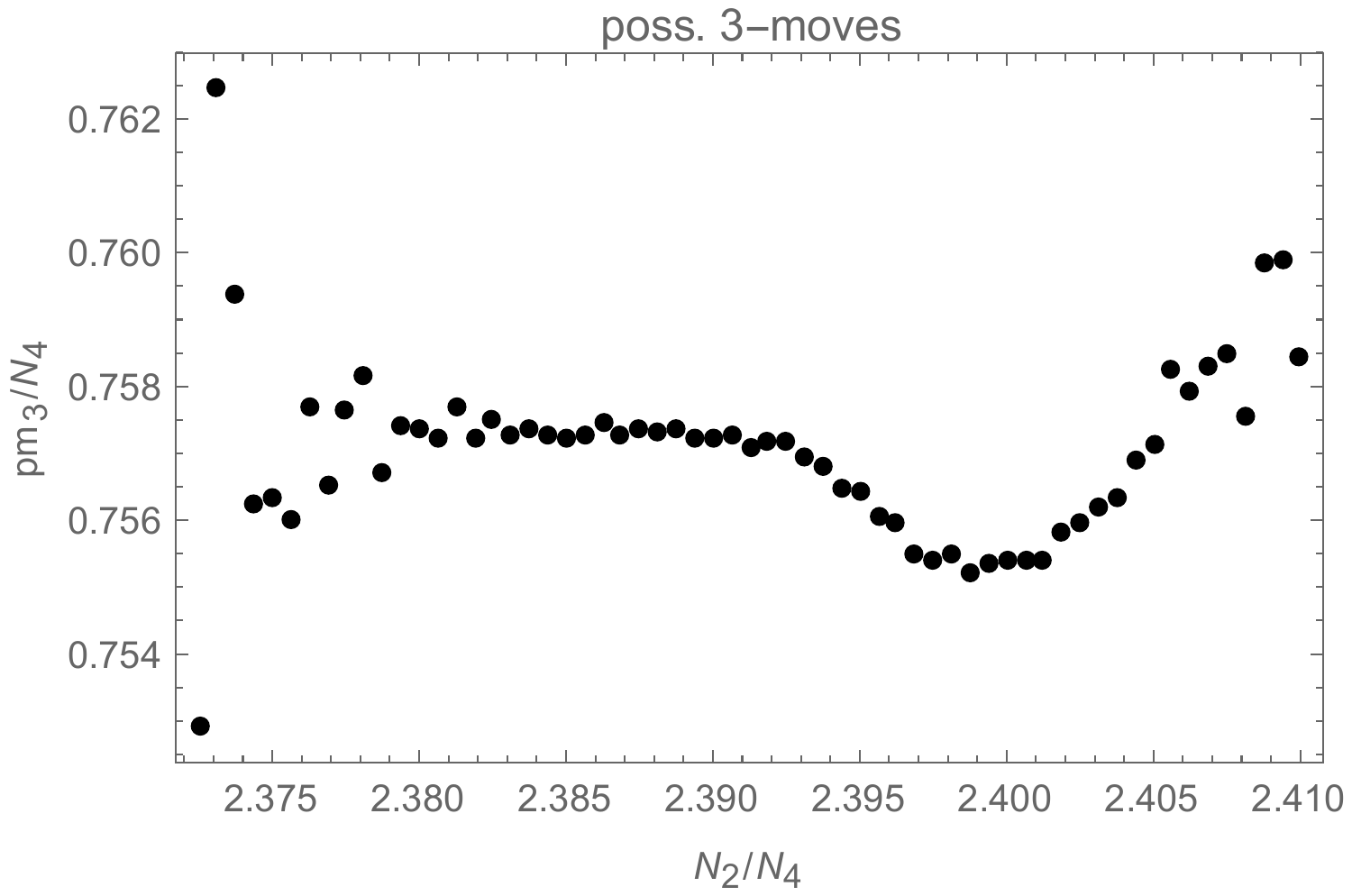}
\includegraphics[width=\linewidth]{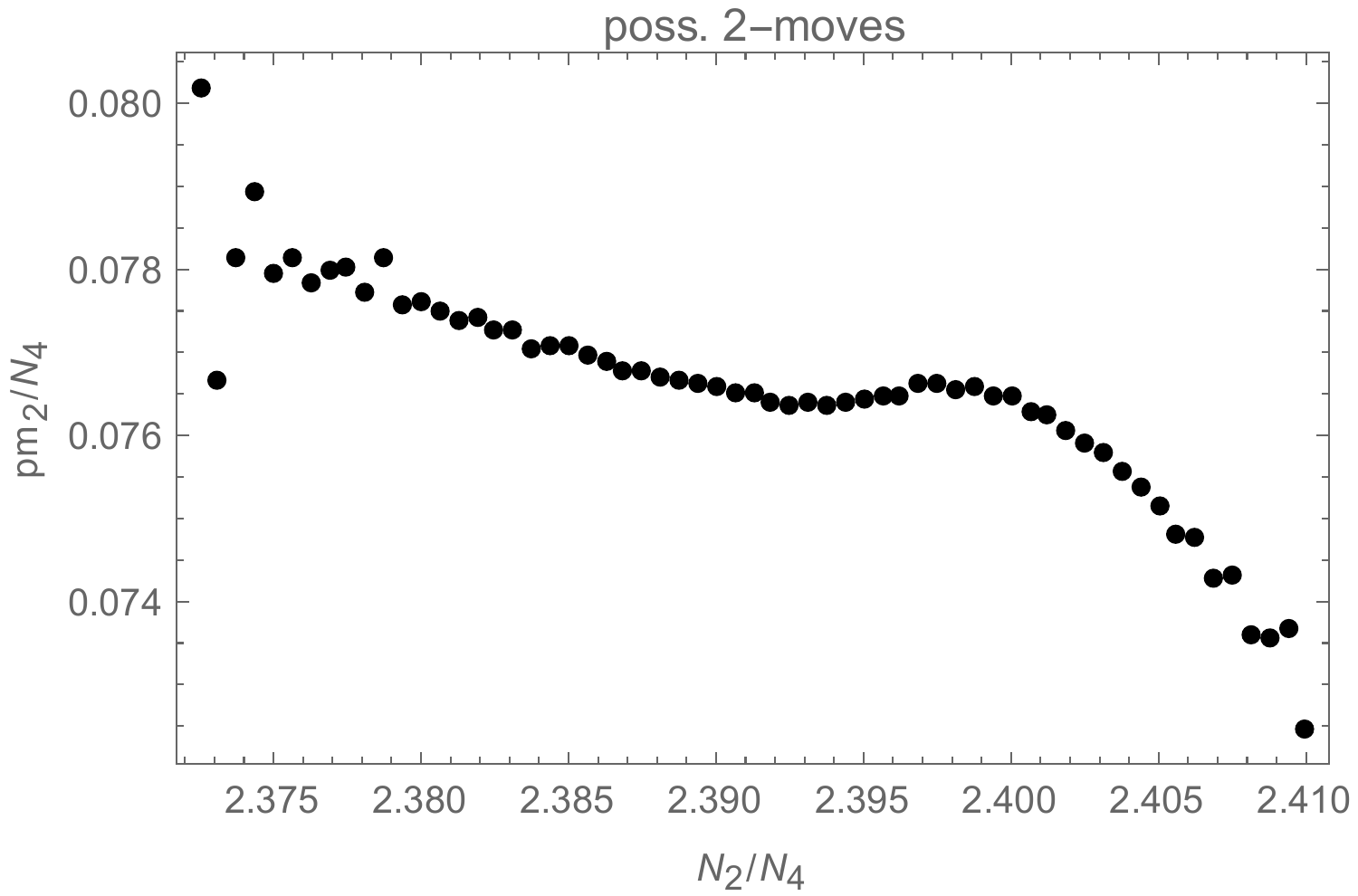}
\includegraphics[width=\linewidth]{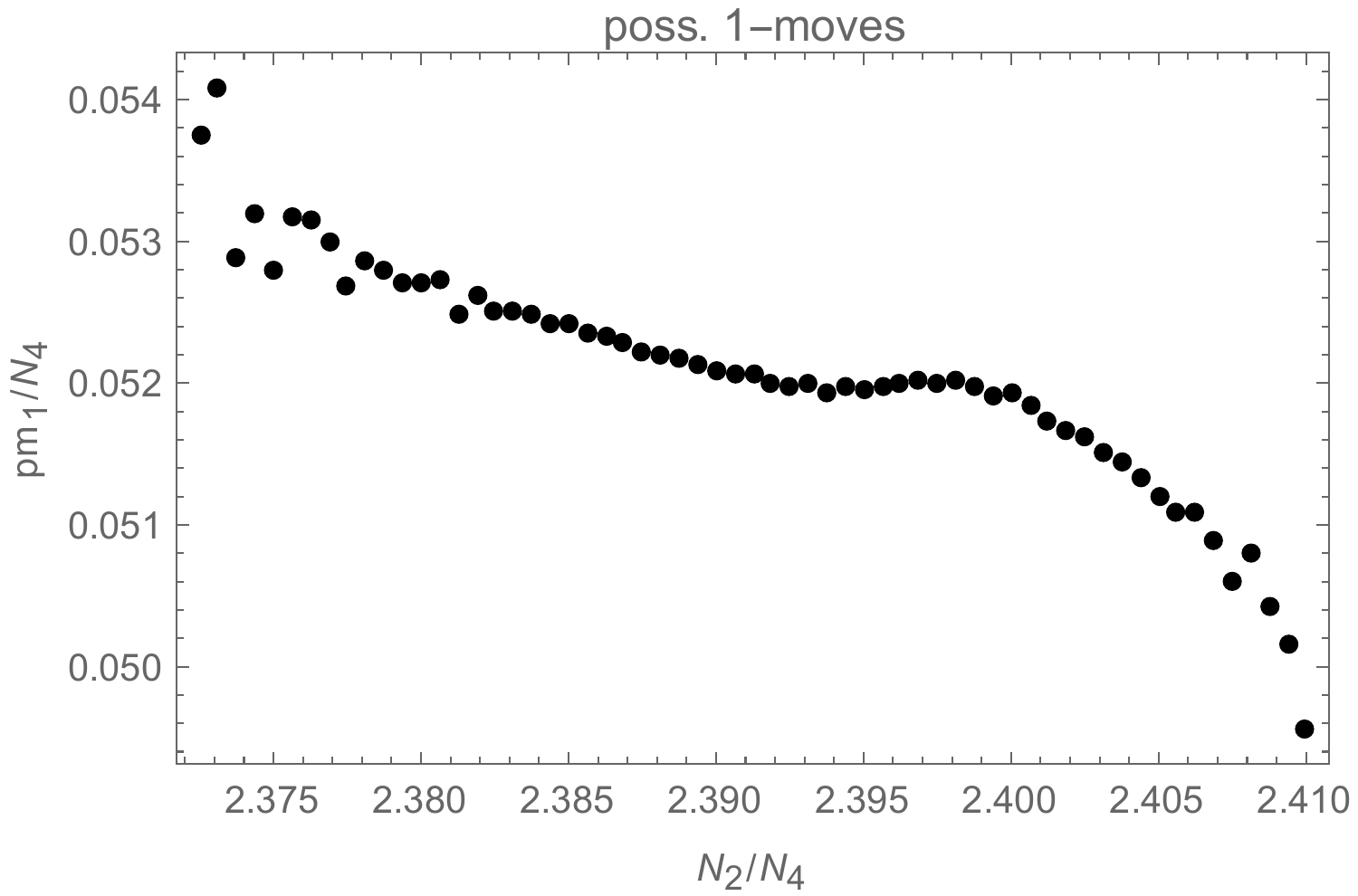}
\includegraphics[width=\linewidth]{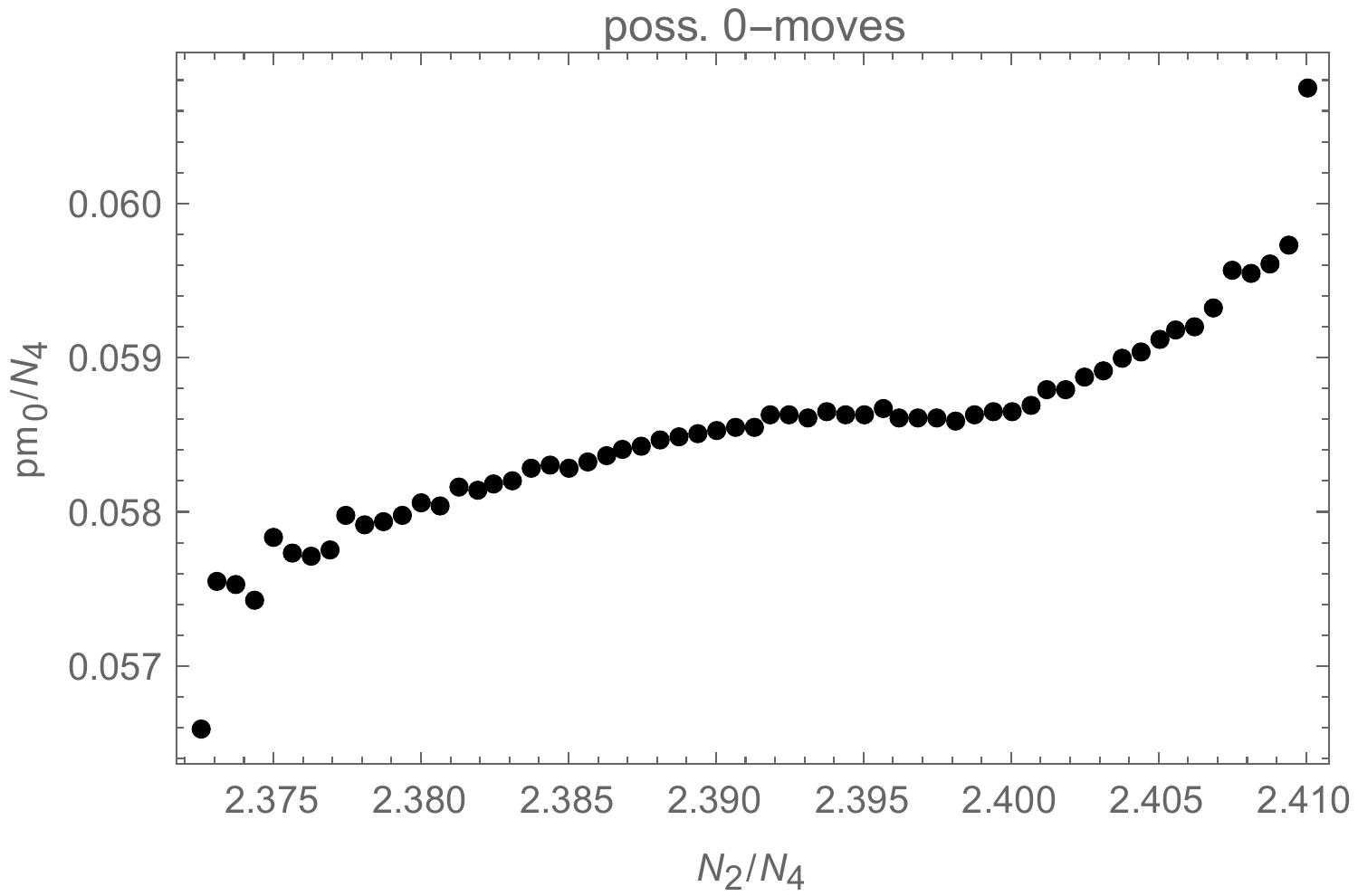}
\end{minipage}\hfill
\begin{minipage}[t]{0.33\linewidth}
\centering
\includegraphics[width=\linewidth]{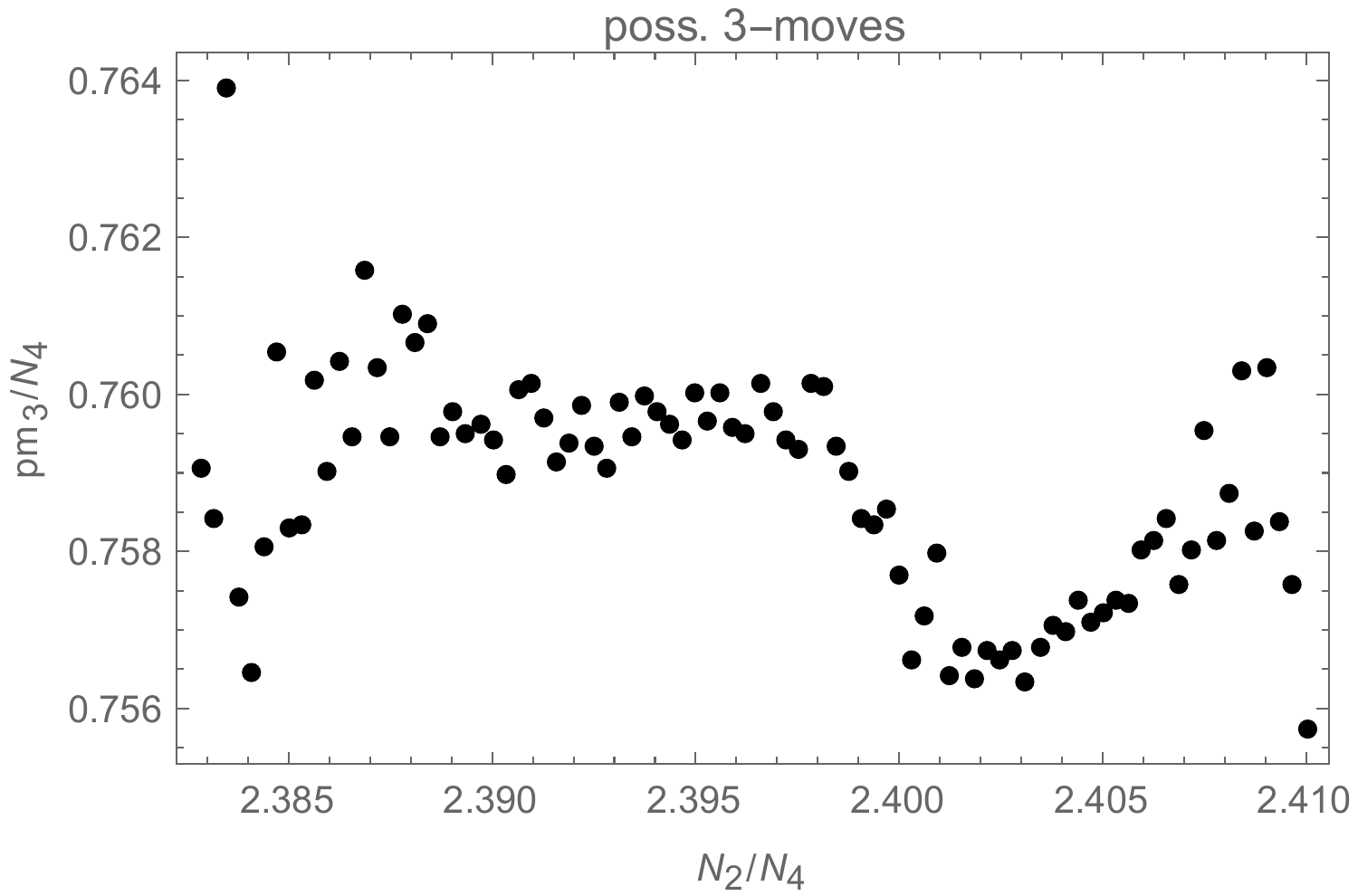}
\includegraphics[width=\linewidth]{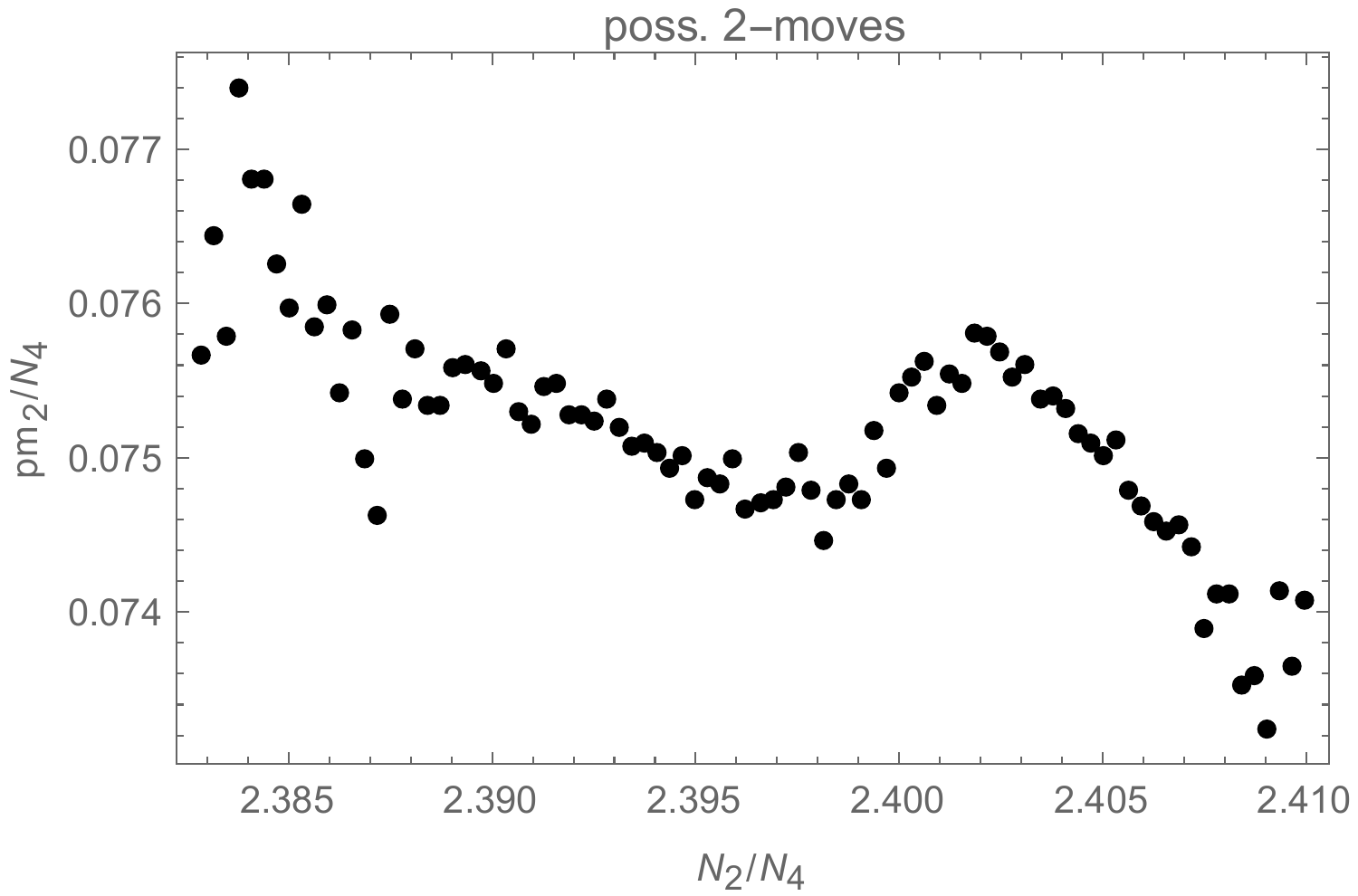}
\includegraphics[width=\linewidth]{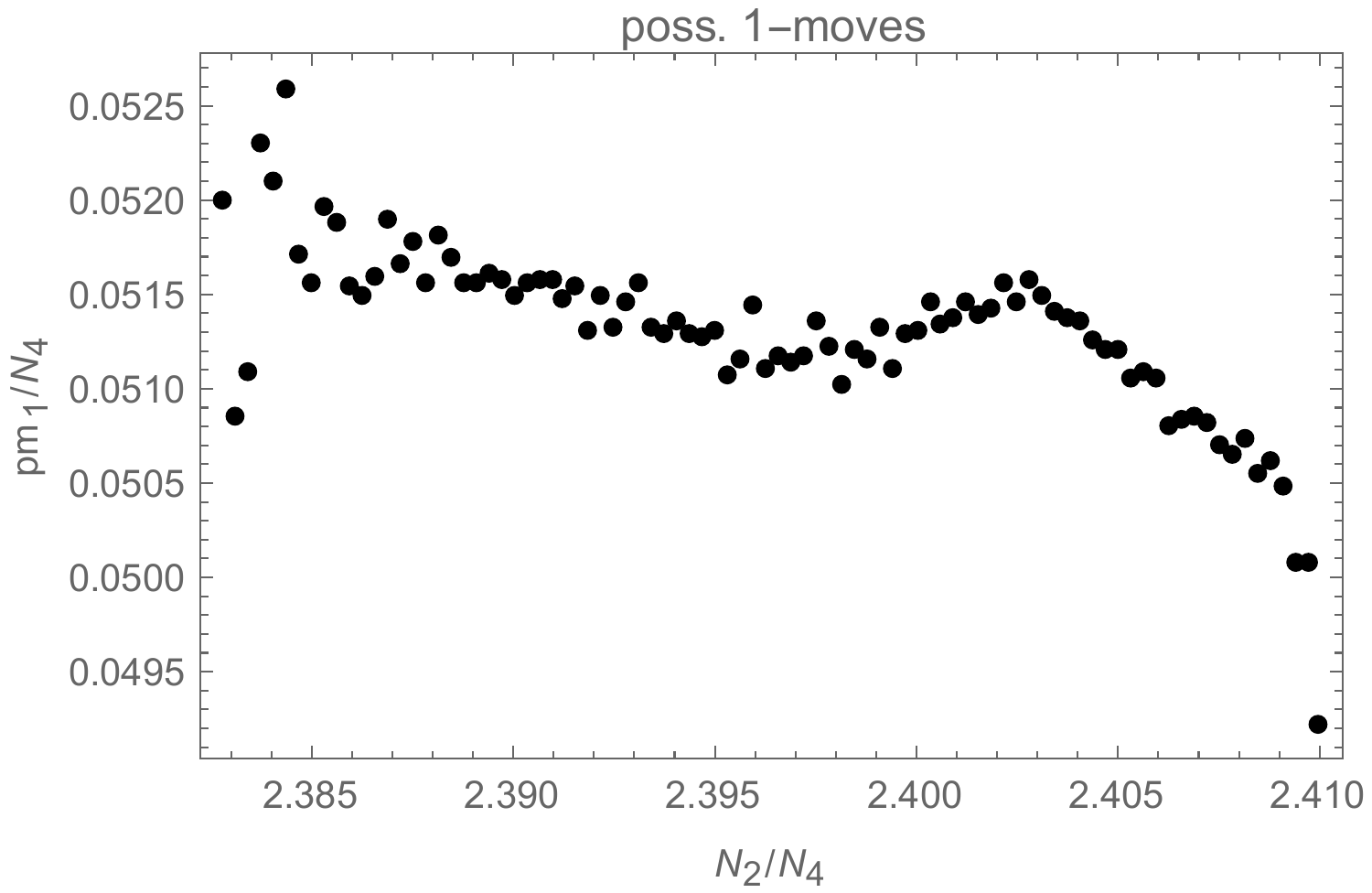}
\includegraphics[width=\linewidth]{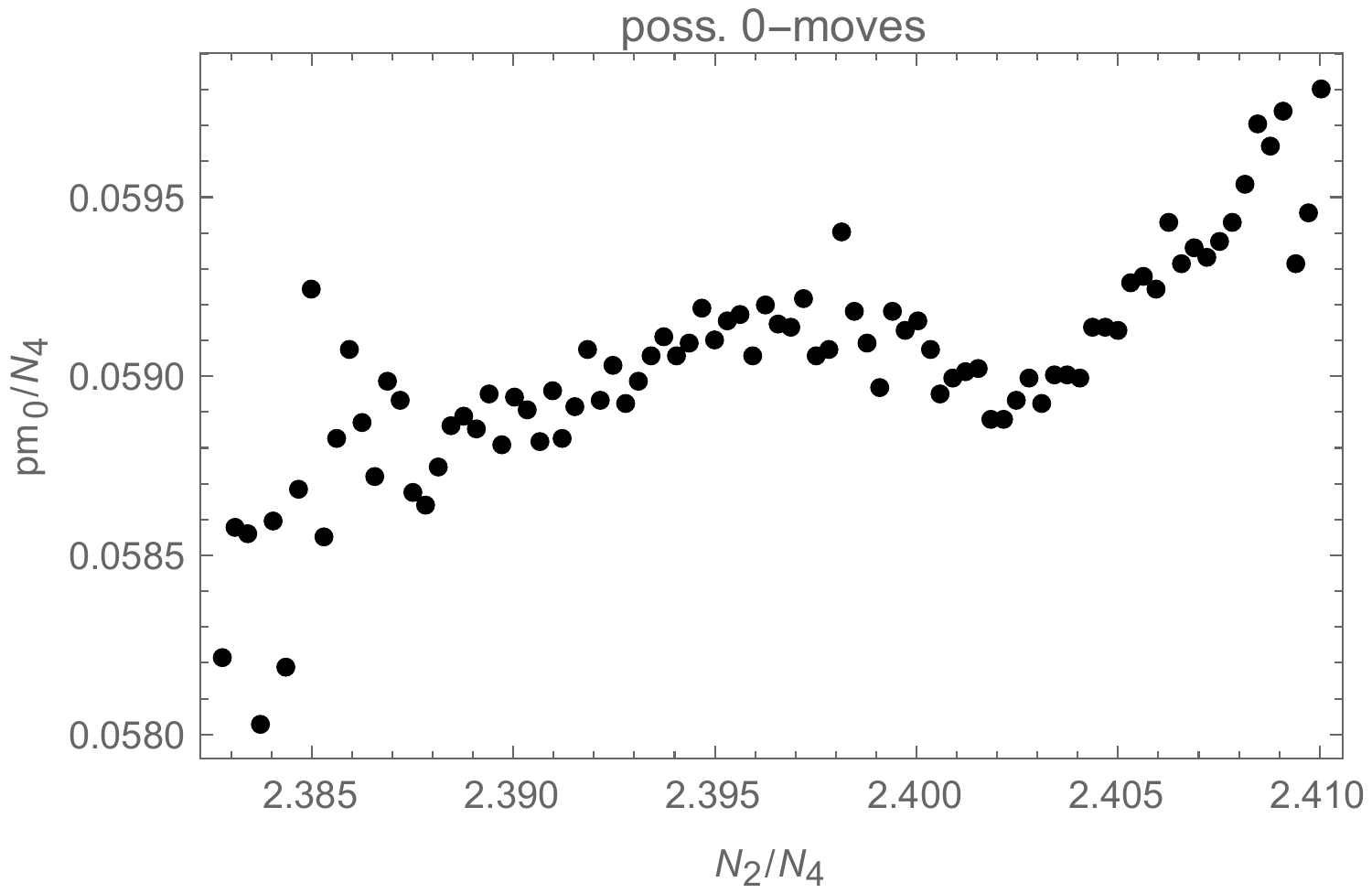}
\end{minipage}\hfill
\begin{minipage}[t]{0.325\linewidth}
\centering
\includegraphics[width=\linewidth]{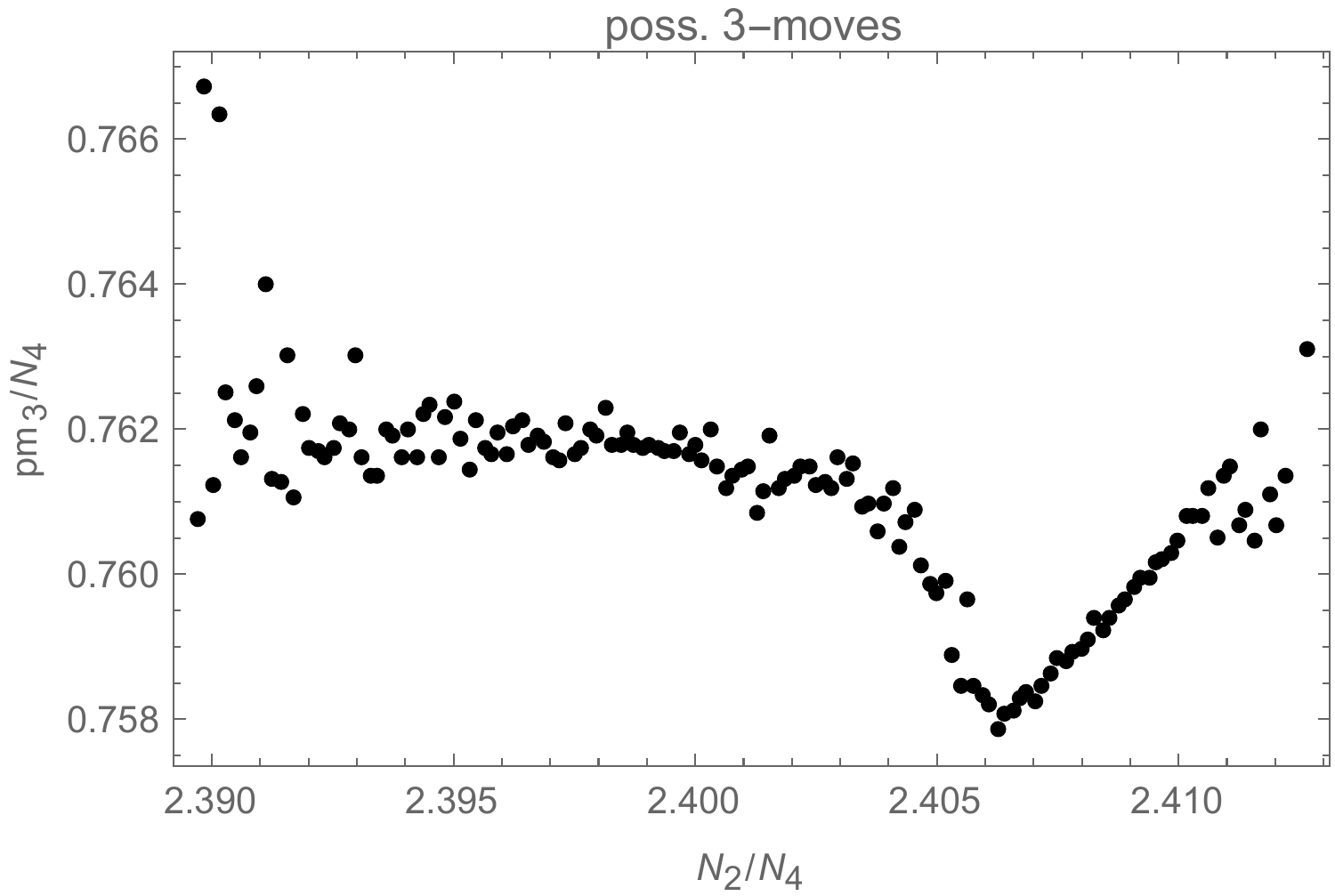}
\includegraphics[width=\linewidth]{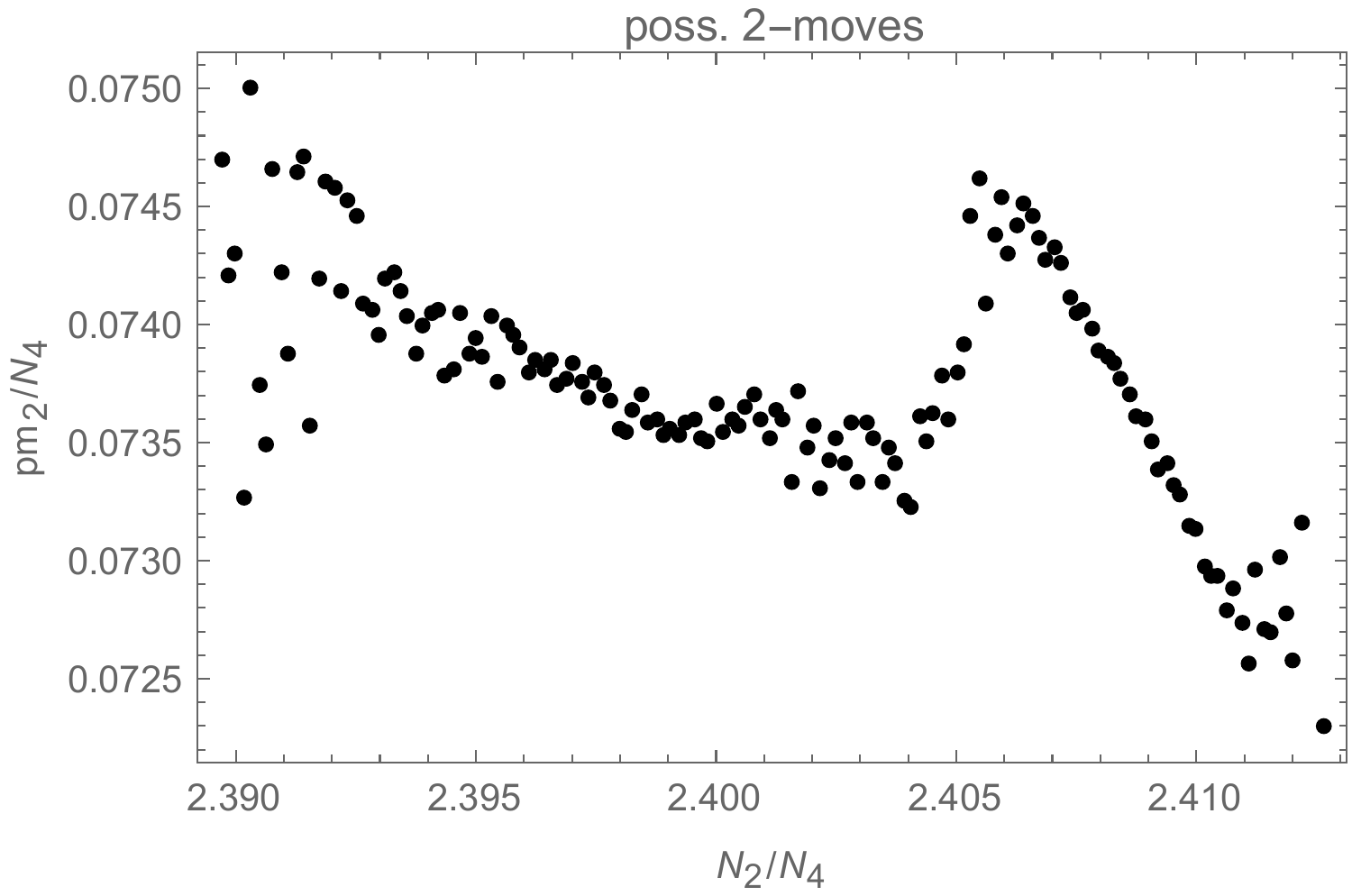}
\includegraphics[width=\linewidth]{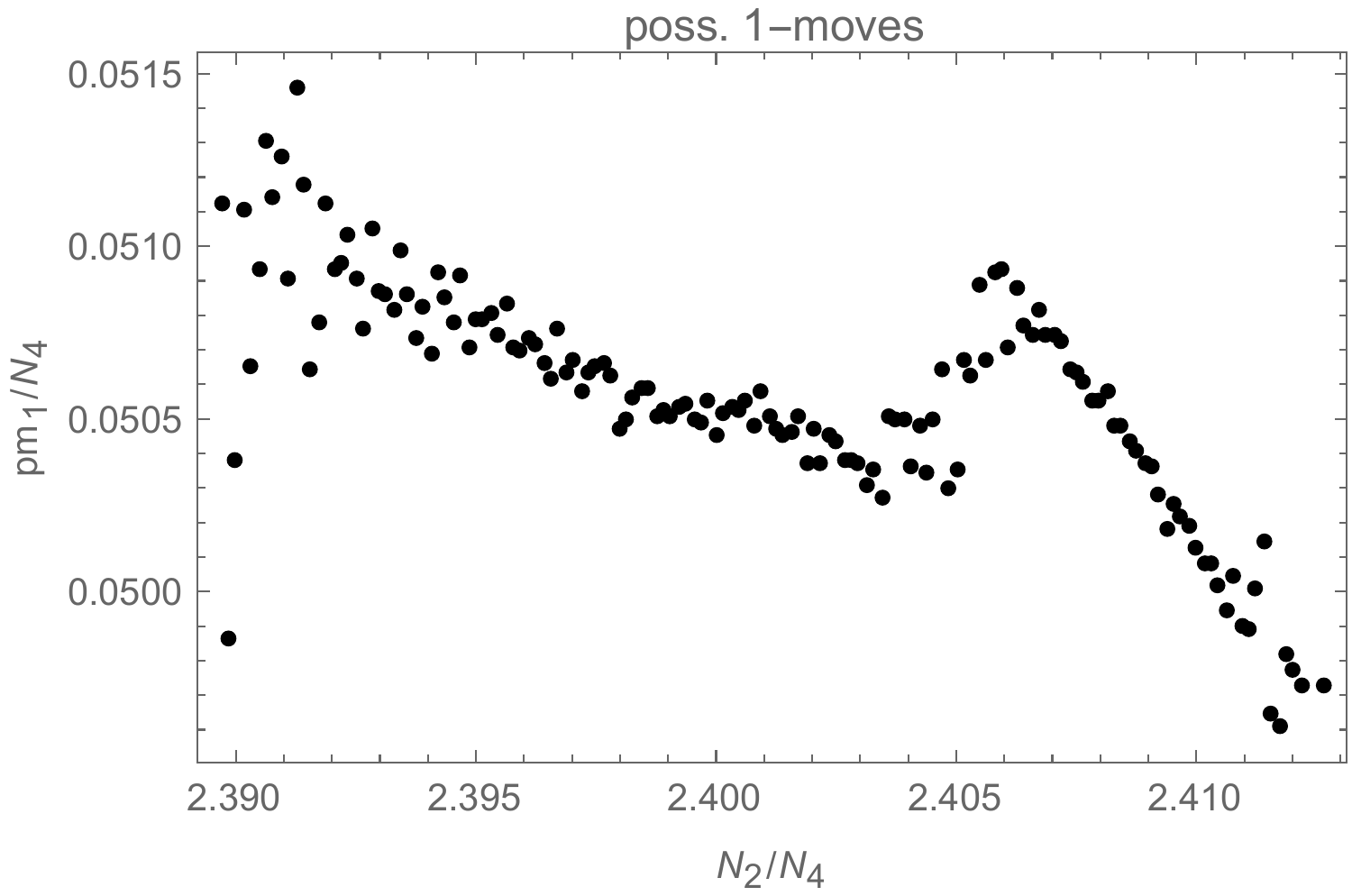}
\includegraphics[width=\linewidth]{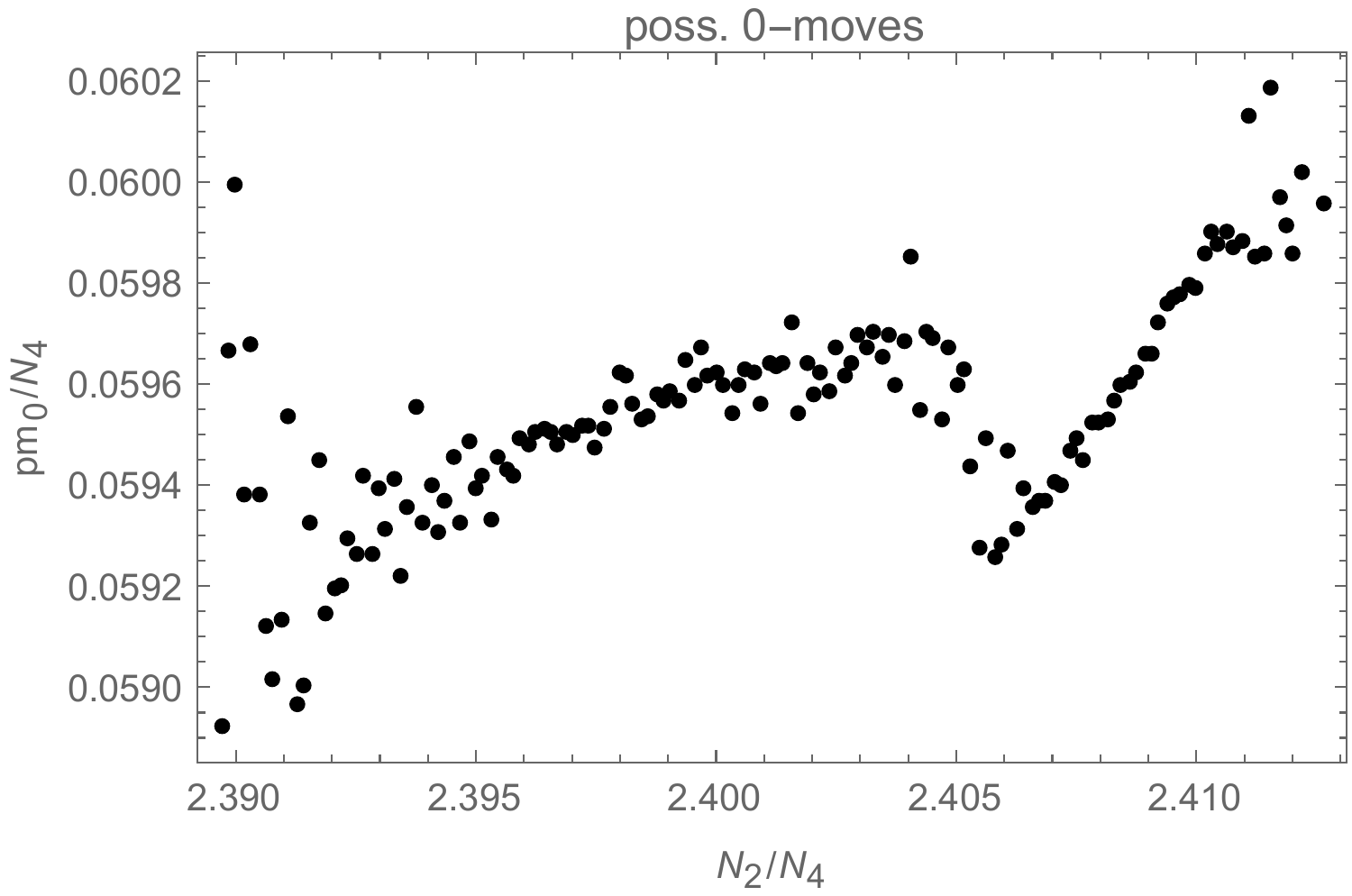}
\end{minipage}
\caption{From top to bottom: average numbers of possible 3,2,1 and 0-moves (normalized by the system size) vs. $N_{2}/N_{4}$ for systems of size (from left to right) $N_{4}=32$k, $64$k and $128$k at the pseudo-critical point. Comparison with Fig. \ref{fig:N2dist} shows, that the strong changes (.e.g. for $N_{2}/N_{4}=\fof{2.398,2.402}$ in the $N_{4}=64$k case) happen at the location of the valley of the corresponding graph in Fig. \ref{fig:N2dist}.}
\label{fig:possmovesvsn2}
\end{figure}

In the pseudo-critical region, in order to change from a rather crumpled to a rather elongated state, the system has to go through the process of producing and growing new baby-universe branches on top of the large "mother universe", until almost the whole volume fits into them while a (distinguishable) "mother universe" disappears. This is illustrated in Fig. \ref{fig:busizehistvsk2} where, as a function of $\kappa_{2}$, it is shown how the total volume of a system with $N_{4}\,=\,32$k is distributed, on average, over bubbles of different sizes.\\
Regarding this process of creating new baby-universe branches, which necessarily starts with the creation of a new size 6 bubble, it is interesting to note that applying a 4-move to a 4-simplex contained in a bubble that consists of only five 4-simplices results in a triangulation which is slightly more restrictive with respect to further application of 3- and 0-moves, as compared to the case where a 4-move is applied outside of such a "volume 5"-bubble. This is explained in more detail in Fig. \ref{fig:doublenecks} for the two-dimensional case,
and in figure \ref{fig:possmovesvsn2}, we show that the effect of this mechanism is indeed observable: the latter figure shows, for different system sizes, the average numbers of possible Pachner $n$-moves at the pseudo-critical point as a function of $N_{2}/N_{4}$. Comparing the graphs in Fig. \ref{fig:possmovesvsn2} with Fig. \ref{fig:N2dist} in order to identify which $N_{2}/N_{4}$-interval corresponds to which phase, we see that the numbers of possible moves undergo an abrupt change precisely in the region where the valleys of the corresponding graphs in  Fig.\ref{fig:N2dist} are located. These abrupt changes are just as one would expect from the phenomenon described in Fig. \ref{fig:doublenecks} which occurs a soon as a 3-move is applied to a 3-simplex that is part of a minimal neck between a "volume 4"- and a "volume 5"-bubble (which leads to a "volume 11"-bubble i.e. a "size 12"-bubble with just one neck): the number of possible 3- and 0-moves is lower and the number of possible 2-moves higher than in a configuration of the same size but without this particular "volume 11"-bubble. 

\begin{figure}[htbp]
\centering
\begin{minipage}[t]{0.49\linewidth}
\centering
\includegraphics[width=\linewidth]{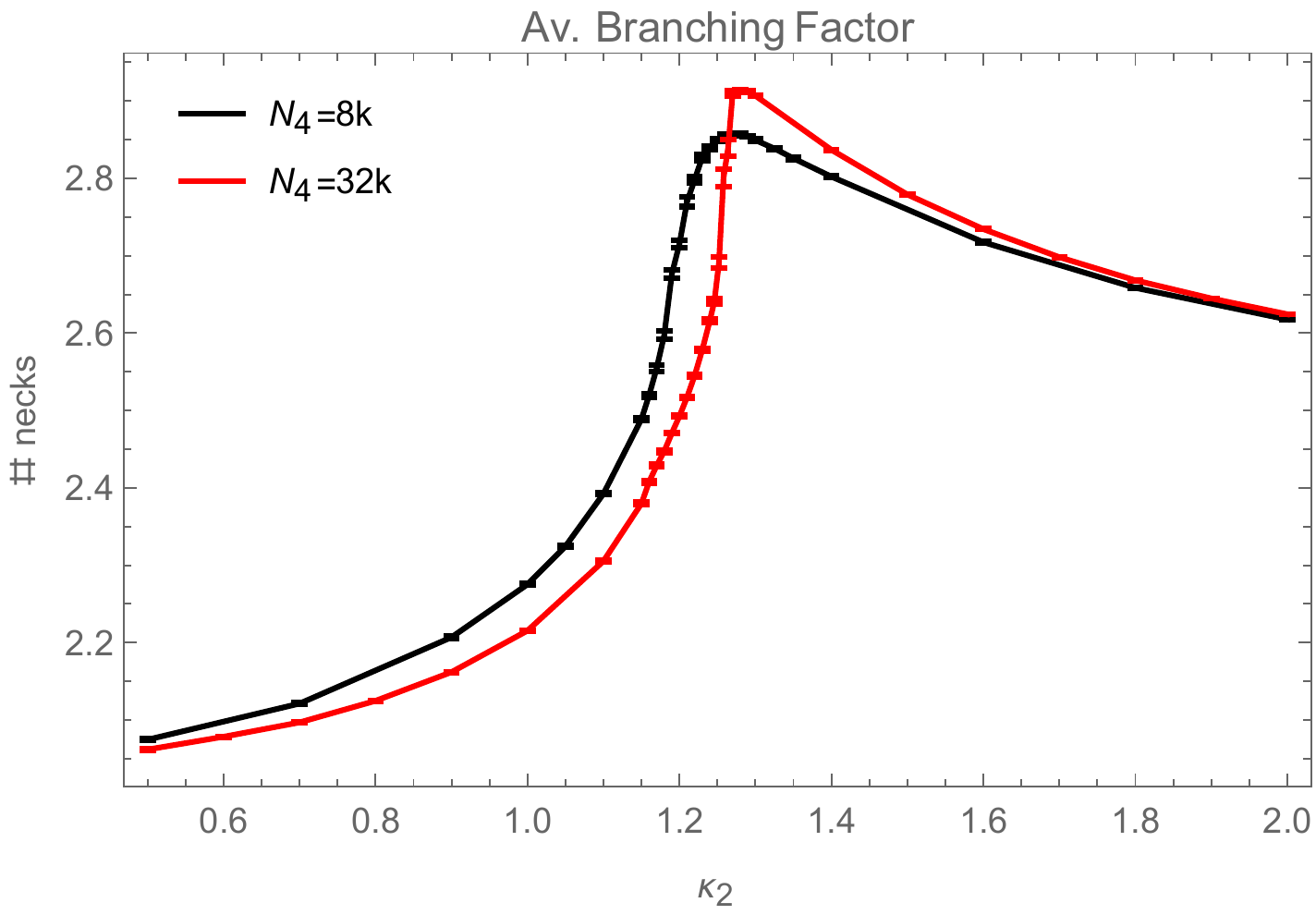}
\caption{The figure shows, as a function of $\kappa_{2}$, the average \emph{branching factor}, i.e. the average number of necks, of bubbles which are neither the largest bubble in the system nor volume 5 bubbles, which are just the terminating leaves of a baby-universe branch. As the branching factor decreases again after the phase transition at $\kappa_{2}\approx 1.258$, a large branching factor alone is not a good indicator for a bubble to be in the elongated phase.}
  \label{fig:branchingf}
\end{minipage}\hfill
\begin{minipage}[t]{0.49\linewidth}
\centering
\includegraphics[width=\linewidth]{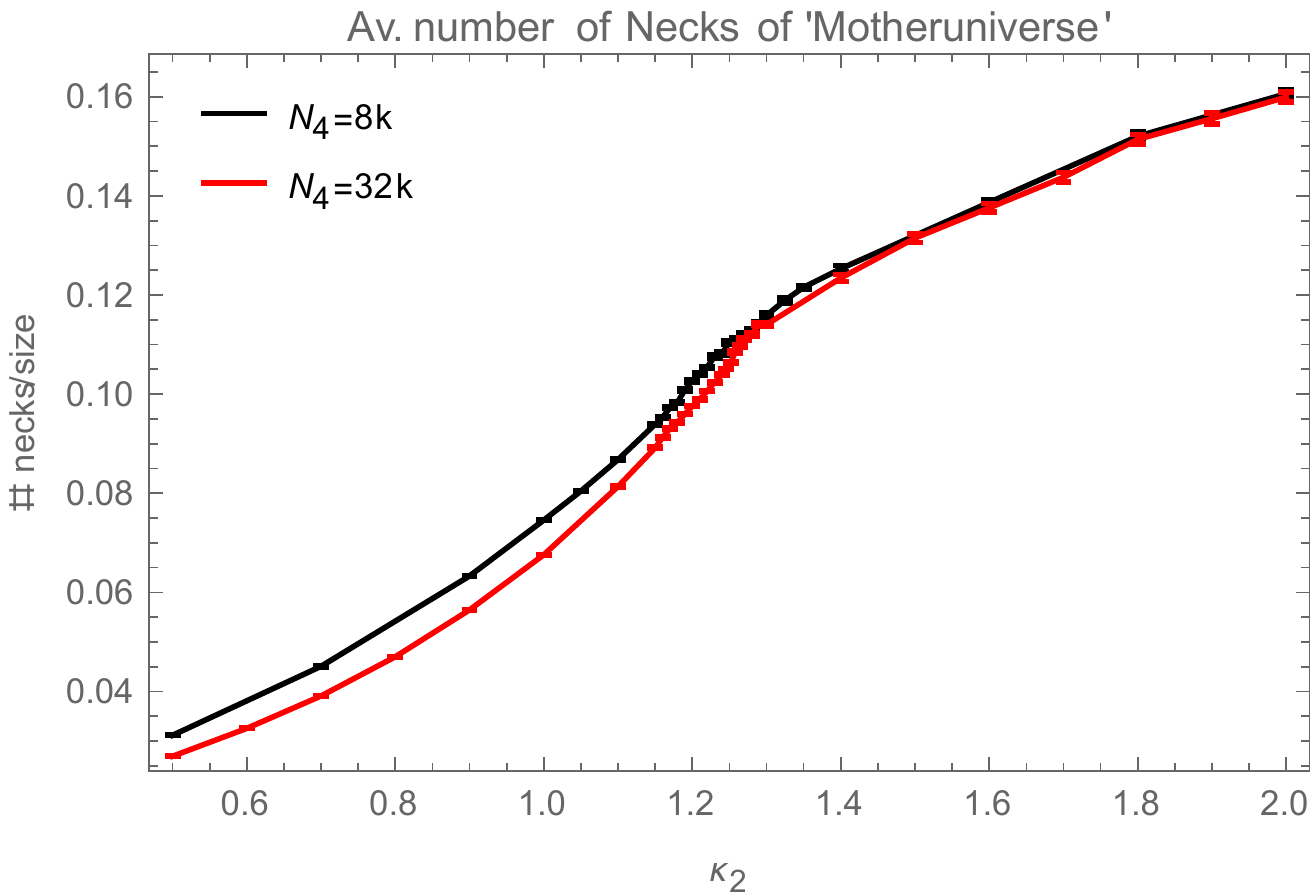}
\caption{The figure shows, as a function of $\kappa_{2}$, the average density of necks of the largest bubble in the system, i.e. "number of necks of bubble"/"size of bubble" (where the size is again given be the sum of the numbers of necks and 4-simplices). As the largest bubble can be assumed to correspond to the crumpled phase for $\kappa_{2}<\kappa_{2}^{pcr}\of{N_{4}}\approx1.258$, but for $\kappa_{2}>>\kappa_{2}^{pcr}\of{N_{4}}$ is just the slightly largest of many almost equally sized bubbles, which are all part of the elongated phase, this shows, that it is the neck-density of a bubble rather than its total neck number which distinguishes between the two phases. Note also that for size 6 bubbles, the neck density is always $\geq 1/6\approx 0.167$ which according to this figure is clearly elongated, as it should be.}
  \label{fig:neckspersizelb}
\end{minipage}
\end{figure}

\begin{figure}[htbp]
\centering
\includegraphics[width=0.8\linewidth]{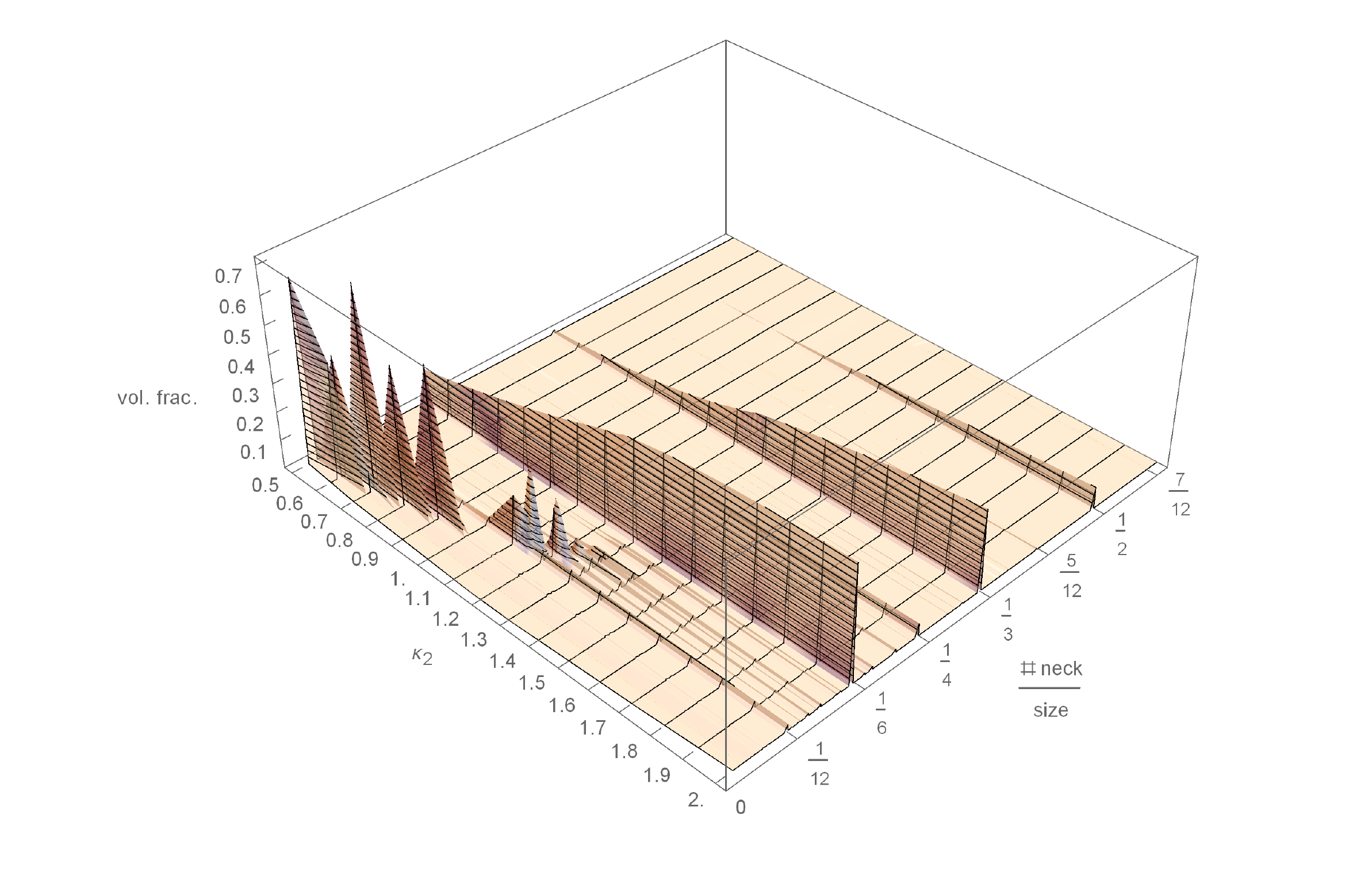}
\caption{The figure shows for a system of size $N_{4}=32$k, how the total volume is distributed over bubbles with different \emph{neck densities}, i.e. different ratios "number of necks of the bubble"/"size of the bubble" and how this distribution changes with $\kappa_{2}$.}
  \label{fig:neckdensdist}
\end{figure}

At the beginning of this section, we mentioned also using the branching factor, at least on an intuitive level, as a criterion to decide if a piece of triangulation is in the elongated or crumpled phase. The branching factor itself turns out not to be a good criterion to distinguish between the two phases as its average value drops again with increasing $\kappa_{2}$ for $\kappa_{2}>\kappa_{2}^{pcr}$ (see Fig. \ref{fig:branchingf}). On the other hand, the related average \emph{neck density} or "branching factor per size" is monotonic and seems to yield a meaningful criterion to distinguish between the two phases (see Fig. \ref{fig:neckspersizelb}). In figure \ref{fig:neckdensdist} we show how the total volume distributes over bubbles with different neck densities but it remains to be understood what the exact value of the critical density, which seems to be around $1/9$, should be and where it comes from.\\


\newpage
\subsection{Balls in Boxes Model}\label{ssec:ballsinboxes}
It is well known that the balls in boxes model \cite{Bialas2,Bialas3} describes nicely the qualitative features observed in EDT simulations. In the canonical formulation the model describes the statistical ensemble of a fixed total number $N$ of balls distributed in a varying number $M$ of boxes:
\[
Z\of{N,\kappa}\,=\,\sum\limits_{M=1}^{\infty}\,\e^{\kappa M}\,\sum\limits_{q_{1},\ldots,q_{M}} p\of{q_{1}}\cdots p\of{q_{M}}\delta_{N,q_{1}+\ldots+q_{M}},
\]
where $p\of{q}$ is the probability for a single box to contain $q$ balls and $\kappa$ is a coupling introduced to control the number of boxes. As shown in \cite{Bialas2}, the qualitative behavior of the model depends only on the sub-exponential factors of the single-box-occupation probability $p\of{q}$, since a redefinition
\[
p\of{q}\,\rightarrow\,p'\of{q}\,=\,\e^{-\kappa_{0}}\e^{\mu_{0} q} p\of{q}
\]
just results in
\[
Z\of{N,\kappa}\,\rightarrow\,Z'\of{N,\kappa}\,=\,\e^{\mu_{0} N}\,Z\of{N,\kappa-\kappa_{0}}.
\]
For power like sub-exponential weight factors
\[
p\of{q}\,=\,q^{-\beta}\quad,\quad q\in\mathbb{N}\label{eq:subexpbweight}
\]
it was shown in \cite{Bialas3} that the free energy
\[
F\of{\kappa}\,=\,\lim\limits_{N\rightarrow\infty}\frac{1}{N}\log\of{Z\of{N,\kappa}}\label{eq:binbfreeenergy}
\]
has a singularity at $\kappa_{cr} = -\log\of{\zeta\of{\beta}}$ for $\beta\in\of{1,\infty}$, where $\zeta\of{x}$ is the Riemann Zeta-function. The phase is called \emph{fluid} for $\kappa>\kappa_{cr}$ and \emph{condensed} for $\kappa<\kappa_{cr}$. For $\beta>2$ the phase transition is $\ford$ whereas for $\beta\in\left(\frac{n+1}{n},\frac{n}{n-1}\right]$ the transition is $\nord{n}$. The order parameter for the transition is the first derivative of the free energy \eqref{eq:binbfreeenergy} with respect to $\kappa$, which yields the average number of boxes divided by the number of balls
\[
r\,=\,\partd{F\of{\kappa}}{\kappa}\,=\,\lim\limits_{N\rightarrow\infty}\frac{\avof{M}}{N},\label{eq:binbordp}
\]
which vanishes in the condensed phase and equals $1$ in the fluid phase.\\

The relation to the 4-dimensional EDT model is normally established by identifying the triangles (or nodes) in the triangulation with boxes and the number of balls in a box with the number of 4-simplices that share the corresponding triangle (node). In this way, the coupling $\kappa_{2}$ of the EDT model nicely takes over the role of the $\kappa$ in the balls in boxes model, the average Regge curvature becomes in the thermodynamic limit the analogue of the order parameter \eqref{eq:binbordp}, and the $\beta$ in \eqref{eq:subexpbweight} is related\cite{Bialas3} to the $\beta$ used in EDT models with a modified measure term \cite{Ambjorn2}. Alternatively, one can identify the bubbles or baby-universes with the boxes and the number of necks of each bubble with the number of balls in the corresponding box \cite{Bialas5}. This yields an effective theory for EDT in the form of a branched polymer model, in which the bubbles are the vertices and the necks correspond to links between the nodes (as in the figures \ref{fig:butrees}, \ref{fig:butreesfixedkappa2}, but ignoring the different sizes of the nodes).\\
While the latter yields just an effective theory, the problem with the former correspondence is, that due to geometric constraints, the interplay between the number of 4-simplices per triangle (or per node) and the number of triangles (nodes) itself is much more involved than the interplay between the balls and boxes in the balls in boxes model. We would therefore like to propose a different correspondence in which the numbers of "balls" and "boxes" are less constrained.\\
To this end, let us focus on the largest bubble of a triangulation which we will from now on also call \emph{base-manifold}: this largest bubble is made up of \emph{elementary building-blocks} consisting of 4-simplices and minimal necks\footnote{Instead of thinking of a minimal neck as a kind of worm-hole to a baby-universe, rather think of the baby-universe as the blown-up interior of a space-time region that has the boundary of a 4-simplex, i.e. a minimal neck.}. Now consider these elementary building-blocks of the base-manifold as boxes and the number of 4-simplices contained in them as balls. An elementary building-block that is an ordinary 4-simplex corresponds to a box containing a single ball, whereas an elementary building-block consisting of a minimal neck corresponds to a box that contains as many balls as there are 4-simplices in the baby-universe branch behind that neck. The 4 dimensional EDT model therefore corresponds to a balls in boxes model with $N_{4}$ balls, where each box contains at least one ball and the number $M$ of boxes can vary from 6 (minimal size for a (combinatorial) simplicial 4-sphere) to $N_{4}$ (no necks in the triangulation). The canonical EDT partition function could therefore be interpreted as the $\kappa=0$ case of the more general partition function
\[
Z\of{\kappa_{2},N_{4},\kappa}\,=\,\sum\limits_{M=6}^{N_{4}}\,\e^{\kappa M}\,Z\of{\kappa_{2},N_{4},M}\,=\,\sum\limits_{M=6}^{N_{4}}\,\e^{\kappa M}\,\sum\limits_{T\in T\of{M,N_{4}}}\frac{1}{C_{T}}\e^{\kappa_{2}\,N_{2}\of{T}},\label{eq:edtbinbpartf}
\]
where $T\of{M,N_{4}}$ is the set of triangulations that possess a largest bubble of size $M$ and consist in total of $N_{4}$ 4-simplices.\\
\begin{figure}[H]
\centering
\begin{minipage}[t]{0.49\linewidth}
\centering
\includegraphics[width=\linewidth]{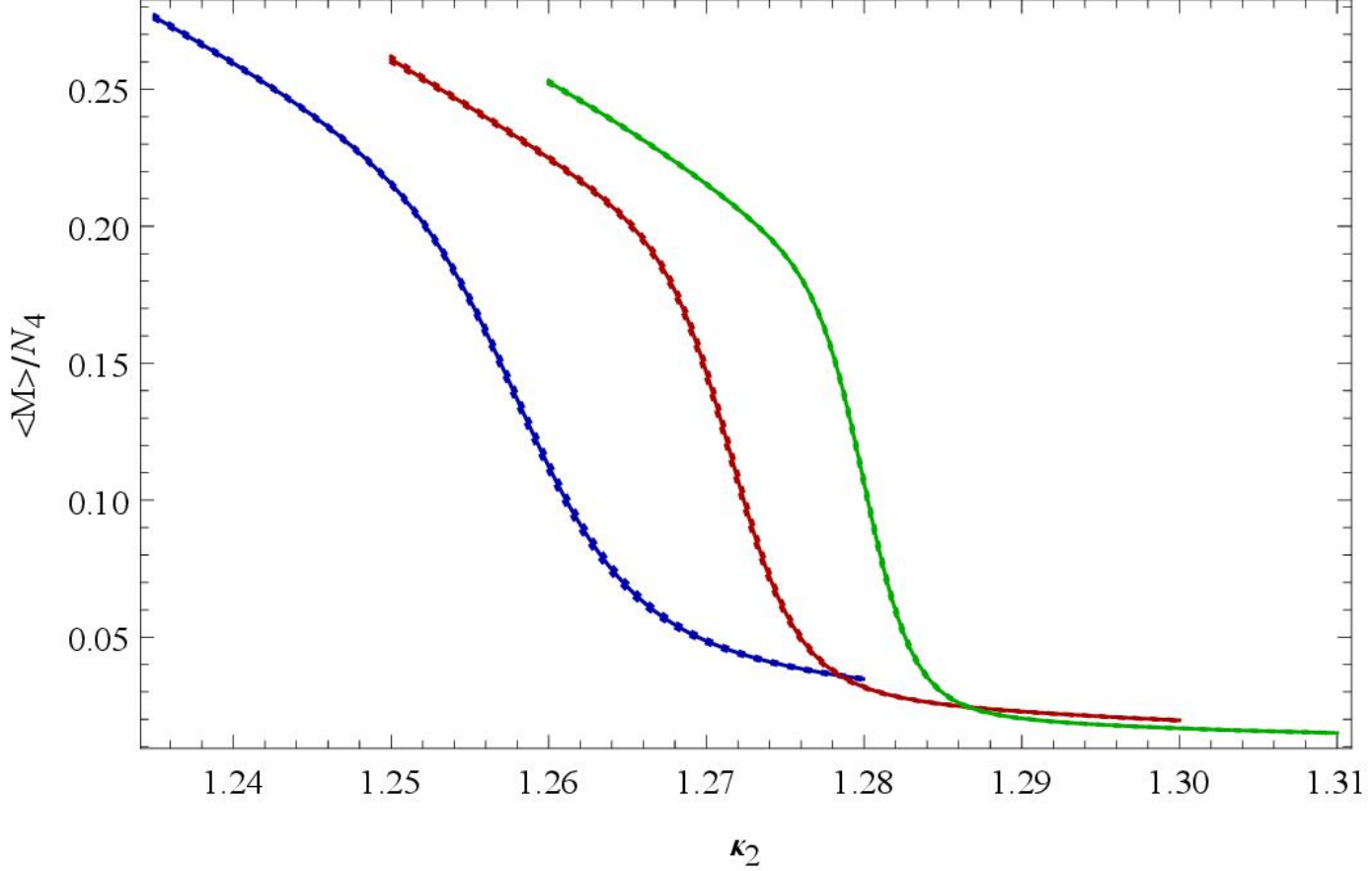}
\end{minipage}\hfill
\begin{minipage}[t]{0.49\linewidth}
\centering
\includegraphics[width=\linewidth]{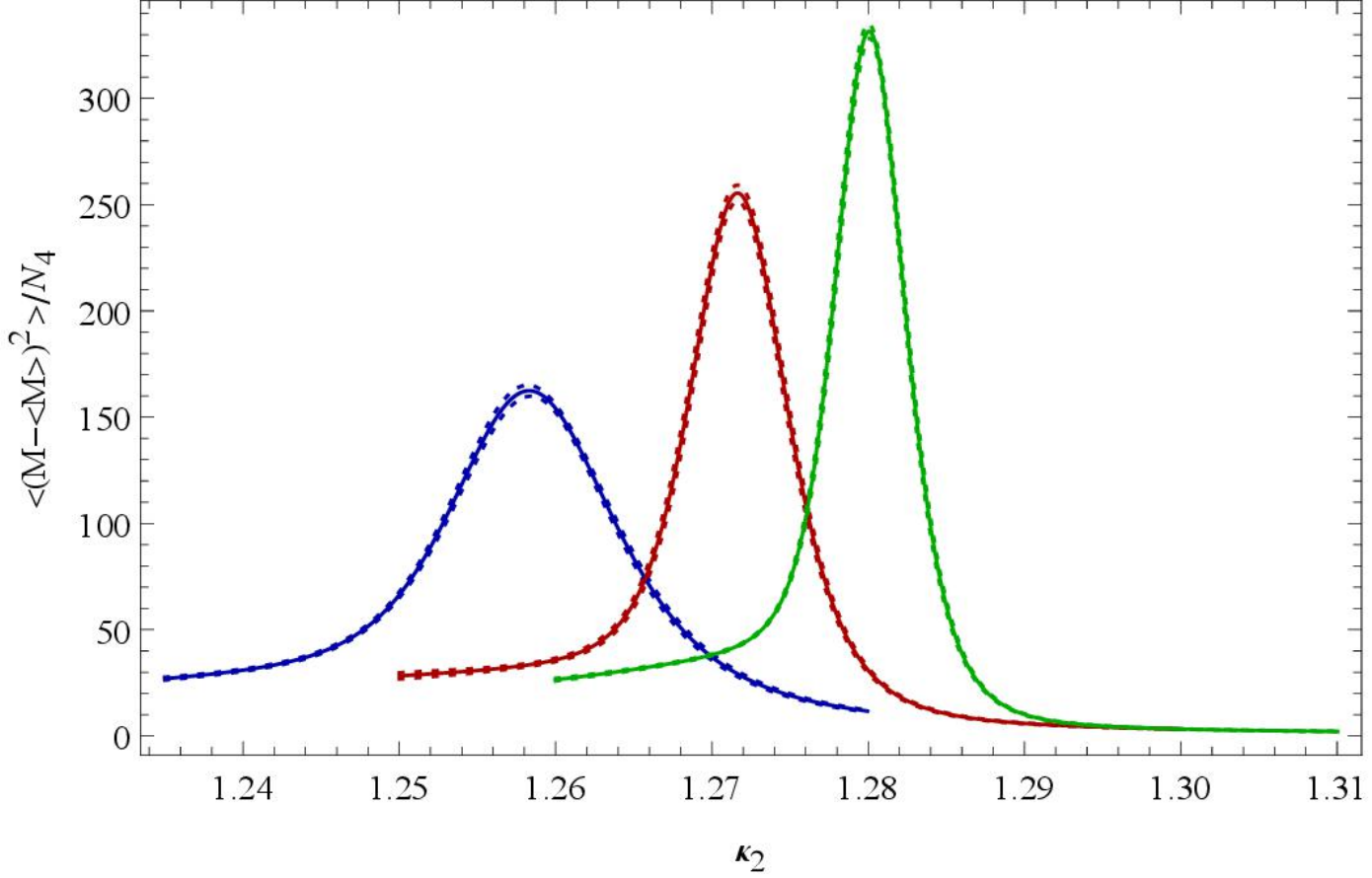}
\end{minipage}
\caption{Average (normalized) number $\avof{M}/N_{4}$ of elementary \emph{space-time building blocks} in the largest bubble (left) and the corresponding susceptibility (right) as a function of $\kappa_{2}$ for systems of total size $N_{4}\approx 32$k
(dark blue), $N_{4}\approx 48$k (dark red)  and $N_{4}\approx 64$k (dark green).}
\label{fig:avmvsk2}
\end{figure}
In terms of \eqref{eq:edtbinbpartf} the average size of the largest bubble and the corresponding susceptibility shown in Fig. \ref{fig:avmvsk2} could for example be expressed as
\begin{align}
\frac{\avof{M}\of{\kappa_{2},N_{4}}}{N_{4}}\,&=\,\frac{1}{N_{4}}\partd{\ln Z\of{\kappa_{2},N_{4},\kappa}}{\kappa}\Big|_{\kappa=0}\\
\frac{\avof{\of{M-\avof{M}}^{2}}\of{\kappa_{2},N_{4}}}{N_{4}}\,&=\,\frac{1}{N_{4}}\partdm{\ln Z\of{\kappa_{2},N_{4},\kappa}}{\kappa}{2}\Big|_{\kappa=0}.
\end{align}
A histogram for the $M/N_{4}$-distribution at the pseudo-critical point is shown in Fig. \ref{fig:mspread} and should be compared with Fig. 2 of Ref. \cite{Bialas2}. As can be seen, Fig. \ref{fig:mspread} looks much more like Fig. 2 of Ref. \cite{Bialas2} than the $N_{2}/N_{4}$-distribution shown in Fig. \ref{fig:N2dist}, which would be the corresponding quantity according to the old identification: triangles $\rightarrow$ boxes, 4-simplices $\rightarrow$ balls. In particular, $M/N_{4}$ is a nice order parameter as it tends to zero in the elongated phase, while $N_{2}/N_{4}$ remains finite.\\
\begin{figure}[H]
\centering
\begin{minipage}[t]{0.49\linewidth}
\centering
\includegraphics[width=\linewidth]{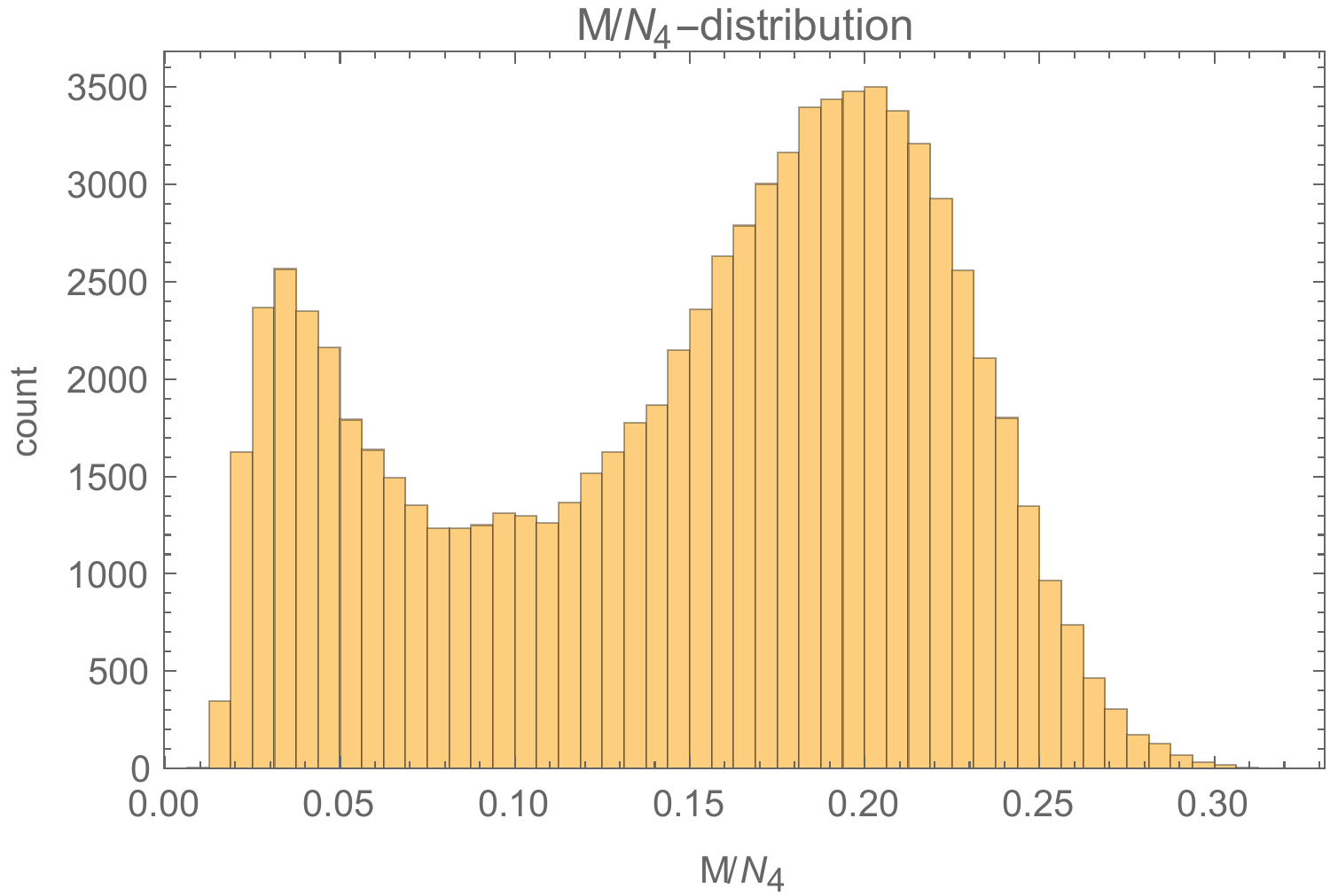}
\end{minipage}\hfill
\begin{minipage}[t]{0.49\linewidth}
\centering
\includegraphics[width=\linewidth]{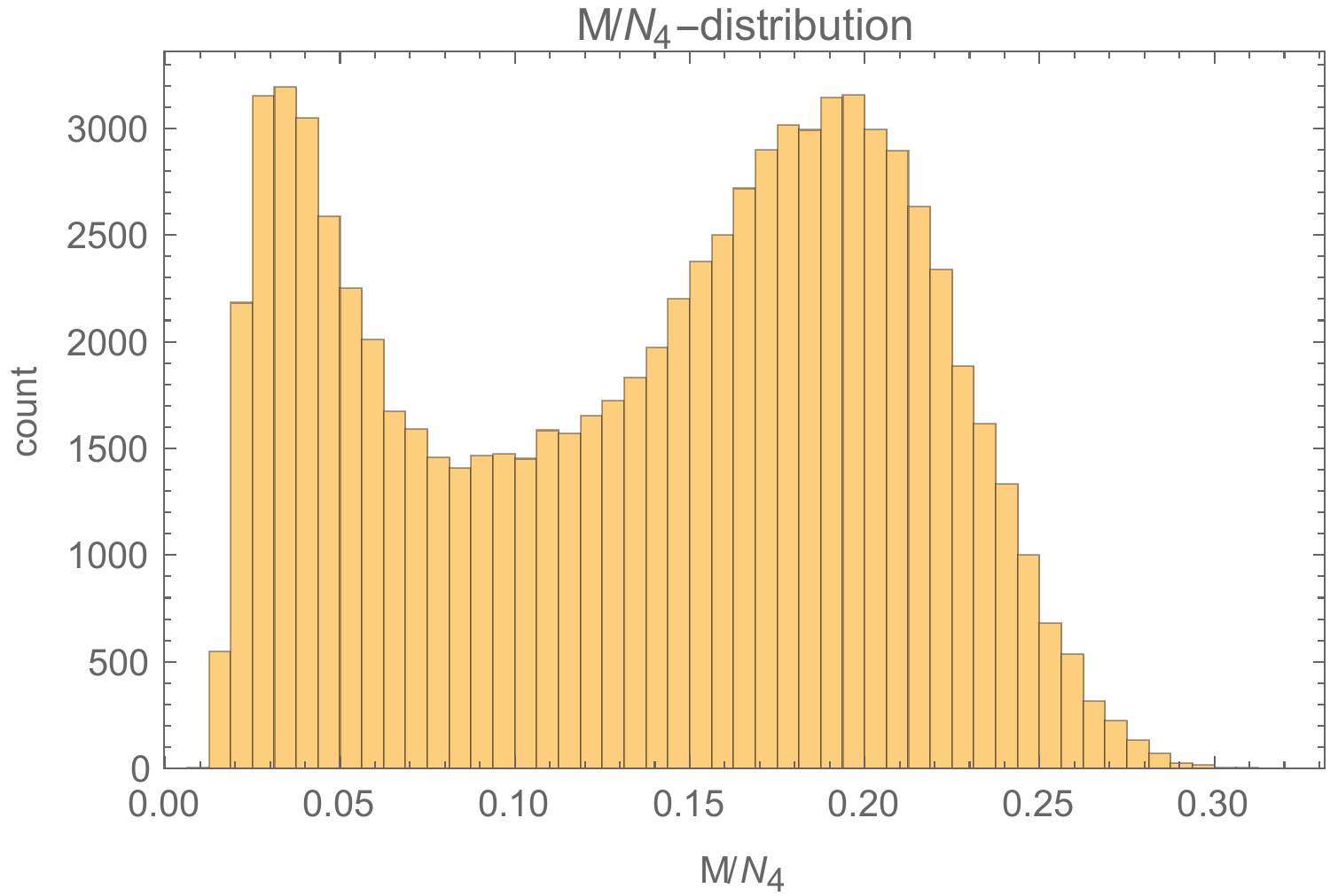}
\end{minipage}
\caption{Histogram for the fraction of 4-simplices contained in the largest bubble of a system of size $N_{4}\approx 32$k at $\kappa_{2}\,=\,1.2577$ (left) and $\kappa_{2}\,=\,1.2587$ (right). This quantity serves as an approximation for $M/N_{4}$, the number of elementary building-blocks of the largest bubble divided by the total number of 4-simplices, for which we would have to take into account also the number of necks of the largest bubble, which would lead to about $12\%$ larger values.}
\label{fig:mspread}
\end{figure}

In what follows, we will show that the $Z\of{\kappa_{2},N_{4},M}$ appearing in \eqref{eq:edtbinbpartf} can be written as
\[
Z\of{\kappa_{2},N_{4},M}\,=\,Z_{0}\of{\kappa_{2},M}\,\sum\limits_{n_{1},\ldots,n_{M}=1}^{N_{4}+1-M}\of{\prod\limits_{k=1}^{M} p\of{\kappa_{2},n_{k}}}\delta_{N_{4},n_{1} + \cdots + n_{M}},\label{eq:edtbinbpartf2}
\]
where the first factor $Z_{0}\of{\kappa_{2},M}$ corresponds to the average number of ways a base-manifold consisting of $M$ elementary building blocks (i.e. minimal necks or ordinary $4$-simplices) is realized at $\kappa_{2}$, and the second factor, the sum, is the corresponding probability for $N_{4}$ 4-simplices to fit into the $M$ elementary building blocks, with $p\of{\kappa_{2},n}$ being the probability for a single elementary building block, to have volume $n$.\\
To write $Z_{0}\of{\kappa_{2},M}$ and $p\of{\kappa_{2},n}$ more explicitly, we need the micro-canonical partition function $Z_{1}\of{N_{2},N_{4}}$ that counts the number of possible triangulations with $N_{2}$ triangles, $N_{4}$ 4-simplices and which have a boundary of the form of a minimal neck. For $N_{4}\geq 5$ each such triangulation can be obtained by removing a 4-simplex from a corresponding triangulation without boundary, that has the same number $N_{2}$ of triangles but consists of $\of{N_{4}+1}$ 4-simplices. Thus $Z_{1}\of{N_{2},N_{4}}$ can be expressed in terms of the ordinary micro-canonical partition function $Z\of{N_{2},N_{4}}$ as\footnote{We neglect complications due to changing symmetry factors $C_{T}$, occurring when removing a 4-simplex from triangulations, as the number of symmetric configurations contributing to $Z\of{N_{2},N_{4}}$ is hopefully negligible for large $N_{4}$. For small $N_{4}$, it might be necessary to take the effect of a changing $C_{T}$ into account.}:
\[
Z_{1}\of{N_{2},N_{4}}\,=\,\begin{cases}
                             1/5!\,&,\, N_{4}\,=\,1\,,\,N_{2}\,=\,10 \\ 
                             \of{N_{4}+1}\,Z\of{N_{2},N_{4}+1}\,&,\, N_{4}\,\geq\,5
                           \end{cases},\label{eq:withboundarypartf}
\]
where $1/5!$ is the symmetry factor of a 4-simplex and $\of{N_{4}+1}$ is the number of possibilities to remove one 4-simplex from a triangulation of size $\of{N_{4}+1}$.\\
The corresponding canonical partition function is then:
\[
Z_{1}\of{\kappa_{2},N_{4}}\,=\,\sum\limits_{T\in T_{1}\of{N_{4}}}\,\frac{1}{C_{T}}\,\e^{\kappa_{2}\of{N_{2}\of{T\setminus\partial T}+\frac{1}{2}N_{2}\of{\partial T}}}\,=\,\sum_{N_{2}}\,Z_{1}\of{N_{2},N_{4}}\e^{\kappa_{2}\of{N_{2}-5}},\label{eq:z1canonicalpartf}
\]
where $T_{1}\of{N_{4}}$ is the set of triangulations with a minimal boundary that consist of $N_{4}$ 4-simplices.\\
We can now express $Z_{0}\of{\kappa_{2},M}$ in terms of \eqref{eq:z1canonicalpartf} and \eqref{eq:edtcpf} by noting that the number of ways in which $M$ elementary building blocks can be glued together to form a base-manifold, is the same as the number of ways to form a triangulation, consisting of $M$ 4-simplices, that does not have any neck. We can therefore write:
\begin{multline}
Z_{0}\of{\kappa_{2},M}\,=\,Z\of{\kappa_{2},M}-5!\sum\limits_{m}Z_{1}\of{\kappa_{2},M-m}\,Z_{1}\of{\kappa_{2},m}\\
=\,\sum\limits_{N_{2}}\of{Z\of{N_{2},M}-5!\sum\limits_{m}\sum\limits_{n_{2}}Z_{1}\of{N_{2}-n_{2},M-m}\,Z_{1}\of{n_{2},m}}\e^{\kappa_{2}\,N_{2}},\label{eq:z0canonicalpartf}
\end{multline}
where the first term within the brackets on the second line corresponds to the number of triangulations consisting of $M$ 4-simplices and $N_{2}$ triangles, while the second term, which is a sum over all possibilities to form a triangulation of size $M$ by glueing two triangulations, each with a minimal boundary, along their boundaries (in \cite{Ambjorn7}, this second term was used to measure the \emph{entropy exponent} by baby-universe counting), subtracts the subset of these triangulations that in addition possess at least one minimal neck\footnote{Again, corrections due to changing symmetry factors (when gluing two triangulations along their minimal boundary) might be necessary in \eqref{eq:z0canonicalpartf}.}, such that the whole bracket yields the number of triangulations with $M$ 4-simplices, $N_{2}$ triangles and no necks.\\ 

The probability distribution $p\of{\kappa_{2},n}$ required for the second factor in \eqref{eq:edtbinbpartf2} is given by $p\of{\kappa_{2},N_{4}}\,\propto\,5!\,Z^{\text{int}}_{1}\of{\kappa_{2},N_{4}}$, where
\[
Z^{\text{int}}_{1}\of{\kappa_{2},N_{4}}\,=\,\sum\limits_{T\in T_{1}\of{N_{4}}}\,\frac{1}{C_{T}}\,\e^{\kappa_{2}N_{2}\of{T\setminus\partial T}}\,=\,\sum\limits_{N_{2}}\,Z_{1}\of{N_{2},N_{4}}\e^{\kappa_{2}\of{N_{2}-10}}\label{eq:internalpartf}
\]
is the canonical partition function for the interior of triangulations at $\kappa_{2}$ that consist of $N_{4}$ 4-simplices and possess the boundary of a 4-simplex.\\
The reason for subtracting the whole boundary from the action in \eqref{eq:internalpartf} is, that these terms are already taken into account in $Z_{0}\of{\kappa_{2},M}$ and we want to avoid over-counting\footnote{Including the boundary terms as in \eqref{eq:z1canonicalpartf} into the action for the elementary building-blocks does not work as now at least three boundaries (not just two) meet at each boundary-triangle. The boundary action of an elementary building-block $T_i$ is therefore given by $S\of{\partial T_i}\,=\,\kappa_{2}\,\sum_{\Delta\in\partial T_i}\frac{1}{n\of{\Delta}}$, where $\Delta$ runs through all triangles in $\partial T_i$ and $n\of{\Delta}$ is the number of boundaries which contain $\Delta$. The $p\of{\kappa_{2},N_{4}}$ would then depend (through the $n\of{\Delta}$) on the connectivity of the base manifold, which is highly undesirable.}.

After having written the EDT partition function in the generalized form \eqref{eq:edtbinbpartf}, i.e. in terms of a base manifold and its elementary volume elements, which can be excited to form "baby universes", some comments are in order:
\begin{enumerate}
\item The terminology "base-manifold" and "elementary building blocks" already suggests that we would like to look at the triangulations, observed in EDT simulations, in a slightly non-standard way. The main reasons for such a re-interpretation are the following:
\begin{itemize}
\item it seems that the base-manifold, with all elementary volume elements in the "ground state" (such that they are just ordinary 4-simplices), can be mapped on a corresponding Lorentzian or causal triangulation,
\item although the boundaries of the elementary volume elements are always minimal, their volume can now change in a discrete manner. This makes EDT to fit a little better into the quantum gravity picture provided by spin-foam models.   
\end{itemize}     
\item The altered physical interpretation suggests, that the thermodynamic limit should be taken by sending $M$, the number of elementary building blocks of the base manifold, to infinity instead of (just) $N_{4}$\footnote{As each elementary building block of the base-manifold contains at least one 4-simplex, $M\rightarrow\infty$ implies $N_{4}\rightarrow\infty$, but the converse is not true.}.
\item According to \cite{Smit}, the phase transition is associated with a change of sign in the effective curvature. In the crumpled phase, the base-manifold (or "mother universe") has negative curvature: there are two \emph{singular vertices}, which could be seen as the centres of two hyperbolic 4-balls (each of them formed by many 4-simplices that are all incident to the same central vertex) that are glued along their boundary to form a topological 4-sphere. Without a term in the action that prevents the base-manifold from shrinking, it seems to be favourable for a triangulation to collapse into a baby-universe tree as soon as the singular vertices disappear. Running simulations at quasi fixed $M$ instead of quasi fixed $N_{4}$ (but with $\kappa_{4}$ sufficiently large), or at a non-zero value of the new coupling $\kappa$, would prevent the triangulation from such a collapse. The base-manifold should then survive the disappearance of the singular vertices and develop a positive effective curvature itself (instead of generating the positive effective curvature by producing many small bubbles), which would give rise to a new phase and a new phase transition that could be of higher than $\ford$.  
\item For finite systems, the role played by the new coupling $\kappa$ in \eqref{eq:edtbinbpartf} is related to that of the anisotropy factor in CDT as $\kappa$ affects the ratio of the average diameter\footnote{The average diameter of an elementary building-block (i.e. the average time needed to pass through it) consisting of $N_{4}$ 4-simplices and $N_{2}$ triangles (counting also the ones in the boundary) can be determined by measuring the average return time of a random walk in systems with $\of{N_{4}+1}$ 4-simplices and $N_{2}$ triangles, according to \eqref{eq:withboundarypartf}.} ($\sim$ average time needed to pass through) and volume of the elementary building-blocks of the base manifold.
\end{enumerate}

In a follow-up paper we will try to verify the above assumptions and study the properties of \eqref{eq:edtbinbpartf} in more detail. An interesting question is of course whether for some values of $\kappa$, \eqref{eq:edtbinbpartf} yields a $2^{\text{nd}}$ or higher order transition in $\kappa_{2}$ (or the fixed $\rho=N_{4}/M$, or fixed $M$ version of \eqref{eq:edtbinbpartf}) and if, when integrating out the volume fluctuations of the elementary building blocks of the base manifold, $\kappa$ and $\kappa_{4}$ can be combined to yield a kind of effective cosmological constant, such that one recovers the form of the original Euclidean Einstein-Regge action. Alternatively one could interpret the additional weight in \eqref{eq:edtbinbpartf} as a measure term.\\

\section{Conclusion}
Our study confirms the qualitative findings of \cite{Bialas,deBakker}: for  $\kappa_{2}\,\approx\,\kappa_{2}^{pcr}\of{N_{4}}$ we find for $N_{4}\,\geq\,32$k a clear double peak structure in the $N_{2}$ distribution, which becomes more pronounced with increasing system size (and there is no sign that the two peaks will eventually merge again in the thermodynamic limit). This is characteristic of a weak $\ford$ transition. A finite size scaling analysis of the $\nord{4}$ Binder cumulant of the $N_{2}$ distribution confirms this further.\\
As the phase transition is $\ford$, finite systems should allow for coexisting phases in a neighbourhood of the pseudo-critical point. It turned out to be difficult to give a precise criterion to distinguish "locally" between the two phases but a candidate could be that bubbles with a \emph{neck density} $\rho_{necks} > \rho_{necks}^{cr}$, can be considered as corresponding to the elongated phase, where $\rho_{necks}^{cr}$ is not yet known exactly but seems to be around $1/9$. Bubbles with $\rho_{necks} < \rho_{necks}^{cr}$ would then correspond to the crumpled phase.\\
Finally we proposed a new correspondence between the EDT and "balls in boxes" models which leads to a generalization of the EDT partition function (with an additional parameter $\kappa$) and a modified interpretation of triangulations contributing to the EDT partition sum in terms of a largest bubble or "mother universe" and its elementary building blocks, which can undergo volume excitations such that their interior could also be interpreted as a baby-universe branch. The additional coupling $\kappa$ enriches the phase structure of the model which could now possibly contain a $\sord$ phase transition line.\\
In the Appendix, we propose and motivate a change in the EDT path-integral measure which introduces tunable parameters ($r_n$). For appropriate choices of the $r_n$, the order of the phase transition of the ordinary EDT model might also change to second order.

\section{Acknowledgement}
We thank J. Smit, A. Goerlich and J. Jurkewicz for helpful discussions.

\newpage
\setcounter{arabicPagenumber}{\value{page}} 
\pagenumbering{Roman}
\setcounter{page}{\theromanPagenumber}
\appendix
\section{Appendix: Geometric Probabilities and Path-Integral Measure}\label{ssec:modmeasure}

It has recently been suggested \cite{Ambjorn2} that the $\ford$ transition of the EDT model could perhaps be changed into a $\sord$ transition by a change of the measure in the partition sum \eqref{eq:edtpf}. However, none of these attempts has proved successful. Here we motivate and derive a new proposal for a measure that could have the desired properties.\\ 

\begin{figure}[H]
\centering
\begin{minipage}[t]{0.495\linewidth}
\centering
\includegraphics[width=\linewidth]{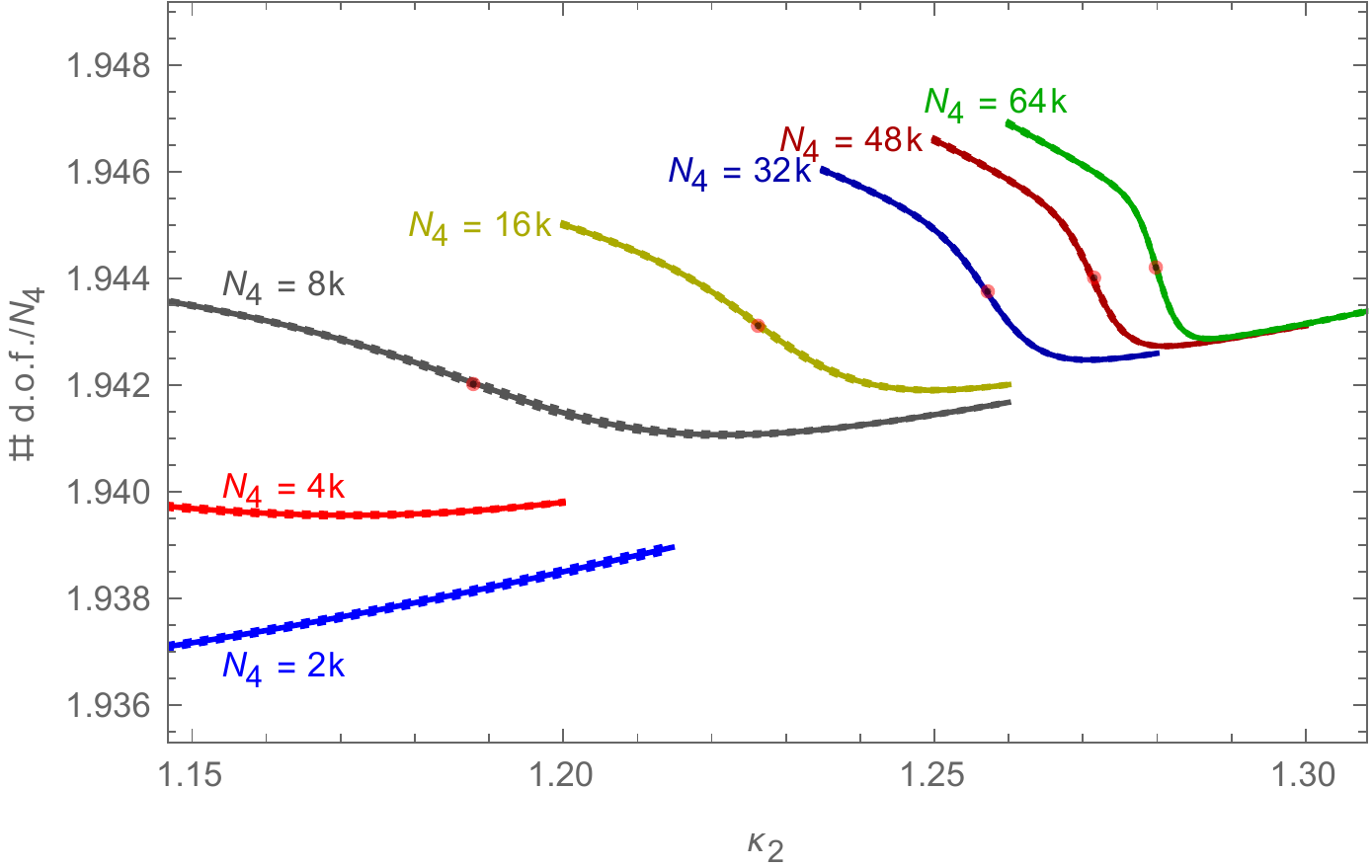}
\end{minipage}\hfill
\begin{minipage}[t]{0.49\linewidth}
\centering
\includegraphics[width=\linewidth]{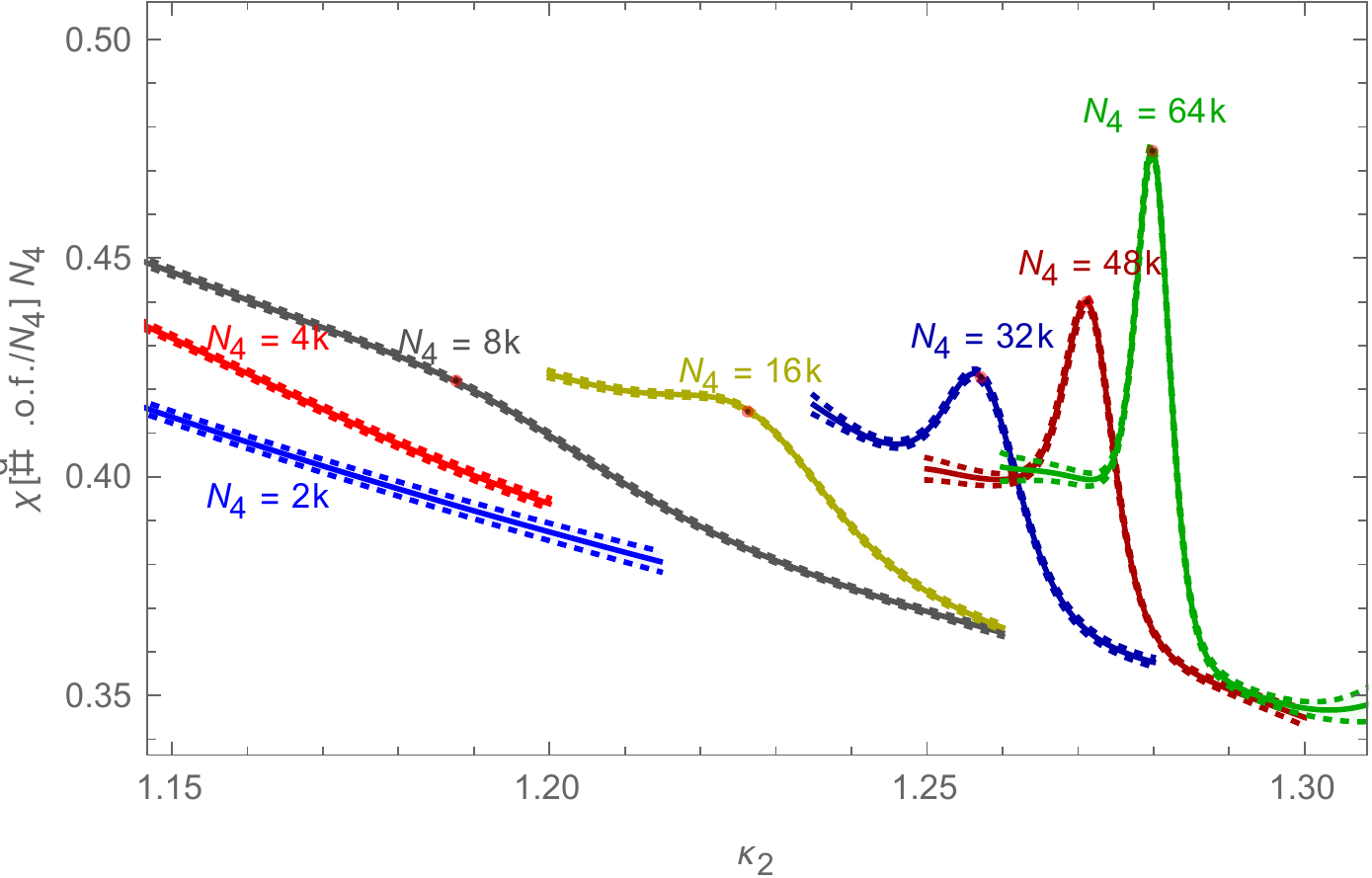}
\end{minipage}
\caption{Average total number of degrees of freedom (i.e. number of possible moves) per volume (left) and the corresponding susceptibility (right) as functions of $\kappa_{2}$ for different system sizes $N_{4}=2,\ldots,64$k. The red dot on each graph marks the corresponding pseudo-critical point/value (i.e. where $\kappa_{2}=\kappa_{2}^{pcr}\of{N_4}$, corresponding to the peak in the $N_{2}$-susceptibility). The simulation was performed with the update scheme described in Sec. \ref{ssec:detailedbalance}.}
\label{fig:avdofs}
\end{figure}

The measure that is normally used is the trivial one for which we have
\[
\rho\of{T}\,=\,\frac{\e^{\kappa_{2}N_{2}\of{T}-\kappa_{4}N_{4}\of{T}}}{C_{T}}\,\approx\,\e^{\kappa_{2}N_{2}\of{T}-\kappa_{4}N_{4}\of{T}}
\]
in the detailed balance equation \eqref{eq:detailedbalance}. But we could also introduce a measure $z\of{T}$ on the space of possible triangulations $T$ such that we would have
\[
\rho\of{T}\,=\,\frac{z\of{T}\,\e^{\kappa_{2}N_{2}\of{T}-\kappa_{4}N_{4}\of{T}}}{C_{T}}\,\approx\,z\of{T}\,\e^{\kappa_{2}N_{2}\of{T}-\kappa_{4}N_{4}\of{T}}.\label{eq:moddetailedbalance}
\]  
To motivate a particular form of the measure $z\of{T}$, assume we are currently in a triangulation $T$ which possesses $f_{n}\of{T}$ locations where a $n$-move could be applied. This means that $T$ has $\scriptstyle\sum\limits_{n=0}^{4}\,\textstyle f_{n}\of{T}$ "neighboring" triangulations. Now think of each location in $T$ where a move can be applied as something similar to a site in an Ising spin system where a spin-flip can occur. But in our case, the "spin-flip" consists of the application of a Pachner-move, e.g. a Pachner $n$-move that flips a piece of triangulation, spanned by $\of{5-n}$ 4-simplices into one that is spanned by $\of{n+1}$ 4-simplices but has the same boundary. As the regions where different moves are possible can overlap, the flip of one region will in general destroy some of the other regions where flips were possible and instead create new ones. We therefore have a fluctuating number of degrees of freedom (see Fig. \ref{fig:avdofs}) and the system is much more involved than an Ising system. 
Nevertheless, one could argue that, as long as the couplings $\kappa_{2}$ and $\kappa_{4}$ are turned off, all possible moves in a triangulation $T$ should be considered as equally likely, just as in the Ising case. More generally: one could assign different probabilities $r_{n}$ to different move types $n\in\cof{0,\ldots,4}$, as long as $r_{n}=r_{4-n}$. For example, one could choose $r_{n}$ proportional to the local volume of a $n$-simplex\footnote{As mentioned earlier: the local volume of a $n$-simplex is the volume of all points which are closer to this $n$-simplex than to any other $n$-simplex.} that allows for a $n$-move (the local volume is the same for $n$ and $\of{4-n}$-simplices which allow for a move).\\
Having such fixed probabilities for the different moves, implies that triangulations with different numbers of locations where moves could be applied, are not equally likely. The corresponding probability weight for a triangulation $T$ can be derived from the balance equation
\[
z\of{T}\,=\,\sum\limits_{n=0}^{4}\sum\limits_{T'\,\in\,\text{nbr}_{n}\of{T}}\,\frac{z\of{T}\,r_{n}}{\sum\limits_{m} r_{m}\,f_{m}\of{T}}\,=\,\sum\limits_{n=0}^{4}\sum\limits_{T'\,\in\,\text{nbr}_{n}\of{T}}\,\frac{z\of{T'}\,r_{4-n}}{\sum\limits_{m} r_{m}\,f_{m}\of{T'}},\label{eq:configbalance}
\]
where $\text{nbr}_{n}\of{T}$ is the set of all triangulations that can be obtained from $T$ by a $n$-move. The detailed balance equation corresponding to \eqref{eq:configbalance} reads
\[
\frac{z\of{T}\,r_{n}}{\sum\limits_{n=0}^{4}\,r_{n}\,f_{n}\of{T}}\,=\,\frac{z\of{T'}\,r_{4-n}}{\sum\limits_{n=0}^{4}\,r_{n}\,f_{n}\of{T'}},
\]
which is obviously satisfied if we have
\[
z\of{T}\,\propto\,\sum\limits_{n=0}^{4}\,r_{n}\,f_{n}\of{T}\label{eq:nmeasure}.
\]
Such measures are particularly simple to implement: by choosing move candidates according to the \emph{selection probability},
\[
p_{n}^{\of{sl}}\of{T}\,=\,\frac{r_{n}}{\sum\limits_{m=0}^{4}\,r_{m}\,f_{m}\of{T}},
\] 
the measure term drops out of the detailed balance equation for the (reduced) transition probabilities $p_{n}\of{T}$:
\[
\rho\of{T}\,p_{n}^{\of{sl}}\of{T}\,p_{n}\of{T}\,=\,\rho\of{T'}\,p_{4-n}^{\of{sl}}\of{T'}\,p_{4-n}\of{T'},
\]
where $\rho\of{T}$ is given by \eqref{eq:moddetailedbalance}, such that there is no need to determine the possible moves of the candidate configuration.

\end{document}